\documentclass[11pt,twoside]{book} 
\usepackage{asp2008n}
\usepackage{times}
\usepackage{lscape}
\usepackage{epsf}

\usepackage{graphicx}
\usepackage{amssymb}
\usepackage{longtable}
\usepackage[figuresright]{rotating}
\usepackage{multirow}

\newcommand{\simless}{\mathbin{\lower 3pt\hbox {$\rlap{\raise 5pt\hbox{$\char'074$}}\mathchar"7218$}}}

\mainmatter

\newlength{\deftabcolsep}
\begin{document}

\title{Star Formation in the Orion Nebula I: Stellar Content}

\author{August Muench}
\affil{Harvard-Smithsonian Center for Astrophysics\\
60 Garden Street, Cambridge, MA 02138, USA}

\author{Konstantin Getman}
\affil{Department of Astronomy and Astrophysics, Pennsylvania State University, \\
525 Davey Laboratory, University Park, PA 16802, USA}

\author{Lynne Hillenbrand}
\affil{Department of Astronomy, California Institute of Technology, \\
Mail Code 105-24, Pasadena, CA 91125, USA}

\author{Thomas Preibisch}
\affil{Max-Planck-Institut f\"ur Radioastronomie, Auf~dem~H\"ugel 69, \\
D--53121 Bonn, Germany}
\affil{Universit\"ats-Sternwarte M\"unchen,
Scheinerstr.~1, D-81679 M\"unchen, Germany}

\begin{abstract}
The Orion Nebula is one of the most frequently observed nearby
($<1$~kiloparsec) star forming regions and, consequently, the subject
of a large bibliography of observations and interpretation.  The
summary in this chapter is bounded spatially by the blister
H\textsc{II} region, with sources beyond the central nebula that are
part of the same dynamical clustering covered in other chapters in
this book.  Herein are discussed panchromatic observations of the
massive OB stars, the general T~Tauri population, the sub-stellar
sources and variable stars within the Orion Nebula.  First, a brief
history of 400 years of observation of the Nebula is presented. As
this history is marked clearly by revelations provided in each age of
new technology, recent ultra-deep X-ray surveys and high resolution
multi-epoch monitoring of massive binary systems and radio stars
receive special attention in this review. Topics discussed include the
kinematics, multiplicity, mass distribution, rotation, and
circumstellar characteristics of the pre-main sequence
population. Also treated in depth are historical and current
constraints on the distance to the Orion Nebula Cluster; a long
standing 10-20\% uncertainty has only recently begun to converge on a
value near $\sim400$ parsecs.  Complementing the current review of the
stellar population is a companion chapter reviewing the molecular
cloud, ionized H\textsc{II} region and the youngest protostellar
sources.
\end{abstract}



\begin{figure}
	[ht!]

	\centering
	\includegraphics[angle=-90, totalheight=0.85\textheight]{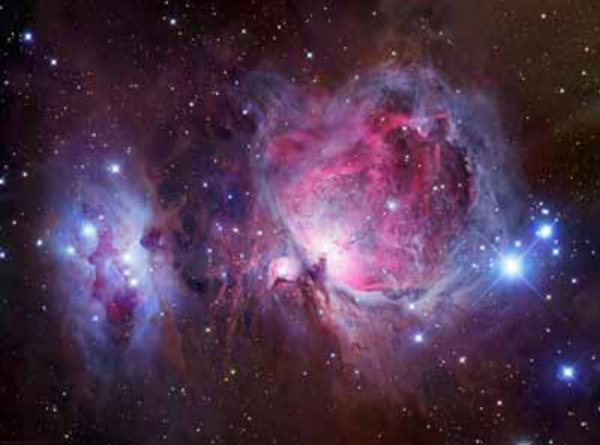}
 \caption{Color composite image of the northern {Orion Molecular
 Cloud.} This ground based image is oriented with equatorial
	North up and East to the left and encompasses the central area
	of the young star distribution shown in Figure
	\ref{fig:omc_stars}. Image courtesy Robert
	Gendler. \label{fig:onc_optical}}

\end{figure}%
\subsection*{Introduction}
 An interesting hypothesis drawn from our knowledge about the {Orion
 Nebula} is that $50,000$ years ago it was invisible to the naked
eye. The ionizing photons of the massive O and B type stars, whose
 projected arrangement yield the namesake {Trapezium,} had not yet burned
away the layers of natal molecular gas out of which they had formed.
 While bright blue stars were visible along the {Sword} of Orion, having
formed continually over the previous few million years, there were, on
a scale perhaps much grander than the present, thousands of smaller
stars hidden from view. In the relatively short interim a blister
H\textsc{II} region created by a newborn $40,000$K O type star
expanded into the molecular cloud, uncovering a large portion of this
embedded star clustering. Nonetheless, star formation continues
vigorously in the remaining molecular cloud today.

Because of the richness of this star clustering $(N_{\star}>2000)$ and
 its relative proximity $(\sim400\mbox{pc})$, the {Orion Nebula} is easily
the most frequently observed nearby ($<1$~kpc) star forming region,
providing an large bibliography of observations and
interpretations\footnote{The introductory sketch of the recent history
 for the {Orion Nebula} is but one hypothesis taken from the breadth of
observational and theoretical studies of the Nebula. A more complete
summary of such models for this history are presented in a companion
 chapter (O'Dell et~al).}. Moreover, the properties of the {Orion
 Nebula} stars, e.g., their masses, evolutionary status, spatial and
velocity distributions, outflows and circumstellar disk properties,
all provide critical tests for theories of molecular cloud evolution
and star formation.

\subsection*{Region Overview}

 Inspecting a magnificent modern large scale visual image of the {Orion
 Nebula} (Figure~\ref{fig:onc_optical}) reveals the major physical
features of this region. From North to South there are a series of
bright emission nebulae, interspersed with dark bands and small
 clusterings of bright stars. {The Orion Nebula H\textsc{II}}
region\footnote{Additional common catalog entries for this region
 include {Messier~42,} {NGC~1976.}} is central to this image and appears to
expand to the southwest from an apex at the location of the O and B
stars.  These apparent alternating nebulae and clusterings have led to
a system of names or designations with boundaries that deserve some
explanation.
\begin{table}
	[!ht]
   \caption{Subregion names in the northern Orion A molecular
     cloud. Papers included: \citet{1954TrSht..25....1P};
     \citet{1964ARA&A...2..213B}; \citet[][M\&L]{1966VA......8...83M};
     \citet{1969ApJ...155..447W}; \citet[][W\&H]{1978ApJS...36..497W};
     \citet[][G\&L]{1998AJ....115.1524G}. Note that the G\&L98
     subclusters F and G, which are outside the immediate field of
     study, appear to correspond to the LDN 1641N and NGC 1999 star
     forming regions, respectively, although the G\&L98 H$\alpha$
     subclusters are shifted $\sim10\arcmin$ to the West. These
     regions are discussed further the chapter by Allen \&
     Davis.\label{tab:omc_subc}} \smallskip
	\begin{center}
		{\small
		\begin{tabular}
			{lcccccc}
			\tableline \noalign{\smallskip} Adopted Name & Parenago & Blaauw & M \& L & Walker & W \& H & G \& L\\
			& 1954 & 1964 & 1966 & 1969 & 1978 & 1998 \\

			\noalign{\smallskip}
			\tableline \noalign{\smallskip}

			Upper Sword & I & Orion Ic &Upper Sword & Group 1 & C1 & A? \\
			NGC 1977 & II & Orion Ic & --- & Group 2 & C2 & A \\
			OMC 2/3 & III & Orion Ic & --- & Group 3 & C3 & B \\
			ONC & IV & Orion Id & --- & Group 4 & D/D1 & C,D \\
			$\iota$ Ori & V & Orion Ic & $\iota$ Ori & Group 5 & C4 & E \\

			\noalign{\smallskip}
			\tableline \noalign{\smallskip}
		\end{tabular}

		}
	\end{center}
\end{table}

 As listed in Table~\ref{tab:omc_subc} the stars along the {Sword} of
Orion have traditionally been segregated into 4 or 5 regions; as shown
in the tabulation the ``names'' for these regions have changed over
time although no new divisions have been made since the 1950s.
Subsequently, the numeral ordering (I,II,...) from Parenago (1954) is
adopted and paired with more descriptive names, e.g., Region II
 corresponds to the {NGC~1977 H\textsc{II}} region.

\begin{figure}
	[t*]

	\centering
	\includegraphics[angle=0, totalheight=4.3in]{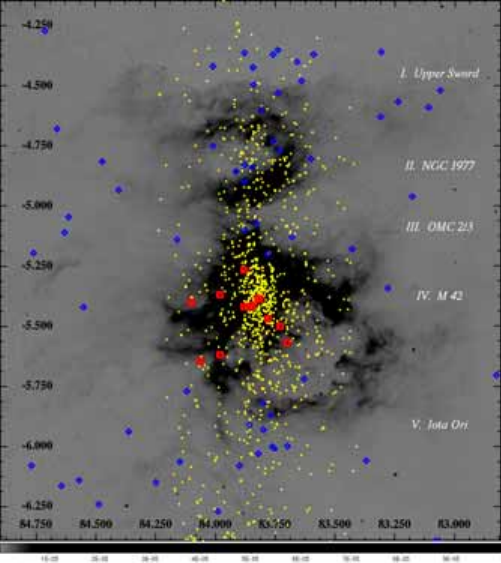}

 \caption{Young star distribution along the {Orion Molecular
 Cloud.} The variable infrared stars identified by
	\citet{2001AJ....121.3160C}(yellow circles) are compared to
 the OB members of the {Orion Ic} (blue diamonds) and Id (red
	squares) \citep{1996PhDT........78B}. The reverse grayscale
	image is the MSX band A (8 micron) image from
	\citet{2003AJ....126.1423K}. Labels (I.,II.,...) mark the five
	subregions of Parenago (Table~\ref{tab:omc_subc}) and their
	more descriptive names. The equatorial coordinate axis
	(decimal degrees) are in equinox J2000. This chapter focuses
 on the stars forming in region IV, the {Orion Nebula.}
	\label{fig:omc_stars}}
\end{figure}

While at optical wavelengths these features seem apparent (and perhaps
 some such as {NGC~1977} are significant) when one views the {Sword} of
Orion at near-IR wavelengths $(2\,\micron)$, which are much less
sensitive to variations in extinction, no boundaries are apparent
between these regions. Figure \ref{fig:omc_stars} presents a map
similar in extent to the previous optical image but where the Nebula
is traced by mid-IR $(8\;\mu\mbox{m})$ emission as observed by the
Midcourse Space Experiment (MSX) \citep{2003AJ....126.1423K}. Infrared
variable stars \citep{2001AJ....121.3160C} extend across all these
groups without clearly demarcating any of them except for the strong
 concentration of sources at the apex of the {Orion Nebula.} Another set
 of frequently used designations concern the OB stars of the {Sword;}
\citet{1964ARA&A...2..213B} segregated the Sword OB stars in the Ic
and Id associations, ordered in part by apparent youth, while
\citet[][]{1978ApJS...36..497W} expanded the Ic association on the sky
but segregated the Sword OB stars into subregions (C1, C2, C3, C4),
which are referred to in this review by the collective designation
Ic*.  While the Orion Id OB ``members'' coincide with the stellar
density maximum in Figure \ref{fig:omc_stars}, the spatial division of
Ic* and Id members appears rather arbitrary.

 Another naming system deserves clarification. The terms {``Trapezium''}
cluster \citep{1931PASP...43..255T}, which
refers to stars immediately
surrounding the asterism that is the arrangement of 4 bright OB stars
 in the center of the Nebula, and {``Orion Nebula Cluster} {(ONC),''} which
dates to \citet{1953ApJ...117...73H}, suggest perhaps that these are
separate entities. This survey of the literature does not reveal any
 physical reason to suppose that the {Trapezium} stars represent anything
distinct about star formation in the Nebula beyond their mass.  While
the study by \citet{1997AJ....113.1733H} arbitrarily divided the
region into three ``radial zones:''
 {Trapezium Cluster} (r$<0.3$ parsec);
 {ONC} (r$<3$~parsec); Orion Ic association, only slight age
gradients between them were found \citep[see
also][]{2004AJ....128..787R}.  It is therefore a secure inference that
the entire region is a single contiguous star forming event that
requires complete description.

\subsection*{Topical Scope of the Review}

This chapter summarizes current knowledge regarding those stars which
 sit within and surround the {Orion Nebula H\textsc{II}} region. A
review of the cold molecular cloud, the hot H\textsc{II} region and
the characteristics of the very youngest stars is presented in a
subsequent chapter (O'Dell et~al.). Further outlying regions in Figure
 ~\ref{fig:omc_stars} e.g. to the North {(OMC 2-3;} {NGC~1977,}
 Upper~Sword) and South {(LDN~1641)} of the Nebula are discussed in other
chapters in this volume (i.e., Peterson \& Megeath and Allen \& Davis,
respectively).  The focus here is on those surveys that provide
constraints on the physical properties of the revealed high mass,
T~Tauri and substellar objects.

The review is organized as follows. First, a brief history of
 important or previously broad reviews of research on the {Orion Nebula}
is presented (Sect.~\ref{sec:history}), followed by an overview of
distance determinations to the Nebula in the past 70 years
(Sect.~\ref{sec:distance}). Individual sections are reserved for
reviewing the T~Tauri members with a focus on summarizing the many
broadband CCD and spectroscopic studies that have occurred during the
past two decades (Sect.~\ref{sec:ttauri}), the O and B type members of
 the Nebula with detailed reviews of each of the {Trapezium} stars
(Sect.~\ref{sec:ob}), and the variable stars
(Sect.~\ref{sec:variables}).

\section{History of Study of the Orion Nebula} \label{sec:history}

\begin{figure}[t]
	\centering
	\includegraphics[angle=135,totalheight=3in]{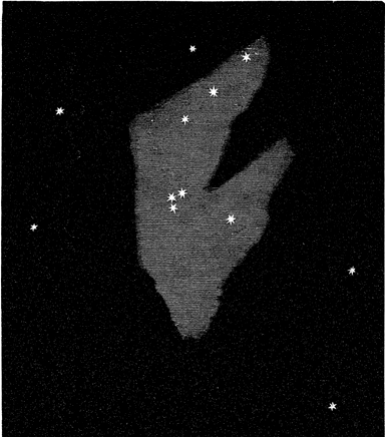}

 \caption{The 1656 sketch of the inner {Orion Nebula} by Huyghens. The figure has been re-oriented from its publication form such that North is up and East is to the left. Image is reproduced from \citet{1882USNOM..18A...1H}.}
	\label{fig:Figures_huyghens}
\end{figure}

During the fifty years after the development of the telescope in 1608
 the nebular nature and stellar content of {Orion Nebula} were
independently discovered by a handful of observers.  The observing
logs of Nicolas-Claude Fabri de Peiresc (1610) and of Johann Baptist
Cysat (at the latest 1618)
~\citep{1854AN.....38..109W,1882USNOM..18A...1H} represent the
 earliest written records that the {Sword} of Orion contained a ``fog''
or ``milky nebulosity,'' which here borrows the words used by
W. Herschel (1802) to describe the region.  The first hand drawn
charts of the region include those of Galileo (1617), who did not
distinguish the Nebula, of Giovanni Battista Hodierna (drawn sometime
before 1654), who did, and the more famous 1656 drawing by Huyghens,
which is the most widely known. His accurate rendition of the central
 nebula surrounding the {Trapezium} provides an origin for the term
``Huyghenian region'' (Figure \ref{fig:Figures_huyghens}).%

\begin{figure}
	[ht*]

	\centering
	\includegraphics[angle=180, totalheight=3.5in]{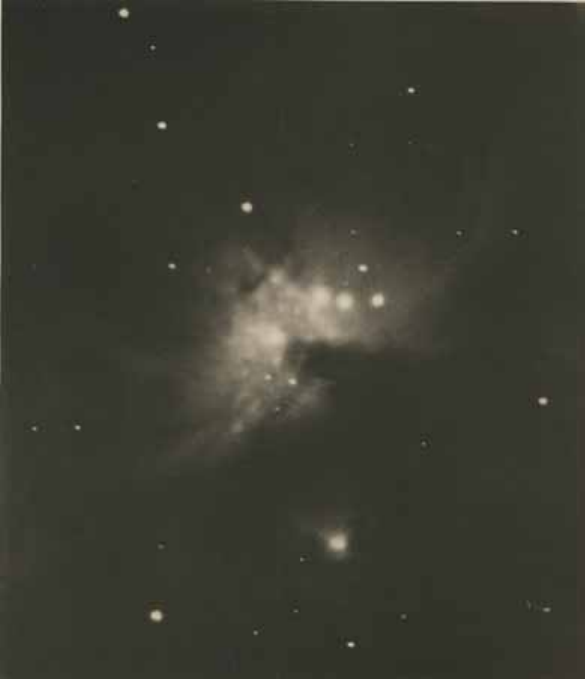}

 \caption{Draper's 1882 photographic image of the {Orion
 Nebula.} North is up and East is to the left. An approximately
	2 hour visible photograph obtained by H. Draper with the 11
	inch Clark telescope. An excerpt from Edward S. Holden's
	``Monograph of the Central Parts of the Nebula of Orion''
	(Washington 1882) p.226: \textit{''The first photograph of the
	nebula of Orion was made by Dr. Henry Draper in September,
	1880, and the unavoidable delay which has occurred in printing
	the present memoir enables me to include an account of the
	astonishing results which he has attained. A wood-cut which I
	had prepared from his first photograph was found to be so
	unsatisfactory that Dr. Draper most generously offered to
	supply the necessary photolitographic reproductions of his
	last negative (taken March 14, 1882) to accompany the brief
	account I had prepared. The full page photolitograph is here
	given...''} \label{fig:onc_draper}}

\end{figure}

A nearly complete journal of 273 years of telescope aided visual
observation of the Nebula is provided in Edward S. Holden's
\textit{The Monograph of the Central Parts of the Nebula of Orion}
(1882). Holden's monograph includes the reproduction of dozens of hand
drawn sketches as well as observing logs each in their original
language.  Variation in the reproduction of the Nebula is
remarkable. The first photographic plates of Orion made by Henry
Draper between 1880 and 1882 were included as an addendum to Holden's
work as well as a discussion of the processes of obtaining these
images.  Figure \ref{fig:onc_draper} is a reproduction of that image
and is captioned with Holden's description.

The next $\sim100$ years of photographic observation of the Nebula
included quantitative studies of its variable stars, the discovery of
a cluster of faint stars in the Nebula's core, very broad censuses and
an expansion in the role of the Nebula as a testbed for new observing
techniques.  Numerous variable star studies were performed using the
Harvard Plates by H. Leavitt (published by Pickering) or confirmed by
M. Applegate (published by Shapley).  In the 1930s, deep red
photographic plates revealed that the Nebula contained a substantial
cluster of fainter stars in addition to the brightest members
\citep{1931PASP...43..255T,1937ApJ....86..119B}.  The broad surveys of
\citet{1935POLyo...1...12B} and \citet{1954TrSht..25....1P} provided
excellent photographic updates to the ~\citet{1867AnHar...5....1B}
visual census of stars in the Nebula.

Parenago, a Russian astronomer, published a major, lengthy analysis of
the region in 1954. A translation of sections of this publication was
undertaken for this review and it indicates that he relied on prior
and concurrent work of female Russian astronomers (e.g., Barkahatova,
Uranova, Kirillova) some of whose names do not reappear in the
literature outside of his book.  His analysis extended to the topics
of parallax, proper motions, astrometry and photometry, covering all
 of the {Sword} of Orion. Most important he found a clear evidence for a
``cloud'' of members lying above the main sequence. Perhaps because of
the Cold War and the lack of translations from his work from Russian
to other languages, his work is extremely poorly cited in the
literature. The disparaging of his work by \citet{1956ApJS....2..365W}
was not unnoticed by the author as revealed in the posthumous
publication, \citet{1961AJ.....66..103P}. Nevertheless, this work was
not considered seriously in most subsequent studies, not appearing,
for example, in the otherwise meticulous study by Goudis (see
below). Even today this significant work has garnered a mere 59
citations\footnote{Derived via the NASA Astrophysical Data Service
circa May 2008 by merging the results of citations to
\citet{1954TrSht..25....1P} and \citet{1954TrSht..25....3P}.}. His
valuable data tables were converted to machine readable formats by
\citet{1992BICDS..40...13M} and ingested into an electronic format in
1997 \citep{1997yCat.2171....0P}.

In 1982, approximately one hundred years after the Holden monograph
and the first photographic images by Draper, two useful summarizing
publications appeared. C. Goudis's \textit{The Orion Complex: A Case
Study of Interstellar Matter,} focused on the structure and nature of
the Nebula itself, details previous approaches to studying the Nebula
and its content, including infrared, radio and spectroscopy, and
provides useful tables of past observations and results. Second, a
 conference was held on the {Orion Nebula} and honoring Henry Draper
\citep{1982NYASA.395.....G}; the conference proceedings include 32
articles on all current aspects of study of the Nebula and records of
the participants discussion about each contribution. In addition there
are a number of articles that provide a history, more detailed than
that of Holden, about Draper's photographic work obtaining these
images of the Nebula as well as his subsequent scientific legacy after
his death in late 1882.

Reviews of the literature over the subsequent $\sim20$~years of CCD
observations of the Nebula include \citet{1989ARA&A..27...41G} and
\citet{2001ARA&A..39...99O}.  The Nebula was also included in a review
that encompassed all of the Orion star forming region
by~\citet{1991lmsf.book....1B}. Its recent study has also been the
focus of a book, \citet{2003onws.book.....O}.

\defcitealias{1918ApJ....47..104K}{a}
\defcitealias{1918ApJ....47..146K}{b}
\defcitealias{1918ApJ....47..255K}{c}
\begin{table}[p]

    \caption{Summary of published distances to the Orion Id
      association. Distances are segregated as corresponding to the Ic
      or Id associations with the notation Ic* used to indicate that
      part of Ic right around Id and to distinguish it from the much
      larger Ic as defined by WH78. Unfortunately the notation Ic* was
      also used by many authors before the Ia,b,c,d subgroups were
      defined to indicate the entire ``Sword'' region; thus some
      ambiguities may remain.  \label{tab:onc_dist} } \smallskip

    \begin{center}
        {\tiny
        \begin{tabular}
            {l@{\hskip5pt}l@{\hskip5pt}c@{\hskip5pt}c@{\hskip5pt}c@{\hskip5pt}c@{\hskip5pt}c@{\hskip5pt}c@{\hskip5pt}c}

            \tableline \noalign{\smallskip} Author Name & Year of & Region & Distance & Error & Stellar & Method & Data & Number \\
            & Pub. & Desig. & Modulus & & Types &   &   & of Stars\\
            \noalign{\smallskip}
            \tableline \noalign{\smallskip}

            \citeauthor{1917HarCi.205....1P} & 1917 & Id+Ic* & 11.5 & & B3 & ZAMS & pg & ? \\
            \citeauthor{1918ApJ....47..104K} & 1918\citetalias{1918ApJ....47..104K,1918ApJ....47..146K,1918ApJ....47..255K} & Ic & 6.34 & & B & PM & pg & ? \\
            \citeauthor{1919PASP...31...86P} & 1919 & Id+Ic* & 8.5 & & B3 & ZAMS & pg & ? \\
            \citeauthor{1929PUAms...2....1P} & 1929 & all & 7.6 & & B & ZAMS & pg & ? \\
            \citeauthor{1931PASP...43..255T} & 1931 & Id & 8.5 & & O9-A2 & ZAMS & pg & 17 \\
            \citeauthor{1946PASP...58..356M} & 1946 & Id & 7.38 & & O7-B1 & ZAMS & pg & 3 \\
            \citeauthor{1949AJ.....54..111M} & 1949 & Id & 8.58 & 0.35 & B1-B3 & ZAMS & pg & 17 \\
            \citeauthor{1952ApJ...116..251S} & 1952,4& Id+Ic & 8.5 & 0.30 & B Stars & ZAMS & pe & 190 \\
            $\vdots$ & 1952 & Id & 8.6 & 0.30 & B Stars & ZAMS & & ? \\
            \citeauthor{1954TrSht..25....1P} & 1954 & Id & 8.0 & & & ZAMS & pg & ? \\
            \citeauthor{1956ApJ...123..267J} & 1956 & Id & 8.0 & & B8-A0 & ZAMS & & ? \\
            \citeauthor{1958ApJ...128...14S} & 1958 & Id & 8.6 & & O6-K2? & PM / RV& plate & 20 \\
            \citeauthor{1962ApJ...136..767S} & 1962 & Id & 8.2 & & B Stars & & $UBV,H\gamma$ & 180 \\
            \citeauthor{1964BAN....17..358B} & 1964 & Id & 8.33 & 0.11 & & ZAMS & 7-filter & 5 \\
            \citeauthor{1965ApJ...142..964J} & 1965 & Id & 7.9 & & & PM / RV & & 21 \\
            \citeauthor{1966VA......8...83M} & 1966 & Ic* & 8.1 & & & ZAMS & & ? \\
            \citeauthor{1968ApJ...152..905L} & 1968 & Ic* & 8.5 & 0.1 & & ZAMS & & 14 \\
            \citeauthor{1969ApJ...155..447W} & 1969 & Id+Ic* & 8.37 & 0.05 & B2-B9 & ZAMS & $UBV$ & 51 \\
            \citeauthor{1973ApJ...183..505P} & 1973 & Id & 7.8 & 0.15 & B stars & ZAMS & $BV$ & 15 \\
            \citeauthor{1975MNRAS.171..219P} & 1975 & Id & 8.1 & 0.13 & B stars & ZAMS & $BV$ & ? \\
            $\vdots$ & 1975 & Id & 7.71 & 0.21 & B stars & ZAMS & $VI$ & ? \\
            $\vdots$ & 1975 & Id & 7.98 & 0.12 & B stars & ZAMS & $V$;SpT & ? \\
            \citeauthor{1977ApJS...34..115W} & 1977 & Id & 8.42 & 0.53 & B stars & & $UBV;ubvy;H_{\beta}$ & 6 \\
            $\vdots$ & 1977 & Ic* & 8.16 & 0.49 & B stars & & $UBV;ubvy;H_{\beta}$ & 44 \\
            \protect{\citeauthor{1981A&AS...44..467M}} & 1981 & Id+Ic* & 8.20 & 0.15 & & ZAMS? & $UBV?$ & ? \\
            \citeauthor{1981ApJ...244..884G} & 1981 & K-L region& 8.41 & 0.40 & H$_2$0 masers&stat. parallax & VLBI $pm+rv$ & $\sim 30$ \\
            \citeauthor{1981ApJ...248..963B} & 1981 & Id+Ic & 8.0 & 0.5 & & ZAMS & $BV$ & ? \\
            \citeauthor{1982AJ.....87.1213A} & 1982 & Ic & 7.87 & 0.09 & B stars & ZAMS & $M_V; H_{\beta}$ & 41 \\
            $\vdots$ & 1982 & Id+Ic* & 8.19 & 0.10 & B stars & ZAMS & $M_V; H_{\beta}$ & 15 \\
            \citeauthor{1990AJ....100.1994W} & 1990 & Ic & 7.7 & 0.50 & B Stars & ZAMS & co-$H_{\gamma}$ & ? \\
            $\vdots$ & 1990 & Id & 8.2 & 0.03 & B Stars & ZAMS & co-$H_{\gamma}$ & 2 \\
            \protect{\citeauthor{1994A&A...289..101B}} & 1994 & Ic & 8.0 & 0.49 & B Stars & ZAMS & $VBLUW$ & 34 \\
            $\vdots$ & 1994 & Id & 7.9 & 0.25 & B Stars & ZAMS & $VBLUW$ & 3 \\
            $\vdots$ & 1994 & cloud & 8.1 & 0.48 & B Stars & Red. & $VBLUW$ & \\
            \citeauthor{1999osps.conf..411B} & 1998 & Ic & 8.32 & 0.17 & B Stars & trig. parallax & Hipparcos & 34 \\
            \citeauthor{1999AJ....117..354D} & 1999 & Ic & 8.52 & 0.25 & B Stars & trig. parallax & Hipparcos & 34 \\
            \citeauthor{2004ApJS..151..357S} & 2004 & Ic* & 7.96 & 0.10 & ec.~binary & Radius & - & 1 \\
            \citeauthor{2005AJ....129..856H} & 2005 & Ib+Ic & 8.23 & 0.08 & B5-F Stars& trig. parallax & Hipparcos & 121 \\
            $\vdots$ & 2005 & Ib+Ic & 7.97 & 0.10 & B5-F Stars& ZAMS & $BV$/Hipparcos & 111 \\
            \protect{\citeauthor{2005A&A...430..523W}} & 2005 & Id+Ic & 8.34 & 0.32 & Stars & Red. & CO/Hipparcos & ? \\
            \citeauthor{2006Natur.440..311S} & 2006a & Ic* & 8.19 & 0.30 & ec.~binary & Radius & - & 1 \\
            \citeauthor{2007MNRAS.376.1109J} & 2007a & Id & 8.22 & 0.16 & G6-M2 &$R\,sin(i)$ & Various & 74 \\
            $\vdots$ & 2007a & Id & 7.97 & 0.17 & G6-M2 &$R\,sin(i)$ & Various & 34 \\
            \protect{\citeauthor{2007A&A...466..649K}} & 2007 & Id & 8.19 & 0.06 & O &dyn. parallax & binary orbit & 1 \\
            $\vdots$ & 2007 & Id & 7.94 & 0.06 & O &dyn. parallax & binary orbit & 1 \\
            \citeauthor{2007PASJ...59..897H}  & 2007 & K-L region & 8.20 & 0.09 & H$_2$O masers &trig. parallax & VERA & 1 \\
            \citeauthor{2007ApJ...667.1161S}  & 2007 & Id & 7.95 & 0.13 & radio star &trig. parallax & VLBA & 1 \\
            \protect{\citeauthor{2007A&A...474..515M}}  & 2007 & Id & 8.08 & 0.03 & radio stars &trig. parallax & VLBA & 4 \\
			\citeauthor{2008arXiv0801.4085M}  & 2008  & Id &  7.96 & 0.06 & B1-A0 & MS Fitting & $VI$ & $\sim$20  \\
         \tableline
        \end{tabular}

		}
    \end{center}

\end{table}

\section{Distance} \label{sec:distance}

\citet{1946PASP...58..356M} opens his re-analysis of the
\citet{1931PASP...43..255T} derivation of the distance to Orion with a
sentence still applicable today, ``All published values of the
 distance of the {Orion Nebula} are open to some criticism.'' At the
time, the range in quoted distances was a factor of 10. While the
spread in acceptable values has decreased over time, uncertainty in
 the distance to the stars in or near the {Orion Nebula} at the 10-15\%
level remains today.

The difficulty in estimating distances is due in part to the complex
geometry and kinematics of the region as a whole, and in part to
characteristics of the youthful member stars themselves. Several
stellar subgroups were identified by \citet{1964ARA&A...2..213B}, each
covering several degrees on the sky. These groups, which appear to
have different ages, overlap along the line of sight with a total
depth of more than 100~pc. This renders membership boundaries and
hence distances to the individual subgroups (as well as ages)
 difficult to distinguish. The molecular cloud containing the {Orion
 Nebula Cluster} {(ONC)} is behind most of the optically visible early
type stars in the larger association. Depending on the sub-group
within the association, the O and some of the B stars are slightly
evolved from the ZAMS while the A and/or later stars may be still in
the pre-ZAMS phase. Reddening is spatially variable and significant,
especially towards the Nebula. Further, the vast majority of nebula
stars are photometrically variable and have other signatures of
circumstellar activity in their photometry and spectra.

Table~\ref{tab:onc_dist} contains relevant distance determinations to
 the {Orion Nebula Cluster} and records, where possible, the method used
to derive the distance and if error estimates were documented. Authors
were found to have often included some mix of Id, Ic*, and Ic stars in
their distance estimates; occasionally authors provided estimates
derived for the outer parts of the Id association but applied to the
inner region.  Therefore Table~\ref{tab:onc_dist} includes all Ic and
Id distance estimates; because it is well established that Ic members
experience less line of sight extinction than their Id counterparts it
is secure to infer that Ic distances provide lower limits to the
distance to the cluster.

Four basic methods provide most of the distance estimates to the Orion
region: zero-age main sequence fitting or similar stellar evolutionary
status methods that provide distance moduli, kinematic methods that
assume a specific dynamical model for the cluster, parallax estimates,
and reddening analyses. Most authors have used B stars as distance
probes since they are bright (enabling good data to be obtained) and
close to the ZAMS (enabling distance modulus determination). Below,
each distance estimation method is discussed in turn, followed by a
 synopsis of the best current constraints on the {ONC} distance, and
 finally a survey of commonly cited references for the {ONC} distance
 that, in fact, do not contain actual analyses of the {ONC} distance.

\subsection{Distances Based on Stellar Evolution}

\subsubsection{Assumption of Zero-Age Main Sequence Stars}
Spectroscopic parallax distance estimates involve comparison of
(de-reddened) apparent magnitudes with absolute magnitudes, which are
assumed based on stellar evolutionary status, to derive a distance
modulus. Several early works (see Table~\ref{tab:onc_dist}) reported
distances ranging from 185 to 2000 pc (falling off the range of
distance versus publication year shown in Figure
\ref{fig:onc_dist}). In the first modern set of distance estimates to
 the {Orion Nebula} region, \citeauthor{1931PASP...43..255T} derived a
distance modulus (DM) of 8.5 magnitudes by comparing absolute
magnitudes \citep[from][]{1930LicOB..14..154T} along the main sequence
with the de-reddened apparent magnitudes of B stars in the
cluster.\footnote{The result of a second method based on the angular
  diameter of the cluster was also reported. He assumed the Orion
  cluster had a size typical of other open clusters.} His method would
be repeated by numerous authors in later studies,
\citeauthor{1946PASP...58..356M} re-calculated the total absorption
towards the three brightest stars in the
 {Trapezium,} used a different
absolute magnitude relation \citep[that of][]{1940CMWCI.631....1W},
and derived a much smaller distance modulus, 7.38 magnitudes. These
first two modern distance estimates are the extremes in distance found
in Table~\ref{tab:onc_dist}.

In later work, \citet{1949AJ.....54..111M} similarly studied early B
stars, used yet another absolute magnitude relation \citep[that
of][]{1946PGro...52....1B}, and found a distance modulus of
8.57. \citet{1952ApJ...116..251S} performed the most extensive survey
yet of early type stars throughout the entire Orion region with
special focus on stars within a few degrees of the
 {Trapezium.} He found
a distance modulus of 8.5 for the ensemble and a slightly further 8.6
for stars near the Nebula. Using the ``Q-method" of photometric
dereddening, rather than spectroscopy, and the
\citet{1953ApJ...117..313J} main sequence, \citet{1954ApJ...119..200S}
reported the same distance modulus of 8.5. \citet{1954TrSht..25....1P}
derived a distance modulus of 8.0 from his vast photographic catalog;
he used A stars rather than the B stars typical of other authors.

\citet{1956ApJ...123..267J} recognized that some luminosity evolution
away from a zero-age main sequence will occur between clusters of
different ages and that this may be the case for the B stars in the
 {ONC.} Using a re-calibration of the \citet{1953ApJ...117..313J} main
sequence, they calculated a distance modulus of 8.0 by de-reddening
the data of \citet{1954ApJ...119..200S} for stars below the assumed
upper MS turnoff.  However, their assertion that A stars are on the
main sequence rather than in the pre-main sequence phase of stellar
 evolution is likely not correct for the young {Orion Nebula Cluster,}
leading to an underestimate of the
distance. \citet{1962ApJ...136..767S} used photoelectric data to
revisit and revise the \citet{1952ApJ...116..251S,1954ApJ...119..200S}
distance downward to 8.2 magnitudes by considering the evidence for
stellar evolution. Additional applications of traditional $BV$ or
$UBV$ zero-age main sequence fitting include
\citet{1964BAN....17..358B}, \citet{1966VA......8...83M},
\citet{1968ApJ...152..905L}, \citet{1969ApJ...155..447W} and
\citet{1973ApJ...183..505P} -- revised by
\citet{1975MNRAS.171..219P}. \citet{2005AJ....129..856H} used
Hipparcos photometry and the main sequence of
\citet{2000asqu.book.....C}. These authors contended with reddening in
different ways and selected to varying extents samples free of
binaries or variable stars.

\citet{1978ApJS...36..497W} first applied narrow band Str\"{o}mgren
photometry of B stars to the Orion distance problem, deriving a
 frequently cited distance to the {Ori Id} region of 435 pc, as well as
distances to the other subgroups and newly defined sub-divisions of
the subgroups. \citet{1982AJ.....87.1213A} revised the
\citeauthor{1978ApJS...36..497W} distance using the same data but a
different $\mbox{H}{\beta}$~calibration and different combinations of
subgroups. In general, the \citeauthor{1982AJ.....87.1213A} distance
estimates are 40-80 pc closer those of
\citeauthor{1978ApJS...36..497W}. \citet{1990AJ....100.1994W} used
$\mbox{H}{\gamma}$ and the Balmer discontinuity to determine
$\mbox{T}_{eff}$, surface gravity and the absolute bolometric
magnitude of B stars in Orion, also deriving distance estimates to all
four subgroups in Orion.

\citet{1994A&A...289..101B} used $VLUBW$ photometry and interpolated
grids of Kurucz models to derive stellar parameters from which ZAMS
fitting techniques could be applied. Distances to each OB association
sub-group were derived and these results display a systematic 0.3~mag
shift (closer) than those derived by \citet{1978ApJS...36..497W}.
More recently, \citet{2008arXiv0801.4085M} used the photometry and
effective temperatures of \citet{1997AJ....113.1733H}, the
\citet{1990ARA&A..28...37M} extinction law, and Geneva-Bessell
isochrones to derive 391~pc; this is closer than previous estimates
using main sequence fitting techniques but consistent with
contemporaneous distance estimates using other techniques.

In summary, distance estimates derived using assumed constraints on
the evolutionary status of the Orion OB stars are widely
scattered. The primary uncertainties arise from sample selection,
reddening corrections, assumptions about the evolutionary state of the
early type stars, and the adopted main sequence which can vary by
several tenths of a magnitude between authors.

\subsubsection{Stellar Rotation}
A characteristic of young stars is their relatively rapid rotation,
which is measured using time-series observations that track the
periodicity of cool spots on a star's surface or through spectroscopic
measurement of the velocity broadening of absorption line
features. Coupling these two observations yields a distance
independent measure of a star's radius convolved with the inclination
of the star's rotational axis on the sky.  Comparing the radii derived
from the cluster stars' rotational properties to that derived from
placing the stars on the HR diagram, one can derive the distance to
the cluster if one assumes that as an ensemble the stars in a cluster
have randomly oriented rotation axes the sky.  This technique was
first developed by \citet{1993MNRAS.265..983H} and has been applied to
the Pleiades \citep{1994MNRAS.268..181O}, and Taurus
\citep{1997A&A...322..825P}.  \citet{2007MNRAS.376.1109J} applied this
 method to the {ONC,} using the large database of periodic stars with
measured rotational properties. The fact that the canonical distance
 (480~pc) to the {ONC} did not yield a randomized $sin(i)$ distribution
was first shown by \citet{2001AJ....122.3258R}, who did not, however,
estimate the amount that the cluster's distance would have to
decrease. Jeffries cataloged 74 young stars in the Nebula having all
of the requisite observations and used them to derive a distance of
$440 \pm 32$~pc. After showing that the accreting stars appear to have
systematically underestimated luminosities (and thus a biased $sin(i)$
distribution), Jeffries derived a distance of $392 \pm 32$~pc using a
subset of 32 non-accreting young stars in the Nebula.

\subsection{Distances Based on Kinematics}

\subsubsection{Proper Motions} Kinematic distance estimates involve
radial velocity and proper motion data combined with kinematic
assumptions. In the simplest model of random motions, the distance is
directly proportional to the ratio of the radial velocity and proper
motion dispersions. More complex models such as expansion,
contraction, or rotation can be employed as
well. \citet{1958ApJ...128...14S} derived the first distance to the
 {ONC} using this method. He combined proper motion data with only a few
radial velocities and used an expansion model to estimate a distance
 of 520 pc to the {ONC} region. \citet{1965ApJ...142..964J} presented new
radial velocities and used the \citet{1958ApJ...128...14S} proper
motions of the same stars to derive a smaller distance of 380 pc,
assuming random motions. Finally, \citet{1983ApJ...271..642W}
presented new radial velocity data and computed both a radial velocity
dispersion and a proper motion dispersion for the same stars from the
data of \citet{1954TrSht..25....1P} but did not carry the analysis
through to a distance estimate.

 The most frequently cited distance to the {Orion Nebula Cluster} comes
not from study of stellar motions, but from proper motions and radial
 velocities of H$_2$O masers in the {Kleinmann-Low} {(K-L)} nebula by
 \citet{1981ApJ...244..884G}. {The K-L} nebula is embedded in the {Orion
 Molecular Cloud} behind the {Orion Nebula,} though it is thought to be
within 1 pc of the front edge of the cloud
\citep[e.g.][]{1973ApJ...183..863Z}. Maser velocities were compared
with a kinematic expansion model for the outflow to derive a distance
of $480\pm80$~pc. That expansion model has undergone subsequent
changes in its inferred orientation on the sky
\citep{1998Natur.396..650G,2004IAUS..221..155G} but the impact of
these model changes on that distance estimate have not been
quantified.

The primary uncertainties in these kinematic methods lie with sample
selection, with assumptions of the kinematic models, e.g. random
motion versus expansion, contraction, or rotation, and with their use
over stellar groupings large enough to be considered unbound
associations rather than bound clusters. The scatter in distances
derived from kinematic methods is comparable to that in distances
derived from zero-age main sequence fitting.

\subsubsection{Double-line Eclipsing Binary Systems} Further distance
estimates can be derived from kinematic analysis of double-line
eclipsing binary systems, which provide empirically constrained values
 of the stellar radii. In the {Orion Nebula} plus Ic* association there
are currently two such systems that have refined
results. \citet{2004ApJS..151..357S} derived a distance of
419~$\pm$~21~pc (or 390~pc adopting a more conventional value for the
 bolometric magnitude of the Sun) for {V1174~Ori,} an M-type pre-main
sequence solar analog system. Partly based on this distance and partly
based on age arguments, these authors consider this star a member of
Orion Ic*. A similar, remarkable analysis of the brown dwarf - brown
 dwarf eclipsing binary {{2MASS~J05352184-0546085}} yielded a distance of
$435 \pm 55$ parsecs \citep{2006Natur.440..311S}; both stars are
projected against the southern reaches of the Nebula. One systematic
that is not constrained by the dynamics of these systems is the line
of sight extinction; in these cases increasing the inferred line of
sight extinction acts to move the star's inferred distances to smaller
values.  Additional eclipsing binaries
\citep{2007MNRAS.380..541I,2008ApJ...674..329C} will, eventually, lead
to further fundamental distance constraints.

\subsection{Direct Parallax Determinations}

\subsubsection{Hipparcos} Hipparcos trigonometric parallax distances
to the individual Orion subgroups were first provided by Brown
et~al. \citep[1998; unpublished preprint; see also discussion
 in][]{1999osps.conf..411B} who found that {Orion Ia} is 50-100 pc in
 front of the other associations {(Orion Ib,} Ic), consistent with the
spectroscopic parallax analyses of \citet{1994A&A...289..101B} and
\citet{1978ApJS...36..497W}. A distance of $462 \pm 36$ pc (DM~=~8.32)
was quoted for the Ic group, and cited as preliminary. The
difficulties in interpreting Hipparcos parallax data in the Orion
region of the sky include 1) mostly radial motion of both members and
field stars due to location towards the solar antapex, which causes 2)
significant membership biases, while 3) Orion is located close to the
upper limit (500 pc) of Hipparcos sensitivities.

The Brown et~al. (1998) results were revised by
\citet{1999AJ....117..354D} who also acknowledged the astrophysical
difficulties of Orion and found a distance of $506 \pm 37$~pc
(DM~=~8.52) to the Ic group, using the same stars as Brown
et~al. (1998). Reasons for the different results for the Ia,b,c groups
are not explained. \citet{1999osps.conf..411B} restated the
\citeauthor{1999AJ....117..354D} results but also noted them as
preliminary.

Finally, \citet{2005AJ....129..856H} recalculated parallax distances
to Orion subgroups using a revised B star membership list selected
according to kinematic and color criteria and partially revised
sub-group designations. These authors find a distance of
 443~$\pm$~16~pc to the combined {Orion Ib} and Ic regions, in agreement
with Brown et~al. and de~Zeeuw et~al. within errors. However, the
distances to these two subgroups are derived together rather than
 independently and so the implications for the distance to {Orion Id} is
unclear.

The primary uncertainty with existing Hipparcos parallax estimates of
the Orion distance is its limited precision at such a large distance
combined with the astrophysical circumstances regarding Orion
kinematics. Future missions should redress the first of these issues
and possibly overcome the second.

\subsubsection{Interferometric Observations of Radio Sources} Very
long baseline radio interferometry provides the astrometric precision
necessary to measure and separate the combined proper motion and
parallax reflex motion for compact objects at a distance of
500~pc. Recent results have utilized the fact that the strong magnetic
fields of pre-main sequence objects cause them to be excellent radio
targets for parallax determinations
\citep{2007ApJ...667.1161S,2007A&A...474..515M}, while another group
 has derived the annual parallax of water masers in the {Kleinmann-Low}
nebula \citep{2007PASJ...59..897H}.

\begin{figure}[ht*]
	\centering
	\begin{minipage}[c]{0.52\textwidth} \centering
		\includegraphics[angle=0,width=\textwidth]{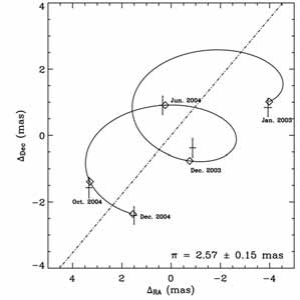}%
	\end{minipage}%
	\hspace{0.04\textwidth}%
	\begin{minipage}[c]{0.42\textwidth} \centering
		\includegraphics[angle=0,width=\textwidth]{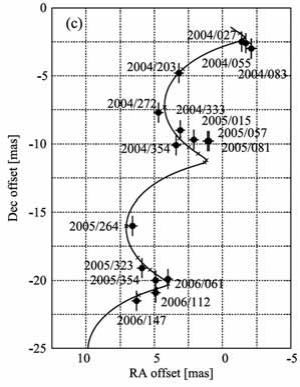}%
	\end{minipage}
	\caption{Space motion of Orion sources observed with very long
	baseline radio interferometry and used for distance
	determinations. Left: space motion of the non-thermal radio
	star GMR A over a 2 year period
	\citep[From][]{2007ApJ...667.1161S}; Right: space motion of a
	water maser spot in the Kleinmann-Low Nebula over a 2.5 year
	period \citep[From][]{2007PASJ...59..897H}. The proper motion
	plus parallax reflex motion best fit is plotted in each
	case. \label{fig:onc_dist2}}
\end{figure}

 Four radio stars in the {Orion Nebula} have been used to derive distance
estimates: GMR A, F, G and 12
\citep[GMR:~][]{1987ApJ...314..535G}. \citet{2007ApJ...667.1161S} used
the star GMR-A, observed it with the Very Long Baseline Array (VLBA)
during the 2003-2004 epoch and found a parallax of
$\pi\;=\;2.57\,\pm\,0.15$~mas or a distance of $389^{+24}_{-21}$~pc
(Figure~\ref{fig:onc_dist2}).  GMR-A is optically obscured by the
molecular cloud associated with the Nebula, and should, thus, provide
an upper limit on the distance to the Nebula. During the 2006-2007
epoch \citet{2007A&A...474..515M} used the VLBA to measure the
trigonometric parallax of GMR-A and three additionial variable
non-thermal radio sources. They found a parallax of
$\pi\;=\;2.390\,\pm\,0.104\;(418.4\,\pm\,18.2\,\mbox{pc})$ for GMR-A,
and a joint solution for all four sources of
$\pi\;=\;2.415\,\pm\,0.040\;(414.0\,\pm\,6.8\,\mbox{pc})$. The
precision of the source positions for such observations are affected
by the time variability of the targets' flux and the fact that these
sources are sometimes resolved (i.e., not concurrently point sources,
e.g., GMR-A). In addition these two VLBA studies used different
calibration techniques.

 \citet{2007PASJ...59..897H} observed water masers in the {Orion K-L}
region during the 2004-2006 period using the VLBI Exploration of Radio
Astrometry (VERA) system in Japan.  Filtering the observed set of
water masers based on signal-to-noise and on $v_{LSR}$, they chose 1
maser spot for which they derived positions at 16 epochs during this 2
year period (Figure~\ref{fig:onc_dist2}).  They derive a parallax of
$\pi\;=\;2.29\;\pm\;0.10$ for this maser spot, corresponding to a
distance of $437\,\pm\,19\,\mbox{pc}$. Their result does assume that
the space velocity of the maser is constant and is not being
accelerated in an outflow or disk.  Similar to the issue for radio
stars, resolvable variations in source structure could impact the
precision of such distance determinations.

\subsection{Distances Based on Reddening}

\citeauthor{1994A&A...289..101B} also considered the results of their
$VLUBW$ data in a traditional $A_{V} \mbox{vs}$ distance modulus plot
to estimate a distance to the Orion A Molecular cloud. In a finding
repeated by the work of \citet{2005A&A...430..523W},
\citeauthor{1994A&A...289..101B} find that the near edge of the cloud
begins to increase the measured $A_V$ values at a distance modulus of
320$\pm$70 mag. From a comparison of $A_V \mbox{vs}\,100\,\micron$
emission from IRAS, \citeauthor{1994A&A...289..101B} find that the far
edge of the cloud is at 500$\pm$30 pc.

\citet{2005A&A...430..523W} report on the work of
\citet{2001PhDT........W} who studied the variation in color excess of
 the Hipparcos stars toward the {Orion A} cloud as a function of their
Hipparcos distance. Finding the distance at which reddening increases
substantially is interpreted as the distance to the cloud. These
authors report an apparent distance gradient, ranging from the
 northern part of {Orion A} where the {ONC} is located to the southern
filaments, which correlates with a gradient in cloud radial velocity.

\begin{figure}[ht*]
    \centering
    \includegraphics[width=\textwidth]{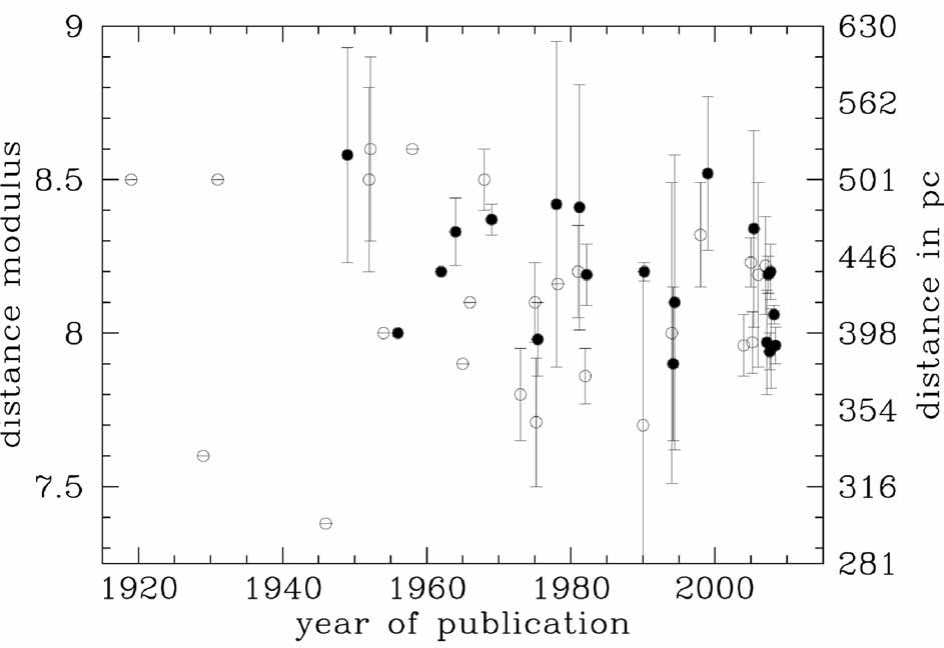}
    \caption{Distance estimates to Orion Id/Ic/Ic* as a function of
    publication date. Error bars are plotted where available. See
    Table~\ref{tab:onc_dist} and text for description of
    symbols. \label{fig:onc_dist}}
    \vspace{-3mm}
\end{figure}

\subsection{Papers Often Cited Inappropriately as Distance References}

This section addresses two issues: first the citation of distances to
 the {ONC} which have no traceable scientific source, and second the
citation of papers in which no distance is quoted or in which the
distance is taken directly from another source.

A variety of distances falling within the observed scatter in measured
distances (Table~\ref{tab:onc_dist}) have been assumed in recent
 studies of the {ONC.} While perhaps valid as estimates guided by
previous literature, it should be pointed out that there are no formal
distance estimates which correspond to often quoted values such as
440~pc \citep[DM = 8.22 as assumed by][]{1986ApJ...307..609H}, 450~pc
\citep[DM = 8.27, as asserted by][seemingly an average of the Ic and
Id distances from Warren \& Hesser 1978]{1989ARA&A..27...41G}, 470~pc
\citep[DM = 8.36 as cited by][for the result of Genzel et~al. 1981,
though not what is quoted in the original
paper]{1989ARA&A..27...41G}.These distances appear to be either round
number estimates of bona fide published values, ad hoc averages of
some sub-set of published values, or unpublished revisions or
restatements.

 Often cited, but inappropriate, references for the distance to the {ONC}
are the following. The original paper segregating the Orion
associations Ia,b,c,d according to morphology, by
\citet{1964ARA&A...2..213B}, quotes a distance of 460~pc (distance
modulus 8.3 mag) which is directly referenced to a work by
\citet{1964BAN....17..358B} making the
\citeauthor{1964ARA&A...2..213B} paper not an original source. Next,
the well-cited comparison of open cluster color-magnitude diagrams and
main sequences performed by \citet{1981A&AS...44..467M} provides a
distance to Orion with no explicit sample, methodology, or error. In
general distances in this work are from zero age main sequence
 fitting; Orion is placed into a group with {NGC 6231} and {NGC 2264} for
this purpose. Finally, the extremely valuable proper motion study of
\citet{1988AJ.....95.1755J} was not, however, a distance estimate to
 the {Orion Nebula Cluster.} These authors simply showed that the
distance of \citet{1969ApJ...155..447W} was consistent with the
rejection of the number of foreground objects expected based on
luminosity function analysis via the proper motions distribution; a
similar analysis and conclusion was drawn by
\citet{1988AJ.....95.1744V}.

\subsection{Final Distance Thoughts}

As evidenced in Table~\ref{tab:onc_dist}, there is not only a large
 range in the distance estimates to the {ONC} region, but most
measurements are accompanied by a large error bar, 15-20\%. It is
 interesting to illustrate the scatter in measured distances to the {ONC}
region by plotting the derived distance modulus and error as a
function of time, as in Figure \ref{fig:onc_dist}. Those distance
estimates that are the most reliable (filled symbols) are
distinguished from the others that either are not exclusively derived
from Id members or use less robust techniques. A notable feature of
this figure is that beyond 1950 the scatter in distance estimates is
relatively unchanged in time. While there is a clear upper bound to
the distance measurements at $\sim525\mbox{ pc}$, there seems to be
some emerging convergence at 400~pc from the many varied techniques
used in recent years.

A problem that persists in more accurately separating the distance to
 the {Orion Nebula Cluster} from that to the Orion Ic* association is
that the two subgroups are projected along the same line of sight,
where the Ic* group is concentrated primarily along and parallel to
 the Orion A Molecular cloud, which contains the slightly younger {ONC}
 and {Trapezium} (See, for example, Figure
\ref{fig:onc_optical}). In
general, there appears to be consistency in the relative distances
between the Orion subgroups amongst those authors quoting values for
various of the Ia, Ib, Ic, Id regions. Although the absolute distances
 have systematic offsets between authors and techniques, {Orion Id} is
typically found to be the furthest.

Indeed, the Id and Ic* subgroups are so aligned that a review of the
membership statistics from the \citet{1997AJ....113.1733H},
\citet{2001AJ....121.1676R}, and \citet{2001AJ....121.3160C}
wide-field studies of this region reveals no morphological signatures
that can separate the two entities. Distributions of infrared excess
 stars are more or less continuous from the {NGC~1977 H\textsc{II}}
 region down to the {NGC 1999} clustering south of the
 {ONC.} Sub-clustering as seen by eye, and in H$\alpha$ surveys are
probably the result of extinction rather than well segregated
clusters. The only significant physical difference between the Ic* and
more embedded stars appears to be differences in their typical
ages. The ages of these subgroups are not included in this discussion,
but the historical age estimates among the spatially defined subgroups
do seem to distinguish themselves
(e.g. \citeauthor{1964ARA&A...2..213B},
\citeauthor{1978ApJS...36..497W}, \citeauthor{1994A&A...289..101B})
with Id the youngest.

\section{Characterizing the T~Tauri Population} \label{sec:ttauri}

\subsection{Optical to Infrared Imaging Surveys} \label{sec:ttauri:survey}

\subsubsection{Photographic and Photoelectric
  Surveys} \label{sec:ttauri:history} There have been a significant
number of photoelectric surveys published before the CCD era that
include tables of source photometry that could have utility in
variability studies. Table \ref{tab:phg} was created for the purpose
of documenting these sources of literature photometry. It is
supplemented with details from a few very large photographic surveys,
e.g., \citet{1981paon.book.....A} with $N_{\star}\,\sim\,15000$.  Data
from some papers, such as that data listed in
\citet{1977ApJS...34..115W}'s primary $UBV$ table, are a merger of a
very large number of published data sources (in that case 35 separate
papers), and the references in these amalgamations are generally not
reproduced here.

\begin{table}
	[!ht]

	\caption{Large scale photographic (pg) and photoelectric (pe) data sources for the Orion Nebula. Ordered chronologically by epoch of publication. No epoch is recorded when no detailed epoch of observation could be determined or the data were the average of many other sources. } \smallskip
	\begin{center}
		{\small
		\begin{tabular}
			{lcllr}

			\tableline \noalign{\smallskip} Paper & Data & Epoch & Filters & $N_{\star}$ \\

			\noalign{\smallskip}
			\tableline \noalign{\smallskip}
			\citet{1954TrSht..25....1P} & pg &    & $m_{pg}$ & 2982 \\
			\citet{1952ApJ...116..251S} & pe & 1951 & $BV$ & 190\\
			\citet{1954ApJ...119..200S} & pe & 1951 & $UB$ & 184 \\
			\citet{1957ApJ...126..134J} & pe & 1954-1955 & $UBV$ & 49 \\
			\citet{1962ApJ...136..767S} & pe &   & $UBV$ & $\sim180$ \\
			\citet{1963VilOB...5...18K} & pg & 1952? & $m_{pg}$ & $\sim2000$\\
			\citet{1966VA......8...83M} & pe &   & $UBV$ & 36 \\
			\citet{1968ApJ...152..913L} & pe &   & $UBVRIJKL$ & 196 \\
			\citet{1969ApJ...155..447W} & pe & 1958-1967 & $UBV$ & $\sim300$ \\
			\citet{1973ApJ...183..505P} & pe & 1970-1971 & $UBVRIJHKL$ & 51 \\
			\citet{1975MNRAS.171..219P} & pe & 1972-1973 & $UBVRIHKL$ & 48 \\
			\citet{1976AJ.....81..375M} & pe & & $UBVRIJHKL$ & 51 \\
			\citet{1977ApJS...34..115W} & pe & & $UBV$ & 526 \\
			--- & pe & 1968 & $UBV$ & $109$ \\
			--- & pe & 1972 & $uvby$ & $492$ \\
			\citet{1981SvAL....7...21S} & pe & 1978-1979 & $UBVRI$ & 117\\
			\citet{1981paon.book.....A} & pg & 1979 & $UBVI$ & $\sim15000$ \\
			\citet{1982PASJ...34..241I} & pg & 1970-1971 & $RI$ & 413 \\
			\citet{1984AJ.....89..399R} & pe & 1982-1983 & $UBVRI$ & 41 \\

		\tableline
\end{tabular}

		}

	\end{center}
	\label{tab:phg}
\end{table}

\subsubsection{H$\alpha$ Surveys} \label{sec:ttauri:halpha} Slitless
optical grism surveys of young star forming regions can be valuable
tools for identifying young stars
\citep[e.g.][]{1988cels.book.....H}. This is because young stars
frequently show strong H$\alpha$ line emission, which is related to
their active chromospheres as well as circumstellar accretion;
however, the very strong hydrogen line emission background of the
 {Orion Nebula} probably result in a significant underestimate of the
true membership if based upon H$\alpha$ statistics
alone. \citet{1953ApJ...117...73H} documented 255 H$\alpha$ stars
within a 3.5 degree region around the
 {Trapezium} while
\citet{1982BITon...3...69P} cataloged 534 H$\alpha$ stars in a
5~degree region. The Kiso Orion surveys
\citep{1989PASJ...41..155W,1989PASJ...41.1195K,1991PASJ...43...27W,1993PASJ...45..643W,1995PASJ...47..889N},
while valuable for their coverage of most of the Orion constellation,
 appear to be very incomplete in the {ONC} as evidenced by the lack of a
strong peak in the stellar density within the Nebula as found by the
subsequent analysis of \citet[][]{1998AJ....115.1524G}; \citet[see
  also][]{1988AJ.....95.1755J}. In their review of the Orion
association, \citet{1991lmsf.book....1B} collated the existing
H$\alpha$ star catalogs into a single list, including 87 new stars
from \citet{1992A&A...265..144W}.

\subsubsection{Modern Optical Surveys} \label{sec:ttauri:optical} A
 review of modern optical CCD surveys of the {Orion Nebula} begins with
the work of \citet{1986ApJ...307..609H}. Their CCD observations were
taken with the 40~inch Nickel telescope at Lick Observatory, had a
pixel resolution of $0.267\arcsec$ and consisted of a mosaic of small
$(\sim2.5\arcmin)$ fields. The authors used narrowband interference
filters to minimize nebular contamination but final photometry was
reduced to and reported in the Johnson-Cousins $VI_C$ system. Their
Table~1 contains photometry for 98 of the 140 sources detected and
uses the \citet{1954TrSht..25....1P} number system except for 30
sources that are labeled ``anonymous.'' The authors used these new
data to construct color-magnitude diagrams and explore the age and age
spread for the cluster, finding most stars to be $\sim1$~Myr or
younger (Section \ref{sec:ttauri:prop}).

 The first {ONC} photometry from the \textit{Hubble} Space Telescope was
published by \citet{1994ApJ...421..517P}. This survey consisted of 11
irregularly mosaicked Planetary Camera fields in the F547M and F875M
filters. Their Table~4 contains aperture photometry for 326 objects,
using an aperture beam of $0.12\arcsec$ and converted into the $VI_C$
passband system. Cross references from their ``PC'' identifier system
to that of Jones \& Walker (JW) and Parenago (P) are
given. Unfortunately, they report their astrometry to be quite poor
$(\sim1\arcsec)$. The high resolution of these data provided excellent
new statistics on visual binaries in the cluster, identifying 35
sub-arcsecond pairs (their Table~6). Additional HST observations were
obtained, reduced and analyzed by \citet{2004ApJ...606..952R}. Their
results include observations in the F336W, F439W filter passbands as
well as data from archived F547M, F791W images. They tabulate the
resulting $UBVI$ data for 40 sources with spectral types from
\citet{1997AJ....113.1733H}.

\begin{figure}[!t]
	\centering
		\includegraphics[width=\textwidth]{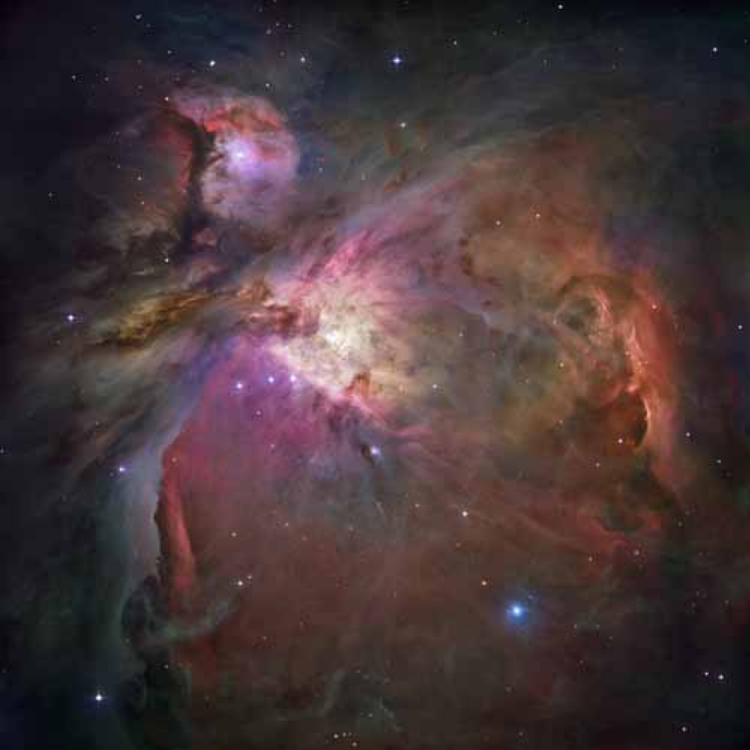}
 \caption{The {Orion Nebula} in optical light as mapped by the
        Advanced Camera for Surveys with the \textit{Hubble} Space
        Telescope. The field is approximately
        $30\arcmin\times30\arcmin$ in size with North oriented up and
        East to the left.  From the Orion HST Treasury program,
        PI. M. Robberto.}
	\label{fig:Figures_ONCtreasury_reduced2}
\end{figure}

 The comprehensive survey of the {Orion Nebula Cluster} by
\citet{1997AJ....113.1733H} included new $VI_C$ photometry in addition
to a large corpus of spectroscopy (Section
\ref{sec:ttauri:spec}). Hillenbrand cataloged 1578 sources including
332 new detections with approximate completeness limits of
$I_{C}\sim17.5$ and $V\sim19$. Her tabulation was constructed from new
data (3 epochs) and literature sources (7); photometry of sources
appearing in multiple catalogs was chosen based on the angular
resolution of the survey (e.g., the Prosser et~al. HST results were
given preference). Their numbering system is a merger of Jones \&
Walker (JW; \#1-1053), Parenago (32 sources, e.g. Parenago 1891 =
 Hillenbrand 1891 = {$\theta^{1}$~Ori~C),} Prosser et~al. HST sources
(9000+PC\#) and new detections: 3000+N for epoch 1993 data, 5000+N for
epoch 1995 data and 6000+N for epoch 1996 data. Most of the global
 stellar properties of the {Orion Nebula} members are derived from this
comprehensive study.

In addition to these published surveys modern telescope archives
contain large quantities of publicly available optical data. The most
significant of these is the 104 orbit Cycle~13 \textit{Hubble} Space
Telescope Treasury Program (PID 10246; PI. M. Robberto) that surveyed
 a $\sim20\arcmin \times 20\arcmin$ region of the {Orion Nebula} with the
Advanced Camera for Surveys (ACS). Observations took place between
October 2004 and April 2005 and the surveyed filters included F435W,
F555W, F658N, F775W and F850LP. Parallel observations were also
obtained with the Wide-field Planetary Camera 2 (F336W, F439W, F656N,
F814W) and NICMOS (NIC3: F110W, F160W); all these data can be obtained
via the Multimission Archive at STScI. Figure
\ref{fig:Figures_ONCtreasury_reduced2} is from their ACS mosaic (Press
release STScI-PR06-01). A similarly large set of ground-based optical
data (PID 273.C-5042(A)) is publicly available from the ESO
archive. It was observed during January 2005 with $UBVI_C$ and
$H\alpha$ filters (Da Rio et al., in preparation).

\subsubsection{Optical Variability Surveys} \label{sec:ttauri:var}
Occasionally, multi-epoch variability surveys publish calibrated
time-averaged photometry for their sources. Sources for such
 photometry in the {Orion Nebula} include \citet{1999AJ....117.2941S} and
\citet{2002A&A...396..513H}. The former provides data for their 254
periodic stars in the $I_C$ passband while the latter provides
narrowband photometry for 1562 objects time averaged over 45 days. The
narrowband filter used by Herbst et~al. was centered at 815.9~nm. In
both cases the reported peak to peak variation of 0.2~magnitudes is
probably a good measure of the typical uncertainty inherent in a
single epoch optical survey of a young cluster. Note that
\citet{2002A&A...396..513H} adopted the same numbering system of
\citet{1997AJ....113.1733H} in their Table~1, extending it to
N$>$10000. The optical photometry for variable and periodic stars
 surveyed in the outer {ONC} by \citet{2000AJ....119.3026R} and
\citet{2001AJ....121.1676R} is single epoch.

\citet{2000AJ....119.3026R} and \citet{2001AJ....121.1676R} presented
 $UVI$ photometry of the ``flanking fields'' of the {ONC.} The Rebull
flanking field surveys cover an area from $1.5>R>0.25$ degree from
 {$\theta^{1}$~Ori~C} out to the limits of the {Orion A} cloud to the north
 (see chapter by Peterson \& Megeath) and south of {$\iota$ Ori} and the
 {OMC-4} clumps. The Tables 1 and 2 of \citet{2000AJ....119.3026R}
include a total of 4792 (1620) sources (candidate members) that were
found in or near the locus of confirmed Orion sources on the optical
color-magnitude diagram. A total of 1564 sources (726 candidate
members) have valuable wide-field $U$ band data in
\citet{2000AJ....119.3026R}.

\subsubsection{Near-Infrared Data} \label{sec:ttauri:infrared}
Near-Infrared observations are necessary to explore embedded
 populations in young regions like the {Orion Nebula} and are sensitive
to re-radiated thermal emission from circumstellar
disks. \citet{1973ApJ...183..505P} performed an optical+infrared
survey of 51 Parenago stars over a $0.5\deg$ region. These
observations included the first and essentially only wide-field
 $3\,\micron$ photometry for {Orion Nebula} members until the year
2000. Additional $HKL$ data for 35 stars were presented in
\citet{1975MNRAS.171..219P}. \citet{1982AJ.....87.1819L} expanded the
 census of sources near the {B-N} object. Although they tabulate only
 sources within $35\arcsec$ of the {B-N} object, larger maps including
 the {Trapezium} are referenced and shown,
including a source map from
\citet{1976ApJ...207..770B}, who surveyed at $2$ and $20\,\micron$ but
did not tabulate any point source
photometry. \citet{1984MNRAS.206..465H} produced a non-chopped
 $2\,\micron$ map that covered the entire {OMC-1} molecular ridge (both
 {B-N/K-L} and {OMC-1S)}
 and tabulated 88 sources with
cross-references to Lonsdale et~al. and Parenago.

 A crucial near-infrared survey of the {Orion Nebula} was performed by
\citet{1994AJ....108.1382M}. These authors obtained two complementary
sets of $K'$ data, covering a total of $82\arcsec\times82\arcsec$ and
 centered on the {Trapezium} stars. The higher
resolution tip-tilt
corrected images had a final spatial resolution of $0.35\arcsec$ and
the authors quote an astrometric precision of $0.06\arcsec$. Their
Table~1 lists 123 detections (48 new stars), including photometry even
for the brightest OB stars and extending to a quoted completeness
limit of $K'=16$. This tabulation is the origin of the ``TCC'' or
Trapezium
 Cluster Catalog identifiers and provides cross-references of
their near-IR data to the catalogs of Jones \& Walker (1988), Parenago
(1954a) and Prosser et~al. (1994) sources, as well as a detailed and
valuable cross-referencing of known VLA radio sources from
\citet{1993A&AS..101..127F}, the proplyds
\citep[e.g.,][]{1994ApJ...436..194O}, H$\alpha$ sources from
\citet{1979A&A....73...97L}, mid-IR sources and structures from
\citet{1994ApJ...433..157H} and their Table~1 of optical/near-IR
sources.
\begin{figure}[t*]
	\centering
	\begin{minipage}[c]{0.48\textwidth} \centering
		\includegraphics[angle=0,width=\textwidth]{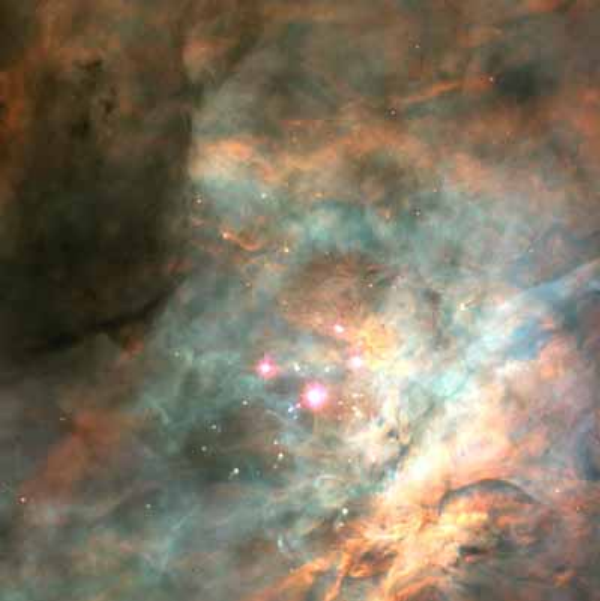}%
	\end{minipage}%
	\hspace{0.04\textwidth}%
	\begin{minipage}[c]{0.48\textwidth} \centering
		\includegraphics[angle=0,width=\textwidth]{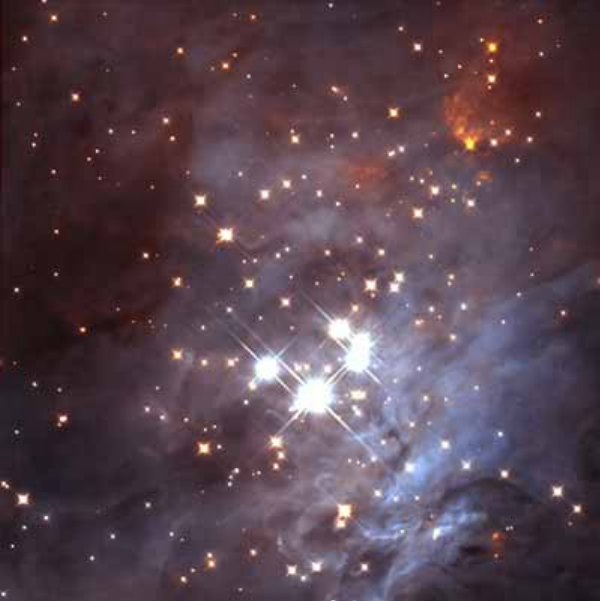}%
	\end{minipage}
 \caption{An HST view of the inner {Orion Nebula}
in the optical \citep[WFPC2;][]{1996AJ....111..846O} and the
near-IR \citep[NICMOS;][]{2000ApJ...544.1044L}. The field is
roughly $3\arcmin$ on a side and each image has a resolution of
order $\sim0.1\arcsec$. \label{fig:onc_ir}}
\end{figure}

Subsequent near-IR surveys can be divided into those which covered
very large areas of the Nebula and those that concentrated on the
 central $\sim5\arcmin$ around the {Trapezium.}
Wider field surveys that
provide large tabulations of near-IR photometry include
\citet{1995AJ....109..709A}, \citet{1998AJ....116.1816H} (multi-epoch)
and \citet{2001AJ....121.3160C} (time series). Concerns about
questionable and probably nebular extractions should be applied to the
results of almost any survey for sources against the bright,
background Nebula. \citet{1998AJ....116.1816H} and
\citet{2000ApJ...540..236H} suggested a large percentage of false and
duplicate detections in the Ali \& Depoy data, while Hillenbrand \&
Carpenter also noted 1 or 2 nebular knots among the sources listed in
the optical catalog of Hillenbrand (1997). Similarly, in a
$17\times17\arcmin$ field centered on the
 {Trapezium,}
\citet{2005ApJS..160..319G} found 1145 sources in the 2MASS
point-source catalog that lacked \textit{Chandra} X-ray detections;
however, only $\sim200$ of these are good quality 2MASS detections;
the rest $(\sim900)$ lack detection in multiple bands and most are
probably spurious. The catalog of \citet{1998AJ....116.1816H} avoids
this problem because it tabulates near-IR photometry only for those
1578 optical sources listed in \citet{1997AJ....113.1733H}; this
tabulation includes new bolometer and array observations supplemented
by literature results. Similarly, the \citet{2001AJ....121.3160C}
2MASS near-infrared variability survey tabulates photometric results
for those 1235 variables (out of 17,808 sources) found in a
$0.84\deg\times6\deg$ region. Variable stars in this catalog were
typically observed 16 times over a 2 year period.

Narrow field, deeper surveys of the region immediately around the
 {Trapezium} have included the $K$-only AO survey of
\citet{1999AJ....117.1375S}, the Keck $(HK)$ survey by
\citet{2000ApJ...540..236H}, the HST-NICMOS $(JH)$ survey by
\citet{2000ApJ...540.1016L}, UKIRT $IJH$ observations by
\citet{2000MNRAS.314..858L}, a multi-telescope $JHKK_{S}L$ survey by
\citet{2002ApJ...573..366M} and a Gemini $JHK$ survey by
\citet{2005MNRAS.361..211L}. A comparison of the regions surveyed by
most of these authors was given by
\citet{2002ApJ...573..366M}. Additional near-IR data are included in
the COUP catalogs and include previously unpublished photometry
derived by McCaughrean. As mentioned previously, archival NICMOS
observations obtained in parallel to ACS imaging of the Nebula provide
a non-contiguous but as yet unpublished data set for future
use. Additionally, the CFHT archive contains a large WIRCAM/UKIRT
 $YJHK_{S}H_{2}$ data set of a large field surrounding the {Orion
 Nebula.}

\subsubsection{Thermal and Mid-Infrared Data} \label{sec:ttauri:midir}
Initial mid-IR scale maps were limited by a combination of the very
strong nebular background and the poor angular resolution of early
mid-IR cameras. Works by \citet{1969ApJ...155L.193N,
  1974A&A....32..231L,1974ApJ...192L..23F,1975ApJ...202L..33G}, which
span the wavelength regime from $20\mbox{ to }100\,\micron$, tell us
little about the overall stellar content or properties of the members
of the star cluster although they do reveal many details about the
structure of the photodissociation region, e.g. the modern study by
\citet{2006ApJ...637..823K}. Even today the $20\,\micron$ flux from
the Nebula overwhelms all but a few bright protostars and, for
example, renders the inner $15\arcmin$ of the Nebula saturated at
$24\,\micron$ with the \textit{Spitzer} Space Telescope.

\begin{figure}[!t]
	\centering
		\includegraphics[width=\textwidth]{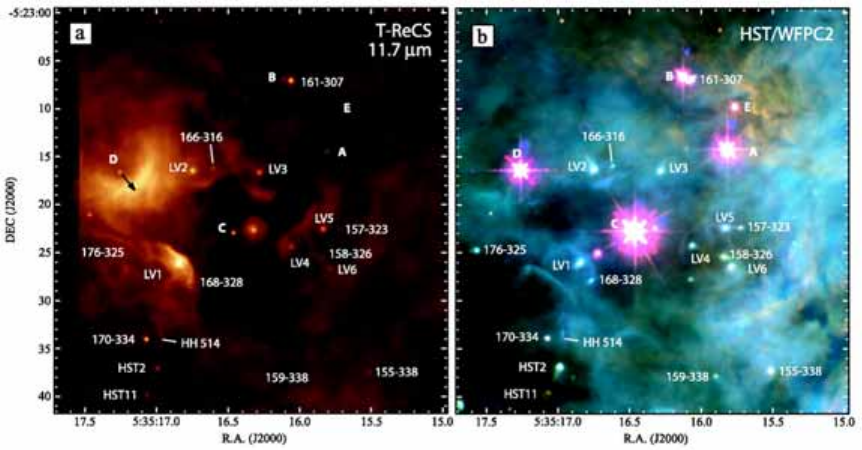}
 \caption{A comparison of the {Trapezium}
        at mid-IR
		$(11.7\,\micron\mbox{, left})$ and optical (HST WFPC2)
		wavelengths. Figure reproduced from
		\citet{2005AJ....130.1763S}. It is important to note
		that at wavelengths longer than these, e.g.,
		$>20\,\micron$, the only prominent point sources that
		can be detected against the strong nebula are located
 near the {B-N/K-L} region \citep{2005AJ....129.1534R}. }
	\label{fig:Figures_trecs_wfpc2_smith}
\end{figure}

There is a multitude of mid-IR surveys
\citep[e.g.,][]{1973ApJ...186L...7R, 1978A&A....62..261B,
1983A&A...127..417L, 1984ApJ...281..172W, 1998ApJ...509..283G,
2004AJ....128..363S, 2004ApJ...605L..57G} that have focused only on
 small embedded regions like the {B-N/K-L.} Full discussion of the
stellar content of these embedded regions is reserved for a subsequent
chapter (O'Dell et~al.; Part II). The $12\,\micron$ image of the
 {Ney-Allen} region \citep{1973ApJ...180..809N} of the central
 {Trapezium}
from \citet{1990ASPC...14..301M} was the first modern mid-IR
\textit{array} observation, revealing many narrow arcs and details of
the prominent structures in this region. \citet{1994ApJ...433..157H}
published 8.8 and $11.7\,\micron$ SpectroCam (SC) images of the
central region, including fluxes for 13
sources. \citet{1994ApJ...429..726H} present additional 10 and
 $20\,\micron$ maps of the central {Trapezium.}

Thermal infrared $3\,\micron$ data for $\sim400$ sources in a
$7\arcmin\times7\arcmin$ region were analyzed by
\citet{2000AJ....120.3162L}, who used these data to estimate the disk
fraction as a function of source mass and to identify a large sample
 of protostars throughout the {OMC-1} cloud. The data tables used in that
work were presented in \citet{2002ApJ...573..366M}. Deeper, higher
resolution $3\,\micron$ data of $\sim400$ sources in a smaller
$5\arcmin\times4\arcmin$ region were published by
\citet{2004AJ....128.1254L}; these data extended measurements of the
disk fraction into the brown dwarf regime and provided new results on
the protostellar population.

Recent longer wavelength observations with the spatial resolution
necessary to detect individual sources against the bright background
include \citet{2005AJ....129.1534R} (10 \& $20\,\micron$;
$4\arcmin\times5\arcmin$; $0.5\arcsec$ resolution; 177 sources) and
\citet{2005AJ....130.1763S} ($11.7\,\micron$;
$2\arcmin\times3\arcmin$; $0.35\arcsec$ resolution; 91 sources), see
Figure~\ref{fig:Figures_trecs_wfpc2_smith}. These works focused their
study on the proplyds, jets and emission structures in the PDR, which
are a focus of the following chapter (O'Dell et~al., Part
II). Publications that include \textit{Spitzer} observations, which
are much lower resolution than any of the ground based mid-IR
observations, include \citet{2006ApJ...646..297R} and
\citet{2007ApJ...671..605C}; both of these works focused on the
relationship between the rotational properties of the young stars and
their disk excess properties (Sect.~\ref{sec:ttauri:disk}).

\begin{figure}
	[ht*]

	\centering
	\includegraphics[angle=0.,width=4.5in]{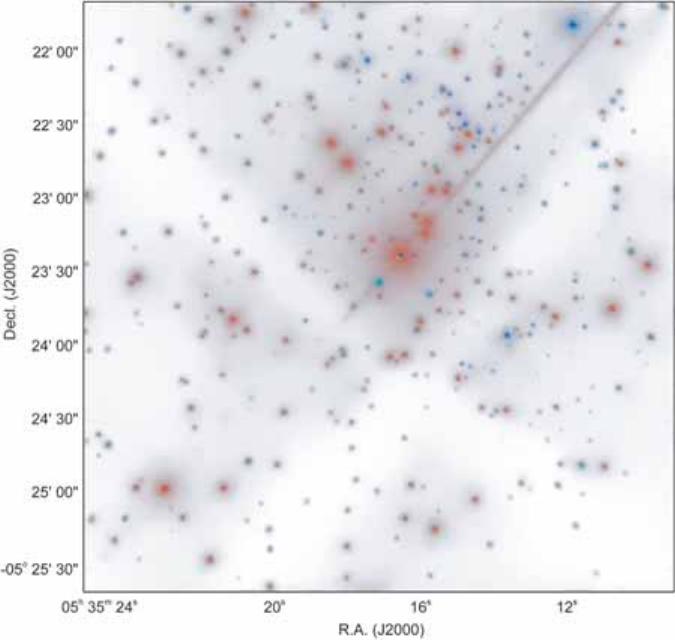}

	\caption{The central $4\arcmin \times 4\arcmin$ COUP ACIS-I
	field shown with $0.25\arcsec$ pixels. Image reproduced from
	\citet{2005ApJS..160..319G}. The displayed image has been
	smoothed from the natural integer version using the
	\textit{Chandra} Interactive Analysis of Observation (CIAO)
	\texttt{csmooth} procedure. Red, green and blue colors
	correspond to counts in the 0.5 – 1.7 keV, the 1.7
	– 2.8 keV, and the 2.8 – 8.0 keV bands,
	respectively. \label{fig:xray:fig1b}}

\end{figure}
\subsection{X-ray Observations of the Nebula} \label{sec:ttauri:xray}

The birth of stars takes place in thermodynamically cold and neutral
media with characteristic energies of $\ll1$ eV per
particle. Paradoxically, those processes associated with star
formation produce and are subject to violent high energy processes
with characteristic energies of $\ga\,10^3$~eV. The principal evidence
for this is X-ray emission from stars throughout their pre-main
 sequence (PMS) evolution. {The ONC} was the first cluster of PMS stars
to be detected in the X-ray band \citep{1972ApJ...178..281G} and
non-imaging studies soon found that the X-ray emission is extended on
scales of a parsec or larger \citep{1979ApJ...228L..33B}. Early
explanations for the Orion X-rays included winds from the massive
 {Trapezium} stars colliding with each other or
the molecular cloud, and
hot corona or magnetic activity in lower mass T~Tauri stars. The
$Einstein$ \citep{1979ApJ...234L..59K}, $ROSAT$
\citep{1995ApJ...445..280G} and $ASCA$ \citep{1996PASJ...48..719Y}
imaging X-ray observatories established that both the massive
 {Trapezium} stars and many lower-mass T~Tauri
stars contribute to the X-ray emission.

 {The ONC} was intensively studied during the first year of the $Chandra$
mission with several instrumental setups: the Advanced CCD Imaging
Spectrometer (ACIS) in imaging mode \citep{2000AJ....120.1426G,
2002ApJ...574..258F, 2002ApJ...572..335F, 2003ApJ...584..911F} and as
detector for the High Energy Transmission Grating Spectrometer
\citep{2000ApJ...545L.135S,2001ApJ...549..441S,2003ApJ...586.1441S,
2003ApJ...595..365S}, and with the High Resolution Imager
\citep[HRI;][]{2003ApJ...582..382F,2003ApJ...582..398F}. While many
valuable results emerged from these early $Chandra$ studies, it was
recognized that more would be learned from a deeper and longer
 exposure of the {Orion Nebula} region. During the fourth year of its
mission $Chandra$ performed an unprecedented $\sim 10$ day (net
 exposure) nearly-continuous observation of the {Orion Nebula,} nicknamed
the Chandra Orion Ultradeep Project (COUP).

\subsubsection{Chandra Orion Ultradeep Project} \label{sec:ttauri:xray:coup}
The COUP study detected more than 1600 X-ray sources, $\sim 1400$ of
which are young stellar
objects \citep{2005ApJS..160..319G}. Figure \ref{fig:xray:fig1b} shows
a ``true-color'' X-ray image of the central
$4\arcmin \times 4\arcmin$ {Trapezium}
region centered on the larger
$17\arcmin\times 17\arcmin$
field of $Chandra$-ACIS-I. Absorbed COUP sources appear blue and
unabsorbed sources appear red. $Chandra$ X-ray studies are
particularly effective in uncovering heavily obscured low-mass cloud
populations (X-rays penetrate up to hundreds of magnitude of
absorption into the cloud) and in discriminating cloud PMS populations
from unrelated older stars (X-ray emission from PMS stars is
$10^{1}-10^{4}$ times elevated above main sequence (MS) levels). Most
of the non-PMS contaminants in the COUP field are extragalactic active
galactic nuclei (AGNs), which can be confused with non-flaring
YSOs. \citep[Only 16 probable field stars with discrepant proper
  motions and NIR colors are present in the COUP source list, which
  are available through ][]{2005yCat..21600353G}. But the long
exposure improves the opportunity for capturing powerful X-ray flares
which are characteristic of YSOs and not AGN. Based on the variability
analysis of heavily absorbed COUP sources without optical/NIR
counterparts and detailed simulations of the extragalactic background
population, \citet{2005ApJS..160..353G} argue that 75 COUP sources are
previously unknown embedded cloud members, of which forty-two are
confirmed by the detection of powerful X-ray flares. These X-ray
discovered stars are spatially clustered within the two well-known
 {OMC-1} cores and the dense molecular filament, which extends northwards
 from {OMC-1} to {OMC-2/3.}

\begin{figure}[bp]
	\centering
	\includegraphics[width=\textwidth]{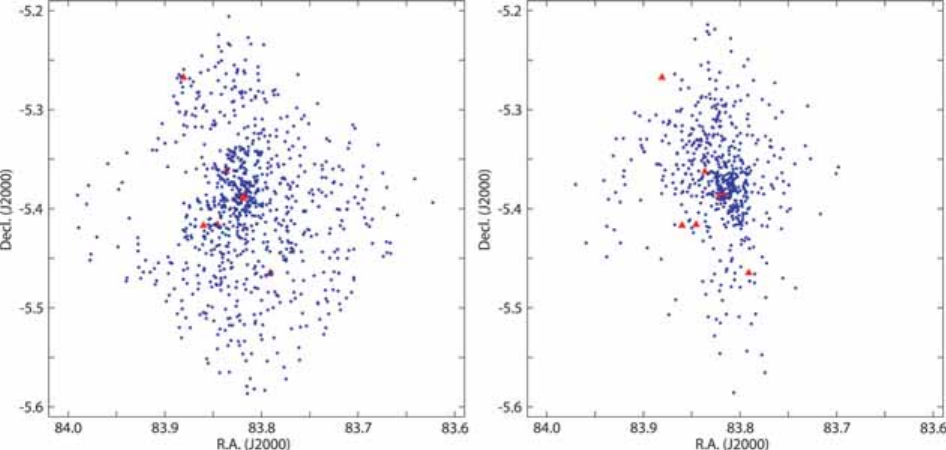}
 \caption{Diagram of the {Orion Nebula} field showing 1408 COUP
 X-ray sources associated with the {Orion Nebula} segregated by
	obscuration. Figures reproduced from
	\citet{2005ApJS..160..379F}. Left: Lightly obscured subsample
	with $N_H <\sim 10^{22}$ cm$^{-2}$. Right: Heavily absorbed
	subsample with $N_H >\sim 10^{22}$ cm$^{-2}$. The large
	triangles show 10 hot O7-B3 stars, while dots show the
	remaining cool member population.\label{fig:xray:fig4}}
\end{figure}

\begin{figure}[htbp*]
	\centering
	\includegraphics[angle=0.,totalheight=3.0in]{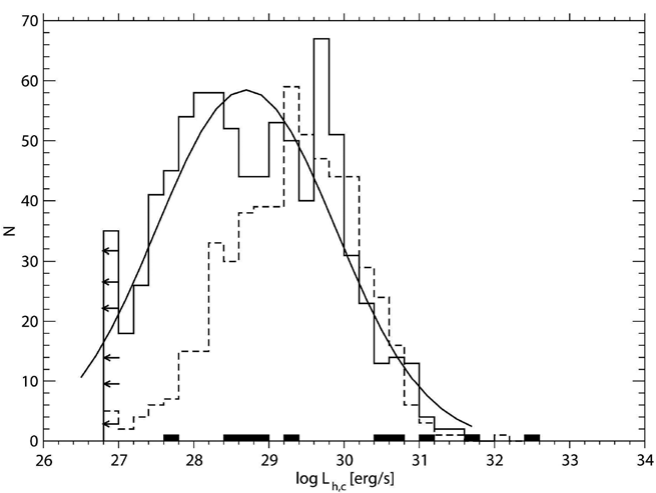}
	\caption{Histogram with the differential distribution of hard
	$(2.0-8.0)$ keV band X-ray luminosity corrected for
	absorption. Figure reproduced from
	\citet{2005ApJS..160..379F}. Solid
	line denotes the COUP unobscured cool star sample with
	Gaussian fit, dashed line show the obscured sample, and the
	black-filled histogram show the 10 hot stars, including
 {$\theta^{1}$~Ori~C.} \label{fig:xray:fig5}}
\end{figure}

The census of COUP sources with confirmed Orion membership includes
1315 stars with known optical/NIR counterparts, 75 new embedded stars,
 16 unidentified likely new lightly obscured members of {ONC}
\citep{2005ApJS..160..319G}, and two faint X-ray sources associated
 with the Herbig-Haro object {HH-210} \citep[COUP \#\,703 and \#\,704;
][]{2006A&A...448L..29G}. Three classes contribute roughly equally to
the integrated X-ray luminosity in the hard $2-8$ keV band: 10
 unobscured hot {Trapezium} stars earlier than
B4, 839 cool (later than B4) lightly-obscured COUP sources with $\log
N_H <\sim 22$ cm$^{-2}$ ($A_V\,\la\,5-6$ mag), and 559
heavily-obscured stars
\citep{2005ApJS..160..379F}. The spatial source distribution for the
cool unobscured (Figure \ref{fig:xray:fig4} left), and heavily
obscured populations (Figure \ref{fig:xray:fig4} right) show a spatial
asymmetry -- a deficit of stars to the east on $0.5-1$ pc scales --
consistent with violent relaxation in the stellar dynamics \citep[see,
however, ][]{2008ApJ...676.1109F} and the concentration of obscured
 sources around both {OMC-1} molecular cores. The X-ray luminosity
function (XLF) of the unobscured cool population is $> 90\%$ complete
down to $M \sim 0.1 M_{\odot}$ and $\sim 50\%$ complete down to $M
\sim 0.03 M_{\odot}$. The XLF shape is roughly log-normal in shape and
the obscured population is deficient in lower-luminosity stars due to
localized circumstellar material (Figure
 \ref{fig:xray:fig5}). One-third of the {Orion Nebula} region hard-band
 emission is produced by the bright O star {$\theta^{1}$~Ori~C,} and half
is produced by lower mass pre-main sequence stars with masses $0.3 < M
< 3\,M_{\odot}$ \citep{2005ApJS..160..379F}.
\begin{figure}[t*]
	\centering
	\begin{minipage}[c]{0.48\textwidth} \centering
		\includegraphics[angle=0.,width=\textwidth]{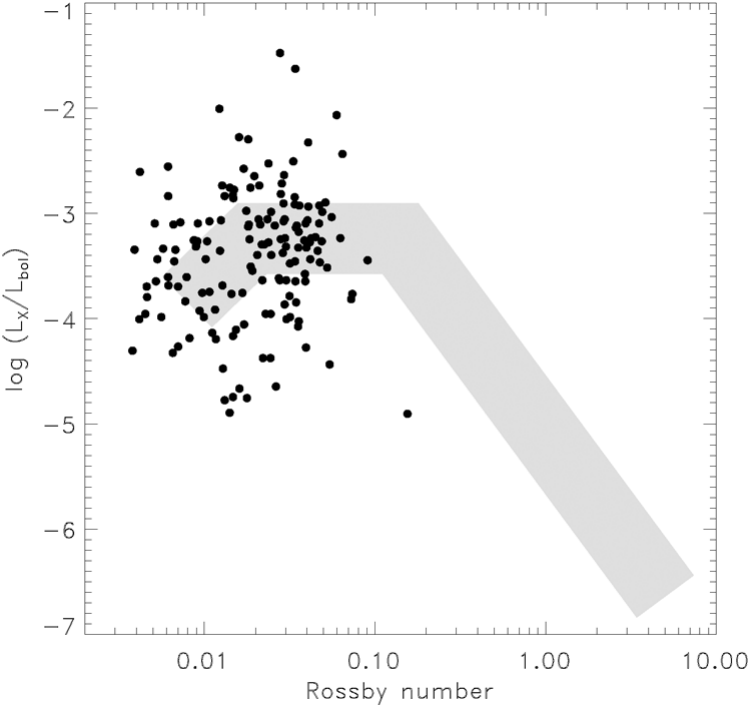}
	\end{minipage}%
	\hspace{0.04\textwidth}%
	\begin{minipage}[c]{0.48\textwidth} \centering
		\includegraphics[angle=0.,width=\textwidth]{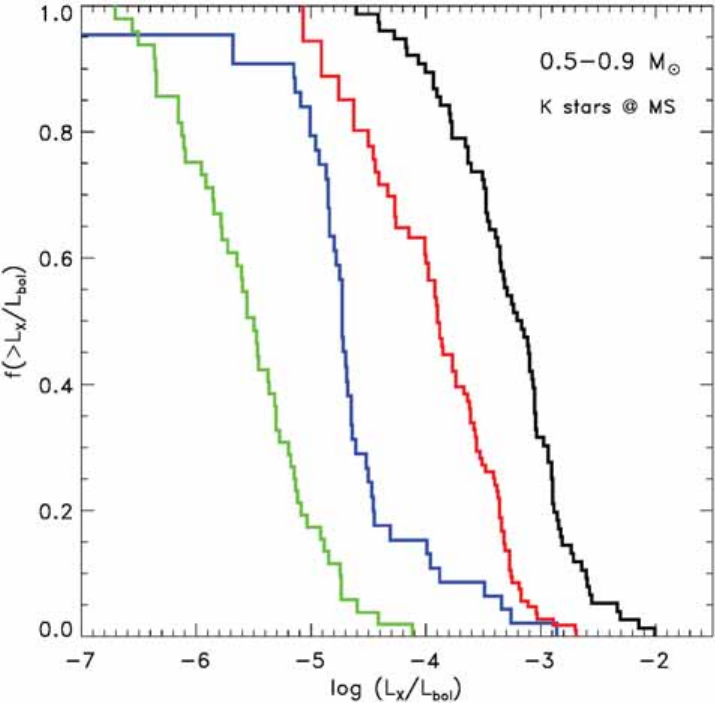}
	\end{minipage}
	\caption{Left: Fractional X-ray luminosity versus Rossby
	number for the COUP T~Tauri stars. Figure reproduced from
	\citet{2005ApJS..160..401P}. The grey shaded area shows the
	relation and the width of its typical scatter found for MS
	stars; Right: The black line shows the integrated distribution
	of X-ray luminosities for COUP Orion stars (ages $\sim 0.1-10$
	Myr) in a narrow mass range. Comparison distributions are the
 Pleiades cluster (red line, age $\sim 100$ Myr), {Hyades}
	cluster (blue line, age $\sim 500$ Myr), and solar
	neighborhood stars (green line, ages $\sim 1-5$ Gy). It is
	clearly shown that the activity-age relation continues through
	the pre-main sequence phases. Figure reproduced from
	\citet{2005ApJS..160..390P}. \label{fig:xray:fig10}}
\end{figure}

With the detection limit of $L_{X,min} \sim 10^{27}$ ergs/s for the
unobscured COUP population, X-ray emission was detected from more than
$97\%$ of the optically visible late-type (spectral types F-M) T~Tauri
 stars (TTS) in the {ONC,} demonstrating that there is no ``X-ray quiet''
population of late-type stars with suppressed magnetic
activity. \citet{2005ApJS..160..401P} show that TTS with known
rotation periods lie in the saturated or super-saturated regime in a
diagram comparing X-ray activity with the stellar interior Rossby
number (Figure \ref{fig:xray:fig10}~left). But the TTS show much
larger scatter in X-ray activity than main sequence stars. This
scatter is partly attributable to accretion: while the X-ray activity
of the non-accreting TTS is consistent with that of rapidly rotating
MS stars, the accreting stars are less X-ray active (by factors of
$\sim 2-3$), perhaps because magnetic reconnection cannot heat the
dense plasma in mass-loaded accreting field lines to X-ray
temperatures. The fact that COUP stars do not show the drop-off in
magnetic activity as stars rotate more slowly may suggest that the
magnetic dynamo process is saturated in some way and/or that a
different dynamo is operative in young stars that is independent of
rotation. \citet{2005ApJS..160..401P} do find that COUP X-ray
luminosities are correlated with stellar mass and volume, which
generally suggests a turbulent convective dynamo model.

For main sequence stars older than $\sim 50$ Myr, it has long been
known that younger stars are more magnetically active than older
stars. \citet{2005ApJS..160..390P} clearly establishes that the
activity-age relation continues through the PMS phases (Figure
\ref{fig:xray:fig10}~right) and find a decay law that is
mass-dependent at young ages. \citet{2005ApJS..160..423W} used a
complete sample of 1 solar mass Orion stars in the COUP field to show
that analogs of the young Sun spend one-fourth of their time in flare
state, exhibit incredibly high levels of magnetic activity with the
median luminosities 2-3 orders of magnitude higher for both
``quiescent'' and peak flare levels compared to the contemporary
Sun. \citet{2007A&A...471..645C} further show that X-ray flare
frequency in young lower-mass $(0.1-0.3M_{\odot})$ stars is
indistinguishable from that of the young solar analogs.  Finally,
\citet{2007ApJ...660.1462M} find that elemental coronal abundances in
X-ray luminous young Orion stars are similar to those of older
magnetically active stars.
\begin{figure}
	[h*]

	\centering
	\includegraphics[angle=0.,totalheight=3.75in]{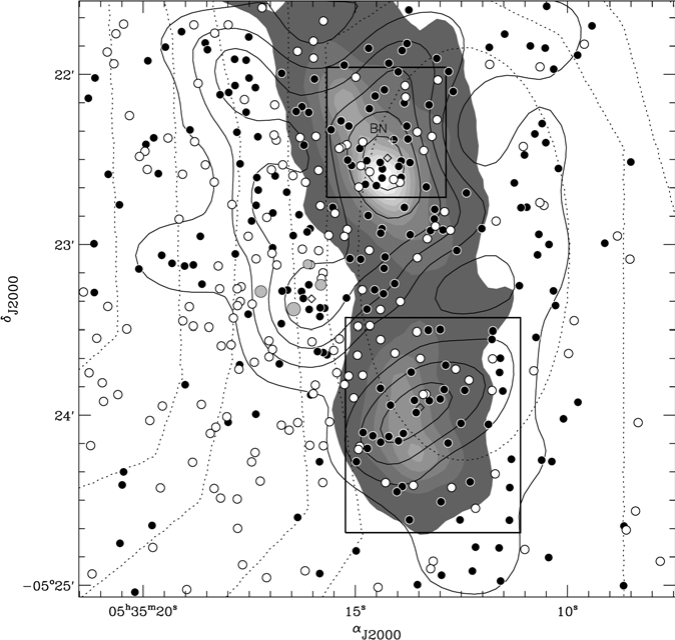}

	\caption{Distribution of COUP sources differentiated by X-ray
	absorption column density and compared to the SCUBA
	$450\,\micron$ map of \citet[][shaded
	contours]{1999ApJ...510L..49J}. Black dots: obscured ($N_{\rm
	H} > 10^{22}$\,cm$^{-2}$) sources; white dots:
	lightly-absorbed ($< 10^{22}$\,cm$^{-2}$) COUP X-ray
	sources. Three major sub-clusterings are differentiated:
 ordered from North to South these include the {B-N/K-L} (boxed),
 the {Trapezium,} and {OMC-1S} (boxed). Figure reproduced from
	\citet{2005ApJS..160..530G}.  \label{fig:xray:fig3}}
\end{figure}

\begin{figure}
	\centering
	\begin{minipage}[c]{0.47\textwidth} \centering
		\includegraphics[angle=0.,width=\textwidth]{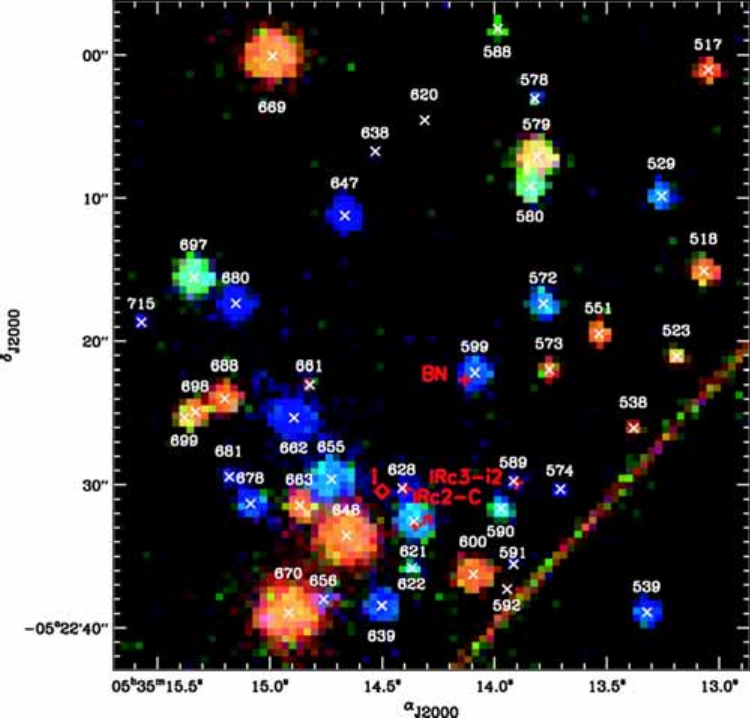}
	\end{minipage}%
	\hspace{0.04\textwidth}	%
	\begin{minipage}[c]{0.47\textwidth} \centering
		\includegraphics[angle=0.,width=\textwidth]{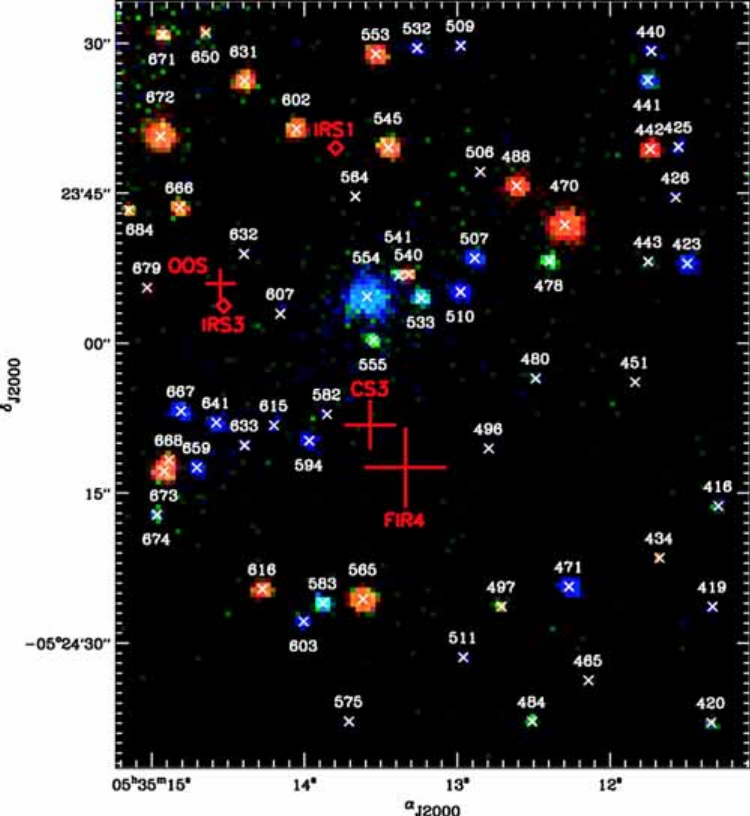}
	\end{minipage}%
	\caption{A comparison of two embedded subclusterings in the
	Nebula as traced by COUP multi-spectral data. The figures show
 the {B-N/K-L} \textit{(left)} and {OMC-1S} \textit{(right)}
	subclusterings, corresponding to the regions outlined in
	Figure \ref{fig:xray:fig3}. Color coding (red, green, and
	blue) correspond to photons in the 0.5-1.7 keV,
	1.7-2.8 keV, and 2.8-8.0 keV bands; red
	markings correspond to well known protostars, hot cores or
	far-infrared source. Figures reproduced from
	\citet{2005ApJS..160..530G}. \label{fig:xray:embedded}}%
\end{figure}

\subsubsection{X-rays from Embedded
  Sources} \label{sec:ttauri:xray:embedded} The spatial distribution
of ``obscured'' COUP sources clearly traces the basic structures of
 the central cluster; each of {Trapezium} core, the {B-N/K-L} and {OMC-1S}
regions appear as over-densities in Figure \ref{fig:xray:fig3}. The
detailed properties of the COUP detected X-ray embedded sources in
 {B-N/K-L} and {OMC-1S} regions (see boxes in Figure \ref{fig:xray:fig3})
are discussed by \citet{2005ApJS..160..530G}. Grosso et~al. found 60
 COUP X-ray sources toward the {OMC-1S} dust continuum core, with more
 than 60\% of them being obscured. In the {B-N/K-L} region 43 sources
were detected and half of these were obscured. Based on comparison of
the X-ray luminosity function of the observed X-ray populations
 embedded in {OMC-1S} and {B-N/K-L} with that of the unobscured {ONC}
population, \citeauthor{2005ApJS..160..530G} estimated total
 populations of 70 versus 80 embedded stars residing inside the {OMC-1S}
 and {B-N/K-L} cloud cores, respectively.

Close-up images of these two regions, scaled to the same physical size
are compared in Figure \ref{fig:xray:embedded}. This is the first
 direct measurement of the low-mass population of the {OMC-1S} cluster
 with 18 new X-ray sources without infrared counterparts. {COUP OMC1-S}
detections include the most embedded X-ray source in the COUP survey,
COUP 632 (= TPSC 1), a protostar with $A_V \sim 500$~mag of visual
absorption. X-ray sources are found close to four luminous mid-IR
 sources {-- B-N,} IRc3-i2, IRc2-C, and Source n -- but their X-ray
variability and spectral properties are typical of coronal activity of
low-mass companions rather than wind emission from massive stars. No
X-ray emission is seen from the radio-bright massive protostar Source
I.

Using the combination of compiled CTIO-ISPI near-IR with Spitzer IRAC
mid-IR imaging data, \citet{2007arXiv0712.2975P} establish the list of
45 protostellar candidates within the COUP field of view: 23
designated as Class~0-Ia with their IR SEDs monotonically rising from
$K$ to $8\;\mu\mbox{m}$ and 22 designated as Class~0-I with SEDs
rising from $K$ up to $4.5\;\mu\mbox{m}$. Of these, $62\%$ have X-ray
counterparts in COUP data. Their tabulations also contain
cross-references to a number of thermal IR surveys of the nebula. The
spatial distribution of these protostellar candidates trace the dense
 molecular filament that extends northward from {OMC-1} to {OMC-2/3} clouds
and is similar to that of 75 likely new embedded cloud members found
in \citet{2005ApJS..160..319G}. However, due to nebular contamination
and crowding in mid-IR, Prisinzano et~al. were not able to classify
 many X-ray embedded sources located in {B-N/K-L} and {OMC-1S} regions
\citep{2005ApJS..160..530G}. In addition, a sub-cluster of seven
 highly embedded X-ray sources in {OMC-1S} (COUP \# 582, 594, 615, 633,
641, 659 and 667) at approximately RA,DEC = 05:35:14.8, -05:24:12
(J2000; see also Figure \ref{fig:xray:embedded}) is simply invisible
in Spitzer data. In regards to the evolution of the X-ray emission,
Prisinzano et~al. find that Class~0-Ia protostellar candidates are
intrinsically less luminous than the Class~II stars.
\begin{figure}
	[htbp] \centering
	\includegraphics[height=3in]{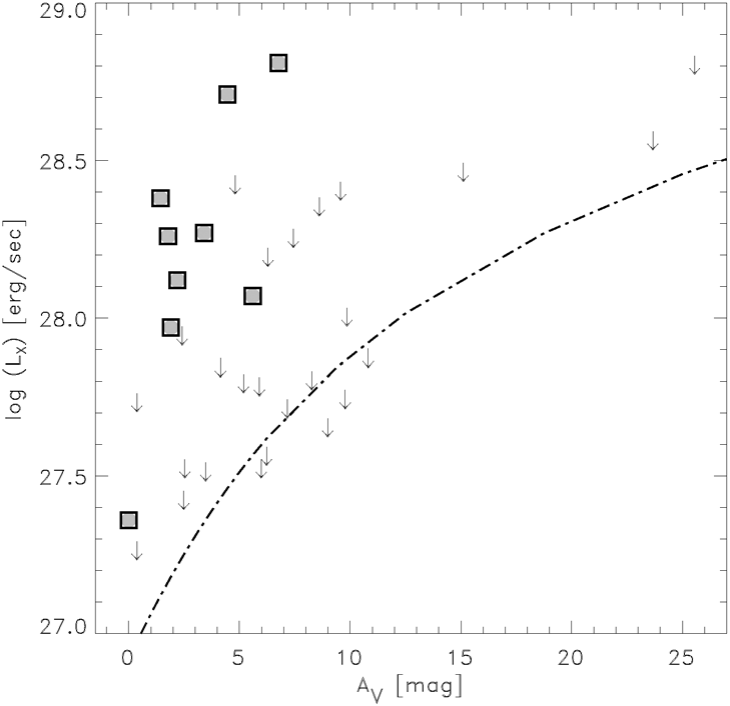}

	\caption{The relationship between optical extinction and X-ray
	luminosity (or its upper limit as marked by arrows) for 34
	Orion brown dwarfs. Figure reproduced from
	\citet{2005ApJS..160..582P}. Detectability of Orion brown
	dwarfs in X-rays by the COUP was limited, apparently, by
	extinction.} \label{fig:Figures_pre_onc_bd_xray}
\end{figure}

\subsubsection{X-ray Properties of Orion Brown
 Dwarfs} \label{sec:bds:xrays} In the core of the {ONC,} 9
spectroscopically-identified brown dwarfs were detected with the COUP
\citep{2005ApJS..160..582P}. The low detection rate is seemingly
related to the substantial extinction to most of these brown dwarfs
(Figure \ref{fig:Figures_pre_onc_bd_xray}). There is no evidence for
changes in the magnetic activity around the stellar/substellar
boundary; instead the X-ray properties of the detected brown dwarfs,
including spectra, fractional luminosities, and flare rates, are
 similar to those of the low-mass stars in the {ONC.} Trends in the
fractional X-ray luminosity and surface flux and a comparison to the
X-ray properties of late type field dwarfs led
\citet{2005ApJS..160..582P} to conclude that the photospheric
temperature of late type stars rather than source mass or surface
gravity controls the X-ray emission mechanism.

\subsubsection{X-ray Observations of Orion Flanking
  Fields} \label{sec:ttauri:xray:flanking} Because the COUP survey
 samples only the inner parts of the {Orion Nebula,} X-ray surveys of
what have been deemed the Orion ``Flanking fields'' are
important. This is because they provide membership information at
large cluster radii, where other methods are ambiguous. The
\citet{2004AJ....128..787R} survey follows the axis of the cloud,
sampling \textit{Chandra} ACIS fields north and south of the immediate
Nebula. While they are much less sensitive than the COUP observations,
their results provide interesting evidence for age gradients along the
 axis of the cloud. New surveys with XMM covering all of the {OMC} and
 {Orion A} clouds are currently being completed and will soon yield more
spatially complete results as well as extensive publicly available
archival data.

\subsection{Spectroscopic Surveys} \label{sec:ttauri:spec}

\subsubsection{Historic Studies} The ``Orion population" stars have
long been of interest to spectroscopists.  Early and numerous studies
using objective prism plates were published by e.g. Herbig, Haro,
Sharpless, Johnson, Walker, and Penston among others.  These authors
were interested in emission-line aspects as well as spectral types
 over the {Orion Ic} and Id regions. Specifically, spectral types for
 stars in the {ONC,} as defined above, were reported by:
\citet{1963ApJ...137..513B};
\citet{1979ApJS...41..743C};
\citet{1946PASP...58..366G};
Herbig as quoted in \citet{1969ApJ...155..447W};
\citet{1965ApJ...142..964J};
\citet{1976PASP...88..712L};
\citet{1977PASP...89..797A};
\citet{1960PASP...72..268L};
\citet{1976AJ.....81..845M};
\citet{1954TrSht..25....1P};
\citet{1973ApJ...183..505P,1975MNRAS.171..219P};
\citet{1958ApJ...128...14S} though referenced mostly to Sharpless;
\citet{1983ApJ...271..237S};
\citet{1931PASP...43..255T};
\citet{1983ApJ...271..642W}.
Many of these studies also include the larger Orion population. In the
majority of cases the literature of this era was focused on
identifying the emission line stars, on characterizing the cluster
sample, and on understanding whether the fainter objects should be
interpreted as reddened, or lower mass, or of older age.

\subsubsection{Modern Low Resolution Surveys}  Modern optical spectra
 of a few tens of {ONC} stars were produced by pioneers such as
\citet{1986ApJ...307..609H} and \citet{1988AJ.....95.1744V}, while
photographic H$\alpha$ plate surveys have also continued (see previous
Sect. \ref{sec:ttauri:halpha}). The largest collection of published
spectral types is contained in \citet{1997AJ....113.1733H}, which
incorporated new data on many hundreds of stars as well as previously
published (those references above plus more modern additions from
e.g. \citet{1993ApJ...406..172D}; \citet{1993AJ....106..372E};
\citet{2004ApJ...601..979W}, and unpublished (e.g. Prosser \& Stauffer
spectra; \citet{1993PhDT.......360S}, PhD thesis, and Hamilton 1994,
MSc thesis) information.  Approximately 950 spectral types were
provided. Since the \citet{1997AJ....113.1733H} publication, however,
approximately 800 more spectral types over the same projected area
have become available; an updated catalog is being prepared
(Hillenbrand et al.).

Relevant sources of new optical spectroscopy include
\citet{2007MNRAS.381.1077R} and spectral types for a few of the
sources in Slesnick et al. (2004). Infrared spectroscopy includes
\citet{2000ApJ...540.1016L}, but has more recently focussed on the
 lowest mass candidate members of the {ONC} with contributions by
\citet{2001MNRAS.326..695L,2006MNRAS.373L..60L}, and
\citet{2004ApJ...610.1045S}. There is also ongoing work of Lada with
the FLAMINGOS multi-object spectrograph. Very late M and perhaps even
 L0 or L1 objects have now been identified in the {ONC} region.

\subsubsection{Modern Echelle Surveys}  Ushering in modern high
 dispersion studies of the {Orion Ic/Id} region were the works of
\citet{1983ApJ...271..237S}, \citet{1990PASP..102..726W},
\citet{1990ApJ...350..348M}, \citet{1991ApJ...367..155A}, and
\citet{1991ApJS...76..383D, 1993ApJ...406..172D} which all focused on
rotational velocities.  \citet{1993AJ....105.1087K},
\citet{1996PASP..108..738D}. \citet{2005ApJ...626L..49P}
\citep[and][]{2007ApJ...659L..41P} subsequently studied lithium
abundances as well as radial velocities for small samples of Orion
stars; \citet{2005AJ....129..363S} derived the same for a larger
sample of several hundred stars, and also provided rotational
velocities.  Additional rotational velocities come from
\citet{2001AJ....122.3258R} and \citet{2004ApJ...601..979W} who both
studied stellar angular momentum.  \citet{1996ApJ...471..847P},
\citet{1995ApJ...452..634C,1998ApJ...493..195C}, and
\citet{2005ApJ...626..425C} published work on abundances including
 several {ONC} stars. \citet{2008ApJ...676.1109F} provide radial
 velocities for a large sample of {ONC} stars in a study of cluster
kinematics.  Most recently, \citet{2007AAS...211.6219Y} have achieved
 the means to study the magnetic fields of {ONC} stars.

At this time, there is a considerable amount of data across the
 stellar mass spectrum on rotational velocities in the {ONC} that,
together with rotation periods, are providing insights into stellar
angular momentum at the earliest stages of stellar evolution.  Lithium
samples are far smaller, but remain valuable.  Abundance information
and magnetic field measurements are intriguing but remain rather
limited at present.
\begin{figure}[ht*]
	\centering
		\includegraphics[angle=-90,width=\textwidth]{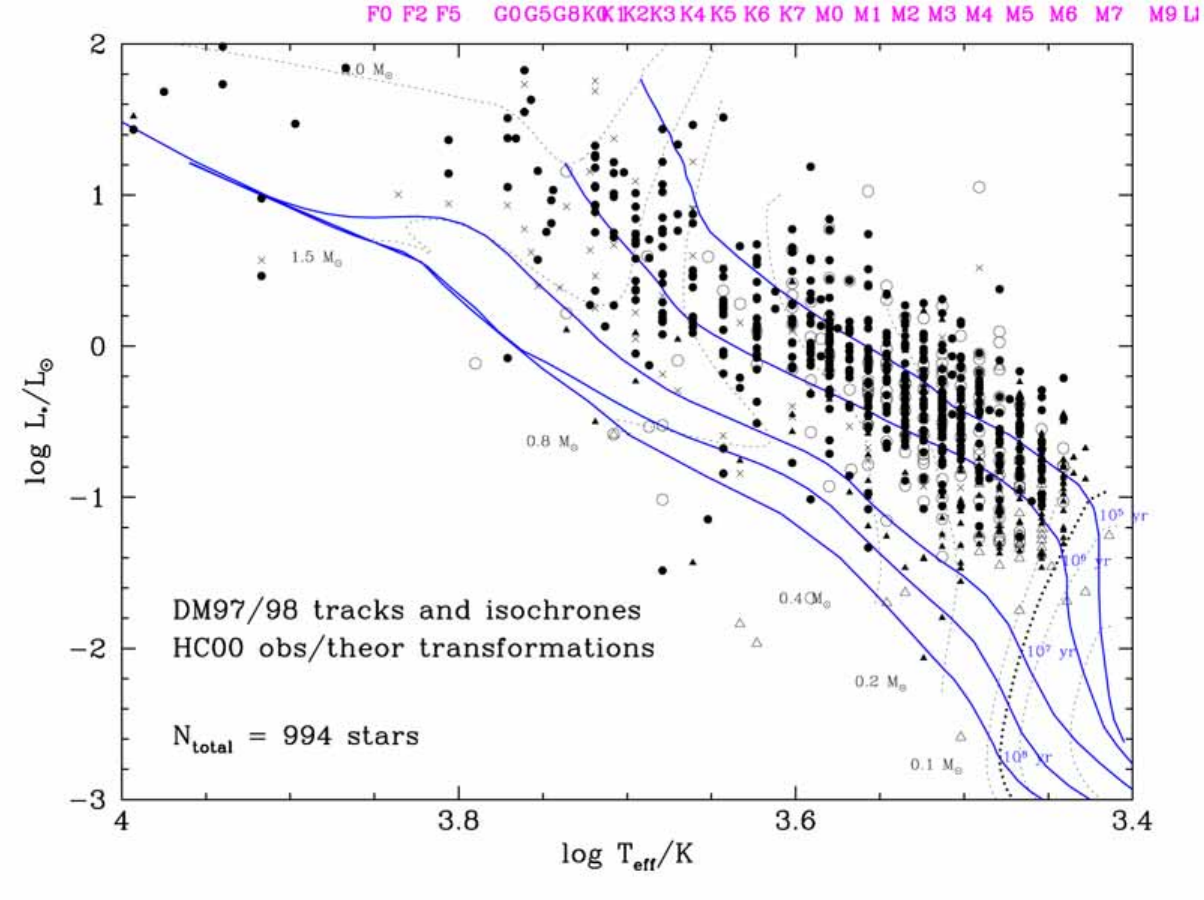}
 \caption{Low mass Hertzsprung-Russell Diagram for the {Orion
 Nebula Cluster} \citep[this is an updated version of that
	constructed in][]{1997AJ....113.1733H}. Isochrones and fixed
	mass evolution tracks are from \citet{1997MmSAI..68..807D}.}
	\label{fig:Figures_lah97_hr}
\end{figure}

\subsection{Stellar Properties} \label{sec:ttauri:prop}

The photometric and spectroscopic surveys described in the previous
two sections have provided the panchromatic data necessary to assess
 the stellar population of the {ONC.} Compared to other such studies of
 older clusters, significant challenges are posed in the {ONC} by the
effects on observed colors of nebular contamination and disk accretion
(both of which cause blueing), and circumstellar and interstellar dust
(reddening). Spectral continuum and certain spectral lines are also
affected. Nevertheless, perseverance has led to understanding of
typical ages and to estimates for individual stellar masses.

\subsubsection{Stellar Ages and Star Forming History}
Notwithstanding subsequent criticism of his data quality by Walker
(1956), \citet{1954TrSht..25....1P} was first to observe a ``cloud''
of subgiants later than spectral type A0 and more luminous than the
 main sequence in the vicinity of the {Orion Nebula.} Along a pathway
 similar to his previous studies of {NGC 2264,} {NGC 6530,} {NGC 6611,} and
 {IC 5146,} \citet{1969ApJ...155..447W} used photoelectric photometry of
sources listed in \citeauthor{1935POLyo...1...12B}'s catalog to
confirm the existence of pre-main sequence stars in the Nebula. Other
photometric $+$ spectroscopic surveys of the 1970's followed suit. It
was \citet{1986ApJ...307..609H} who quantitatively demonstrated the
 youth of the {ONC} stars by comparing their dereddened color-magnitude
diagram to the theoretical evolutionary tracks of
\citet{1985ApJS...58..561V}. They found that the vast majority of
stars were younger than 1~Myr, and therefore seemingly inconsistent in
their age distribution with the canonical 10~Myr life times of
molecular clouds \citep{1980ApJ...238..148B}. Improvements in the data
and in the theory resulted in similar conclusions being drawn by
subsequent workers, e.g. \citet{1994ApJ...421..517P},
\citet{1997AJ....113.1733H} and \citet{2004ApJ...610.1045S}.
Figure~\ref{fig:Figures_lah97_hr} is an update of the
\citet{1997AJ....113.1733H} HR diagram illustrating this star forming
history.  \citet{1997AJ....113.1733H} further suggested a mild age
gradient, featuring young median ages closer to the cluster core and a
slightly older population in the outer nebula region about 2.5~pc from
the core.

However, these later studies also highlighted the existence of a
seemingly rogue population of apparently older stars, that is, those
located well below the main distribution (See the HR diagram of low
mass stars in Figure \ref{fig:bds:imf}~left). These have been
variously interpreted as sources that are coeval with the others but
affected by circumstellar material that renders them visible only in
scattered light, or sources that truly are as old as they appear and
therefore offer evidence for large age spreads in star forming
 regions. Although the typical age of {ONC} members is widely agreed to
be $\sim$1-2~Myr (modulo the systematic effects caused by distance or
by adoption of various among plausible sets of theoretical pre-main
sequence evolutionary tracks), there is still considerable debate
regarding the meaning of the apparent luminosity spread. The problems
illuminated in detail by e.g. \citet{1976AJ.....81..845M} remain. On
the one hand, there is significant evidence that the error budget for
individual stellar luminosities is underestimated due to photometric
variability, difficulties with extinction corrections, unaccounted for
multiplicity, etc.; these effects are in addition to known systematic
problems with the effective temperatures. On the other hand, there are
also known processes such as inefficient convection, large amplitude
magnetic fields, accretion of new material, and perhaps rotation, that
pertain to pre-main sequence evolution. Together these phenomena
suggest caution in any literal interpretation of apparent luminosity
spreads in HR diagrams as actual age spreads.

There is, however, evidence from \citet{2005ApJ...626L..49P,
2007ApJ...659L..41P} regarding spreads in the lithium depletion of
 young low mass stars in the {ONC} that seemingly supports the range in
ages inferred from the luminosity spread in the HR diagram. Ages of
10-30 Myr are derived for a small fraction of the (notably a
moderately to heavily veiled) sample. Further,
\citet{2007MNRAS.381.1169J} used a combination of rotation periods and
$v\sin{i}$ values to infer a distribution of $R~\sin{i}$ values and
hence a statistical distribution of stellar radii $R$ ranging over a
factor of 2-3, which they also argue is inconsistent with an age
dispersion less than 0.3 to 0.5 dex. Clearly more work is needed in
order to resolve the debate regarding the duration of star formation
episodes relative to cluster crossing times.

\begin{figure}[t*]
	\centering
	\includegraphics[angle=0,totalheight=3.5in]{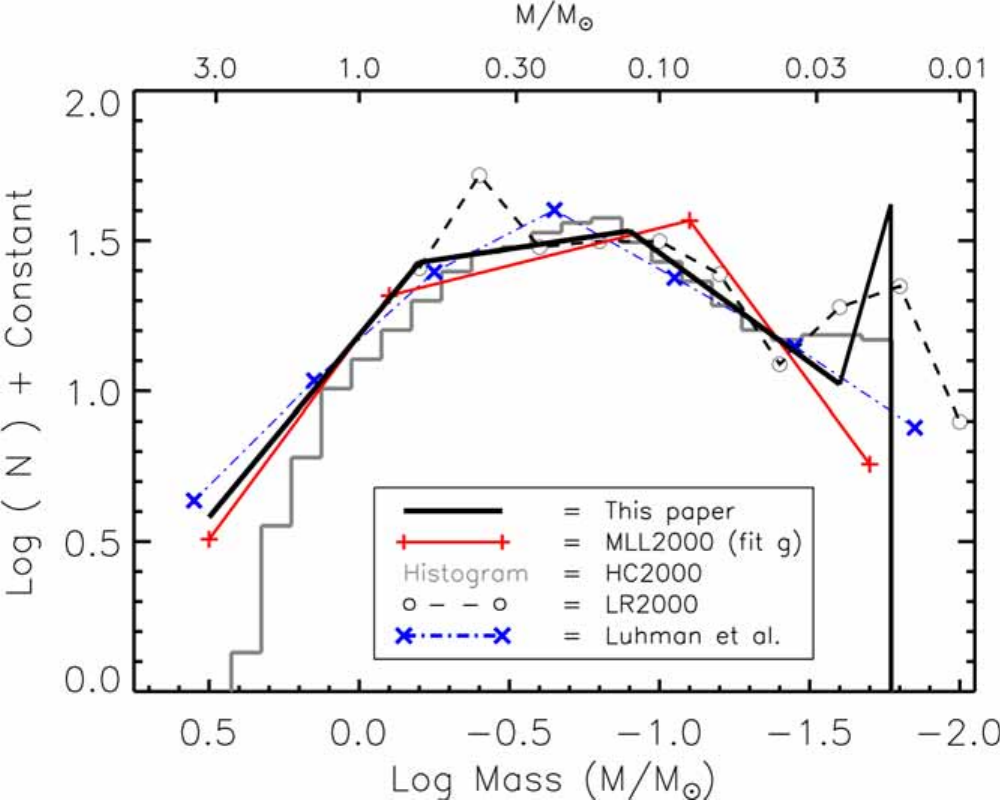}
	\caption{Comparison of IMFs derived for the central $5\arcmin$
 surrounding the Trapezium OB stars in the {Orion Nebula.} Figure
	reproduced from
	\citet{2002ApJ...573..366M}. \label{fig:imf_comp}}
\end{figure}

\subsubsection{Stellar Masses and the Initial Mass Function}  The ONC
was an important test case regarding ideas of ``bi-modal" star
formation in which high-mass and low-mass stars were suggested as
incapable of forming in the same place and at the same
time. Empirically, processes similar to those used to assess stellar
ages from HR diagrams were also used to assess stellar
masses. Evolutionary tracks such as those by
\citet{1966ApJ...144..968I} and \citet{1985ApJS...58..561V} showed
early on that such a bimodal scenario was not applicable to the
 {ONC.} Yet it was not until the 1990's that quantitative measurements of
 the initial mass function were claimed for the {ONC.}

The first efforts involved use of the surrogate luminosity function,
and stellar models translated to the empirical plane with some strong
assumptions regarding age spreads and multiplicity. \citet[][see also
\citet{1993prpl.conf..429Z}]{1991MmSAI..62..761Z} were the first to
apply these techniques to newly available $2~\mu$m survey data of the
 {ONC.} These efforts were followed by others including
\citet{1994AJ....108.1382M}, \citet{1995AJ....109..709A} and
\citet{2000ApJ...533..358M, 2002ApJ...573..366M} that established the
existence of stars as low in mass as the hydrogen burning limit
(0.073~M$_\odot$), as well as a substantial number of likely brown
dwarfs (notably before the confirmation of Gl229b as the first bona
fide brown dwarf!). \citet{2005MNRAS.361..211L} have extended these
arguments to the deuterium burning threshold (0.005 M$_\odot$ or 5
M$_{Jupiter}$). \citet{1994ApJ...421..517P} carried out similar
analysis using $V$ and $I$ data from HST. As an intermediate step
between one-dimensional luminosity functions and two-dimensional HR
diagram methods, \citet{2000ApJ...540..236H} applied a statistical
technique to $H$ and $K$-band color-magnitude diagrams to determine
stellar mass probability functions which could be summed to form an
initial mass function.

Hillenbrand (1997) produced the first ``forward modeling'' method to
 derive an initial mass function in the {ONC,} making use of the full HR
diagram (Figure~\ref{fig:Figures_lah97_hr}) established from optical
photometry and spectroscopy, and the evolutionary tracks of
\citet{1994ApJS...90..467D} and
\citet{1994ApJ...425..286S}. \citet{2000ApJ...540.1016L} and
\citet{2004ApJ...610.1045S} used infrared photometry and spectroscopy
to push the investigation to lower masses, across the brown dwarf
limit.

All of the above studies are consistent with a mass function that
rises in a Salpeter like fashion from the highest masses
 \footnote{\citet{2006MNRAS.373..295P} argue, however that the {ONC}
suffers a deficit in stars more massive than 5$M_{\odot}$.} to the
sub-solar regime, then begins to flatten around 0.5-0.6 M$_{\odot}$
with a peak around 0.2-0.3 M$_{\odot}$ and then turns over into the
brown dwarf regime. The exact details depend on the methods, the
samples, and the models though, as demonstrated by \citet[][see Figure
\ref{fig:imf_comp}]{2002ApJ...573..366M}, there is remarkable
agreement considering possible sources of variance.

\subsubsection{Binarity of T~Tauri Stars in the Nebula} The Orion
Nebula Cluster offers an excellent opportunity to study the frequency
of binarity among young low-mass stars in a clustered environment and
to determine the shape of the binary separation distribution function
at an early age. The first major survey for subarcsecond visual PMS
 binaries in the {ONC} was carried out by \citet{1994ApJ...421..517P}
using the \textit{Hubble} Space Telescope. On their high-resolution
optical images they found 35 new visual binaries with separations up
to 1~arcsec. \citet{1997ApJ...477..705P} and
\citet{1998ApJ...500..825P} found another 7 and 4 binaries,
respectively. The former study used archived $V$-band HST images to
 study an area $R>35\arcsec$ from the {Trapezium} while the latter used
speckle holography to reconstruct ground-based near-infrared $HK$
images within the central $R<30\arcsec$. Additional visual binaries
were found by \citet{1999AJ....117.1375S}~$(R<\sim1.5\arcmin)$ and
\citet{2006A&A...458..461K}~$(15>R>5\arcmin)$, both based on adaptive
optics K-band observations. The results of \citet{2008A&A...477..681B}
illustrate how continued improvements to AO instrumentation will add
new binary systems to the Orion census. Finally, the COUP X-ray survey
 identified several binaries in the {ONC} \citep{2005ApJS..160..353G}.

Recently, \citet{2007AJ....134.2272R} carried out a major survey of
 the {ONC} using H$\alpha$ images obtained with the HST ACIS. In a region
extending from 1 to $\sim20\arcmin$ ($\sim0.1\mbox{ to }2\;\mbox{pc}$)
 surrounding {$\theta^{1}$~Ori~C} they found 78 multiple systems, of
which 55 are new discoveries, with separations less than
 $1.5\arcsec$. Because of the high stellar density in the {ONC,} it can
be statistically determined that 9 of these must be line-of-sight
associations (see also the analysis of \citet{1997ApJ...482L..81S} and
\citet{1998MNRAS.297.1163B} regarding the impact of high stellar
density on true binary fraction). When correcting for this, a binary
fraction of 8.8\%$\pm$1.1\% is found in the limited range from 67.5 to
675~AU. In the same range, the field binary fraction is a factor 1.5
times higher \citep{1991A&A...248..485D}, and the binary fraction in
 loose associations is a factor 2.2 higher than in the {ONC}
\citep{1993A&A...278...81R}. The separation distribution function
shows unusual structure, with a steep decrease in the number of
binaries for separations larger than $0.5\arcsec$. Moreover, the ratio
 of wide to close binaries across the {ONC} shows a major depression
towards the central region, indicating that wide binaries are
destroyed as they pass through the central potential well, as
theoretically expected \citep[e.g.][]{1999NewA....4..495K}.

\subsection{Sub-stellar Objects in the Nebula} \label{sec:bds}

Optical and near-IR imaging surveys (Sect.~\ref{sec:ttauri:survey})
have revealed at least 100 sources within the central 0.5~pc of the
 {Trapezium} that, according to pre-main sequence theory, have
luminosities at an age of 1~Myr that are consistent with their being
substellar objects. While similarly deep images of the outer nebula
exist, photometric censuses have not yet been published. Purely
photometric selection techniques are, however, less precise at
securely identifying brown dwarfs, since their luminosity is a
function of age. For example, \citet{2004ApJ...610.1045S} reported a
number of warmer sub-luminous objects that masquerade as lower
luminosity brown dwarfs (see Figure \ref{fig:bds:imf}~right). Whether
these sources are older or their fluxes contaminated perhaps by
scattered light is not yet clear. Thus, the identity of an individual
young stellar object (YSO) as substellar rather than pre-stellar is
best established using measures of that source's effective surface
temperature and/or surface gravity. The precise boundary at which a
YSO can never attain sustained nuclear burning comes from theoretical
evolutionary models, with a working consensus
\cite[e.g.,][]{1998ApJ...493..909L} that includes young ($<\sim5$~Myr)
objects cooler than $T_{eff}<\sim3500$~K or a spectral type later than
``M6.'' According to this criterion, there are $\sim35$ such objects
in the most recent version of the \citet{1997AJ....113.1733H} wide
field catalog.
\begin{figure}
	\centering
	\begin{minipage}[c]{0.52\textwidth} \centering
		\includegraphics[angle=0,width=\textwidth]{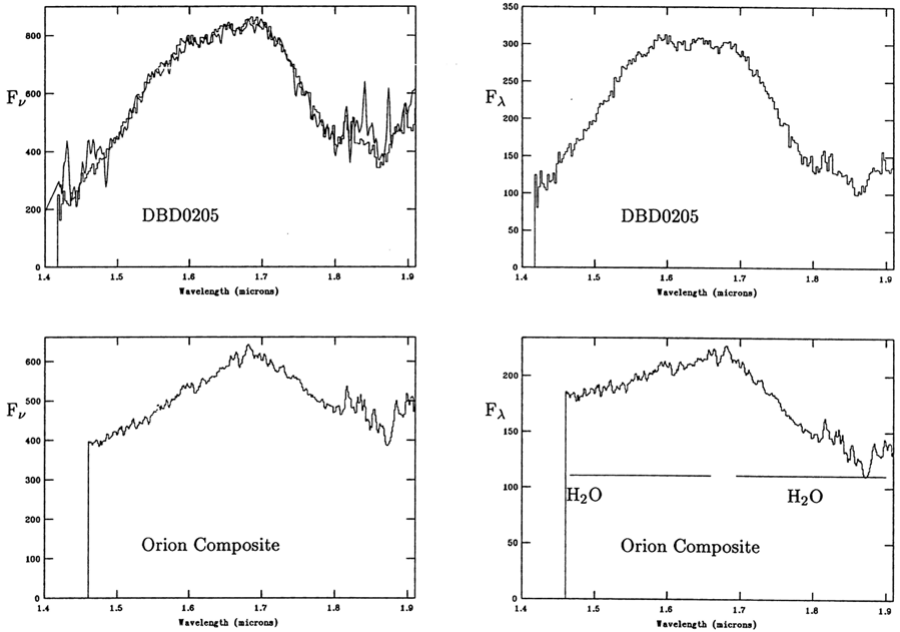}%
	\end{minipage}%
	\hspace{0.04\textwidth}	%
	\begin{minipage}[c]{0.42\textwidth} \centering
		\includegraphics[angle=0,width=\textwidth]{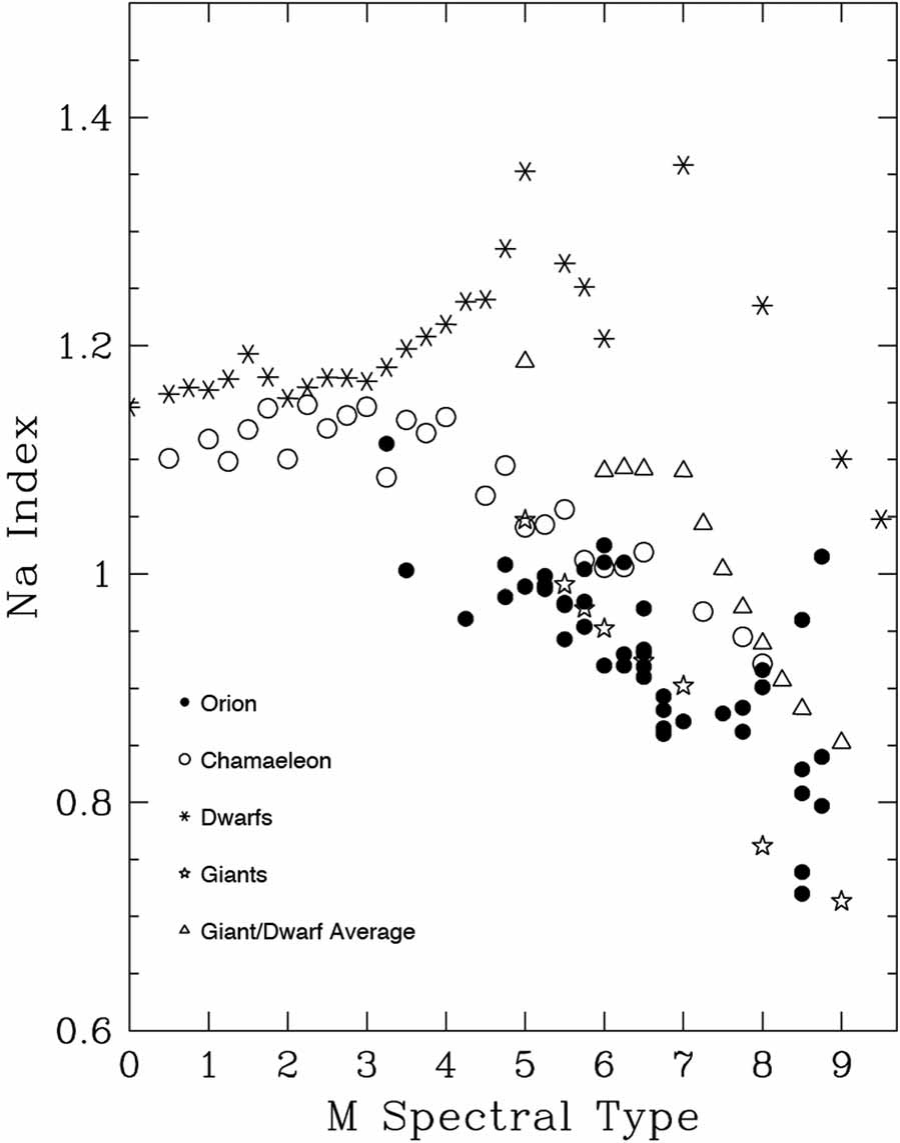}%
	\end{minipage}%
	\caption{Surface gravity of young Orion brown dwarfs. Left:
	sharpening of the $H (1.6~\mu m)$ band spectra compared to
	field dwarfs with high surface gravities and theoretical
	models. Reproduced from \citet{2001MNRAS.326..695L}; Right: a
	comparison of the optical Na I 8183/8195\AA\ index for Orion
	sources, young stars in other regions and field dwarfs or
	giants. Reproduced from \citet{2007MNRAS.381.1077R}
	\label{fig:bds:optical}.}
\end{figure}

\subsubsection{Surface Gravity}  Especially over the last five years,
the utility of spectroscopic surface gravity diagnostics (e.g. NaI,
CaI, KI, TiO, VO, CO, steam) has become apparent, especially towards
the lowest masses, and can in principle be used to distinguish bona
fide cluster members from both faint red foreground dwarfs and
reddened background giants.  Examples of such surface gravity
 sensitive observations of {Orion Nebula} sources are given in Figure
\ref{fig:bds:optical}, including NaI (8183/8195\AA) and steam in the
near-IR $H$ band. Relevant studies in the optical include \citet[][see
  also Figure \ref{fig:bds:optical}]{2007MNRAS.381.1077R}. Near-IR
studies include \citet{2000ApJ...540.1016L}, \citet[][see also Figure
  \ref{fig:bds:optical}]{2001MNRAS.326..695L}, which were the first
observations to reveal the sharp ``triangle'' shape of the $H$ band
spectra for low surface gravity late M stars,
\citet{2004ApJ...610.1045S}, \citet{2005ApJ...625.1063S}, and
\citet{2006MNRAS.373L..60L}.  However, there are not yet well
established methods at all spectral types and the published data on
these Orion sources reflect a lack of uniform spectral
typing. Furthermore, the intermediate-gravity nature of young pre-main
sequence stars is not readily apparent in all cases.  Specifically,
many stars with strong evidence from both emission lines and infrared
excesses for membership are not distinguished as such based on surface
gravity alone. Establishment of cluster membership may need to rely on
kinematic association in addition to surface gravity indicators.

\begin{figure}[h*]
	\centering
	\begin{minipage}[c]{0.42\textwidth} \centering
		\includegraphics[angle=0,width=\textwidth]{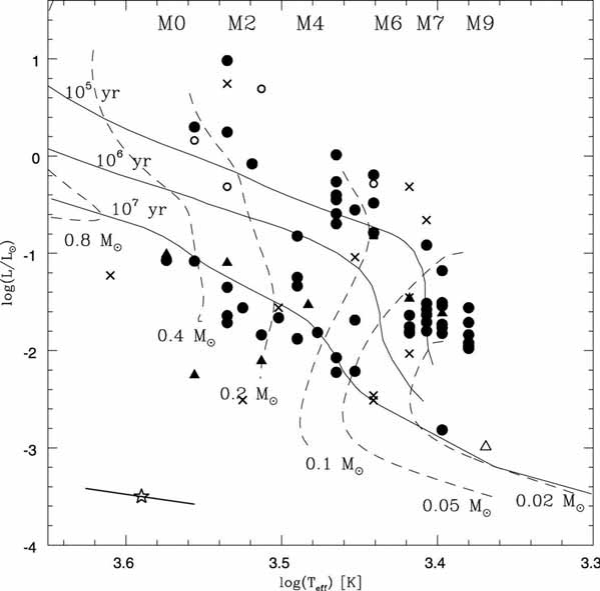}%
	\end{minipage}%
	\hspace{0.04\textwidth}%
	\begin{minipage}[c]{0.52\textwidth} \centering
		\includegraphics[angle=0,width=\textwidth]{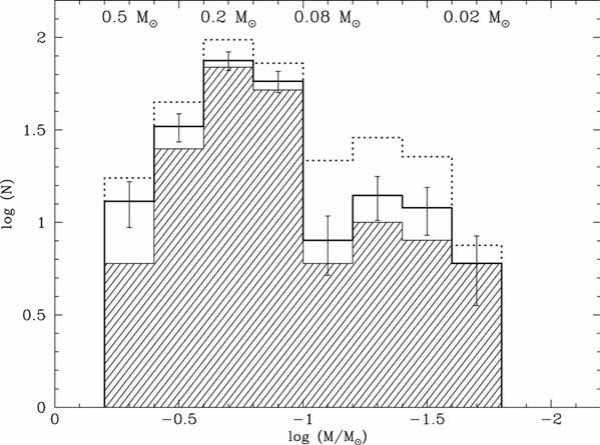}%
	\end{minipage}
	\caption{Left: Central Orion Nebula M~dwarf Hertzprung-Russell
	diagram; Right: resulting spectroscopically derived
	IMF. Reproduced from
	\citet{2004ApJ...610.1045S,2005ApJ...625.1063S}
	\label{fig:bds:imf}.}
\end{figure}
\subsubsection{Masses} By number, brown dwarfs constitute between $20$
 and $30\%$ of the total {ONC} membership
\citep{2002ApJ...573..366M,2004ApJ...610.1045S, 2006ApJ...646.1215L},
although the precise
value depends in part upon the assumption of a luminosity and
effective temperature for a star at the hydrogen burning limit
\citep{2006ApJ...646.1215L}. By mass this is negligible $(<\sim1\%)$
compared to the overall conversion of gas to stars in this cloud. The
reported substellar IMF alpha (per unit log mass) is $\sim0.6$
\citep{2000ApJ...540..236H, 2002ApJ...573..366M}, however, this
function is poorly described by a single power-law as seen in Figures
\ref{fig:imf_comp} and \ref{fig:bds:imf} (right). Additionally, the
 near-IR luminosity function of the central {ONC} displays structure at
the faint end that is greater than expected purely from field star
contamination but is not consistent with a simple declining power-law
IMF given current PMS theory \citep{2002ApJ...573..366M}.

\subsubsection{Interesting Very Low Mass
 Sources} \label{sec:bds:special} The object {2MASS J05352184-0546085}
was found to be a unique brown dwarf-brown dwarf double line eclipsing
binary system  by \citet{2006Natur.440..311S}. The combination of
photometric and spectroscopic monitoring allowed individual masses
$(\sim0.04\,M_{\odot})$ and radii~$(\sim0.6\,R_{\odot})$~to be
determined.   Interestingly, the higher mass component has a cooler
surface temperature than the lower mass component
\citep[][]{2007ApJ...664.1154S,2008ApJ...674..615S}; more recent
evidence of strong surface activity on the primary
\citep{2007ApJ...671L.149R} indicates this temperature reversal could
be explained by the inhibition of convection by strong magnetic fields
\citep{2000ApJ...543L..77D}. The sustained efforts of deep monitoring
projects in Orion are beginning to uncover additional low mass double
line eclipsing systems, which can provide improved constraints on very
 low mass pre-main sequence evolutionary theory. These include {JW~380}
 \citep[][$M\sim0.2\,M_{\odot}$]{2007MNRAS.380..541I} and {Parenago~1802}
\citep[][$M\sim0.4\,M_{\odot}$]{2008ApJ...674..329C}.

\subsection{Circumstellar Matter} \label{sec:ttauri:disk}

With a rich stellar population spanning the mass spectrum all the way
 from O-type massive stars to late-M-type brown dwarfs, the {ONC} is an
obvious arena for investigations of circumstellar disks.  Known for
decades as a rich collection of emission line objects
 \citep[e.g.,][]{1953ApJ...117...73H,1982BITon...3...69P}, the {ONC} was
also an early target of infrared investigations. In particular, the
single channel photometer measurements of \citet{1984AJ.....89..399R},
\citet{1981ApJ...248..963B}, \citet{1976PhDT........20S},
\citet{1976AJ.....81..845M},
\citet{1973ApJ...183..505P,1975MNRAS.171..219P},
\citet{1973ApJ...180..809N}, and \citet{1968ApJ...152..913L}
 established that many {ONC} stars had excess emission at near- to
mid-infrared wavelengths. The frequency of circumstellar disks around
 brown dwarfs in the {Trapezium} appear to be a smooth continuation of
this property for higher mass objects \citep{1998AJ....116.1816H,
2000AJ....120.3162L, 2001ApJ...558L..51M, 2004AJ....128.1254L}.%

The advent of infrared arrays led to the studies of
\citet{1994AJ....108.1382M, 1996AJ....111.1977M},
\citet{1995AJ....109..709A}, \citet{1994AJ....107.2120J},
\citet{1998AJ....116.1816H}, \citet{1999AJ....117.1375S},
\citet{2000ApJ...540..236H}, \citet{2000AJ....120.3162L,
2004AJ....128.1254L}, \citet{2001ApJ...558L..51M},
\citet{2000MNRAS.314..858L}, which provided a census of $JHKL$
excesses, an assessment of the prevalence of circumstellar disks with
stellar mass, and correlations with stellar environment within the
 {ONC.} Most recently, $Spitzer$ data has improved the detailed
knowledge of these disks at mid-infrared $(3.5-24\;\mu\mbox{m})$
wavelengths with papers by \citet{2006ApJ...646..297R} and
\citet{2007ApJ...671..605C} adding to ground-based mid-IR work, e.g.,
\citet{2001AJ....121.1003S}, in investigating disk versus stellar
rotation paradigms.  Additional mid-infrared work focussing on
proplyds is that of \citet{2005AJ....130.1763S}. Towards longer
wavelengths, millimeter investigations including those of
\citet{2006ApJ...641.1162E}, \citet{2008arXiv0803.3217E} and
\citet{2005ApJ...634..495W} have together established the prevalence
 of minimum-mass solar nebula disks in the {ONC,} and the existence of
several much more massive systems.

\subsubsection{Disk Accretion} Studies of the accretion signatures
 typically associated with such disks are difficult in the {ONC} due to
the bright continuum and emission-line backdrop of the
Nebula. However, studies of the Ca\textsc{II} ``infrared" triplet
\citep{1993PhDT.......360S, 1998AJ....116.1816H}, which is the only
strong optical line found in accretion disks  but not in H\textsc{II}
regions, avoids issues with strong and variable background in more
typically observed lines such as H$\alpha$.   Using high dispersion
spectroscopy, \citet{2005AJ....129..363S} and
\citet{2008ApJ...676.1109F} were able to separate nebular and stellar
emission at H$\alpha$ and, focussing on the line wings, claim to find
many sources (including those lacking thermal excess) that indicate
accretion rather than winds.  Finally, studies of the accretion
continuum at UV wavelengths have included those of
 \citet{2004ApJ...606..952R} in the inner {ONC} using HST and
 \citet{2000AJ....119.3026R} in the {ONC} ``flanking fields" from the
 ground. In summary, $80-100\%$ of the stars in the inner {ONC} are
suspected accretors. A median accretion rate of
$10^{-9}\,M_{\odot}/\mbox{yr}$ has been derived, with some evidence
for a dependence of accretion on stellar mass. Deep spectroscopic
studies of such accretion signatures are lacking for the brown dwarf
population.

\subsubsection{X-ray Clues to Disk
  Evolution} \label{sec:ttauri:disk:xray} 
\begin{figure}
	[h*]

	\centering
	\includegraphics[angle=0.,totalheight=4in]{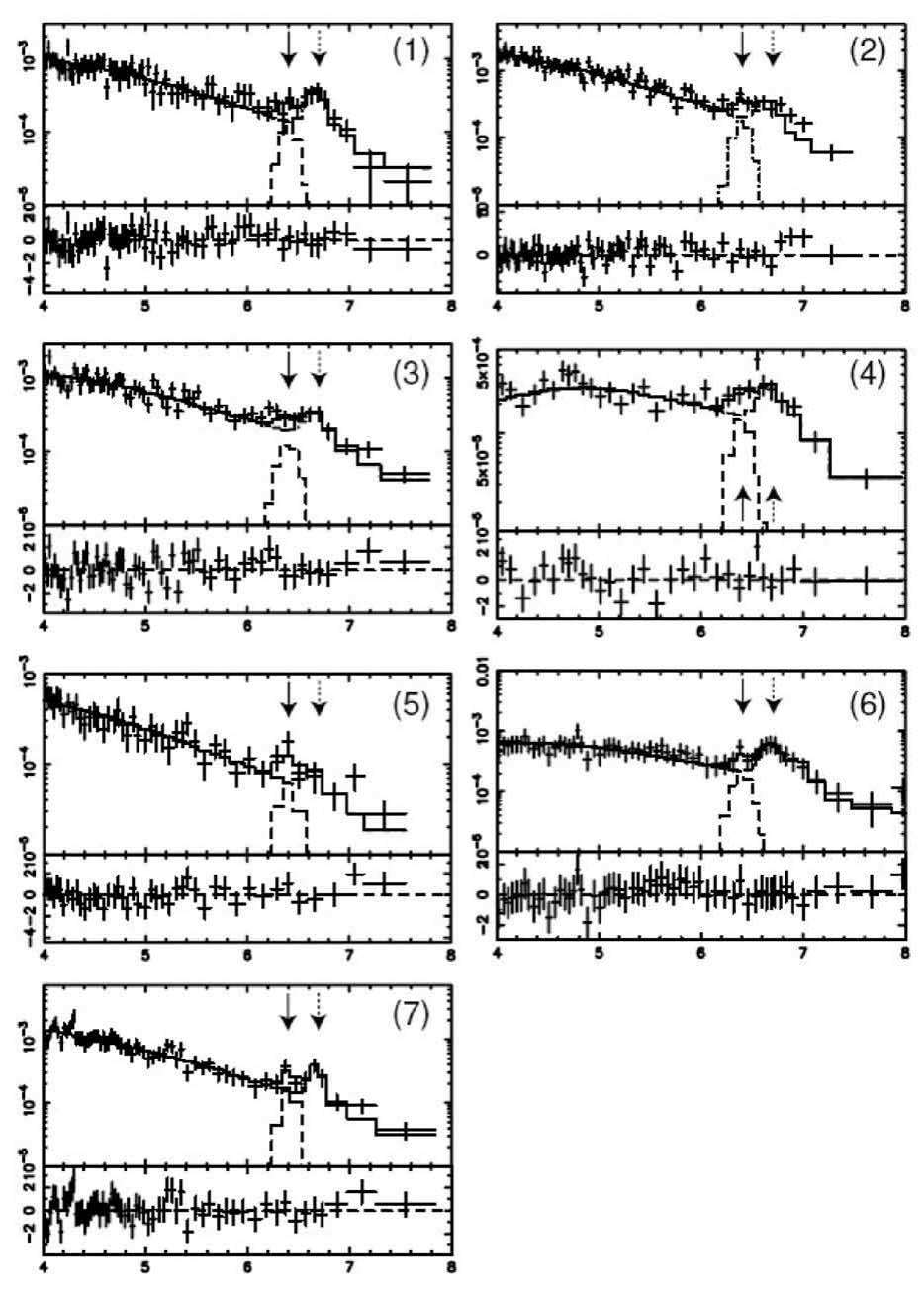}

	\caption{The signature of disk irradiation by X-rays from powerful flares of 7 COUP sources. The presence of a fluorescent emission line from cold iron atoms at 6.4~keV is seen next to the hot plasma line around 6.7~keV (both marked in the figure with arrows). Reproduced from \citet{2005ApJS..160..503T}. \label{fig:xray:fig11}}
\end{figure}

The high X-ray intensity and
 hard spectra found for {ONC} stars indicate that the ionization of disk
gases by stellar X-rays dominates ionization by cosmic rays or other
sources by a large factor ($\sim 10^8$ for 1 M$_\odot$ stars). COUP
provides two lines of direct evidence for establishing where the disk
irradiation by X-rays originates. First, \citet{2005ApJS..160..511K}
find that the X-ray absorbing column of the COUP stars surrounded by
dusty ``proplyds'' imaged with HST increases with disk
inclination. The soft X-rays being absorbed must ionize disk
gas. Second, the COUP results\citep[and $Chandra$ studies of other YSOs; e.g.][]{2001ApJ...557..747I} include the detection of the
fluorescent emission line from cold iron atoms at 6.4 keV, which is
seen next to the hot plasma line around 6.7 keV \citep[Figure \ref{fig:xray:fig11}, ][]{2005ApJS..160..503T}. The equivalent
widths of the 6.4 keV line observed in the COUP sample are compatible
with the fluorescence originating in a centrally illuminated disk
observed face-on, and cannot be attributable to the fluorescence by
interstellar or circumstellar matter along the line of sight. Recent
theoretical studies have predicted that the behavior of protoplanetary
disks and the processes of planet formation will be significantly
altered if the disks are slightly ionized \citep[see reviews by][]{2005astro.ph..1223F, 2005ASPC..341..165G}. For example, the
planets would form in a turbulent and lumpy disk rather than a smooth
disk, which may prevent Earth-like planets from rapidly migrating
through the disk towards the young star. Thus the X-rays from young
stars may have important implications for the formation of planets
around these stars.

\subsection{Kinematics and Cluster Structure} \label{sec:ttauri:dyn}

\subsubsection{Proper Motion Studies}  Proper motion studies in the
literature are greatly muddled by the quality of and by systematics
evident in the observational data. This has yielded conflicting
results about the relative motions of stars and whether they display
systematic expansion or contraction.  Authors including,
\citet{1954TrSht..25....1P}, \citet{1958ApJ...128...14S} and
\citet{1977ApJ...217..719F}, have all produced evidence for the
expansion or contraction of the cluster, with evidence offered by
\citet{1962AJ.....67..699V, 1971ApJ...167..537V} and
\citet{1974RMxAA...1..101A} indicate that these claims are due to
observational error.  Two classic proper motions studies are
\citet{1988AJ.....95.1744V} and \citet{1988AJ.....95.1755J}. The
\citeauthor{1988AJ.....95.1744V} proper motion study used  plates from
three observatories over a 77 year period. The internal error
estimates are very small for the high mass members ($V<12.5$) and they
derived a velocity dispersion of 0.7 milliarcsec/yr (1.3~km~s$^{-1}$
at a distance of 400~pc) for 48 members. \citet{1988AJ.....95.1755J}
used 47 deeper red plates taken over a 20 year timescale to derive
proper motions and membership probabilities for nearly 1000 stars near
the Nebula. Focusing on somewhat lower mass objects, they find a
velocity dispersion of 2.5~km~s$^{-1}$, which they point out is
similar to the clump to clump velocity dispersion of the gas. They
also confirmed van~Altena et al.'s finding of a smaller velocity
dispersion for the bright stars. This suggested to Jones \& Walker a
relationship between velocity dispersion and mass that could also
constrain the dynamical state of the cluster.  Reviewing these
results, \citet{1998ApJ...492..540H} find that any variation in
velocity dispersion with mass appears to be too small to be consistent
with the equipartition of energy occurring during significant
dynamical evolution. The long baseline and high resolution of data in
the HST archive should provide good astrometry for updating these
proper motion results \citep[e.g.,][]{2005ApJ...633L..45O}.

\begin{figure}[htb*]
	\centering
	\includegraphics[angle=0.,width=\textwidth]{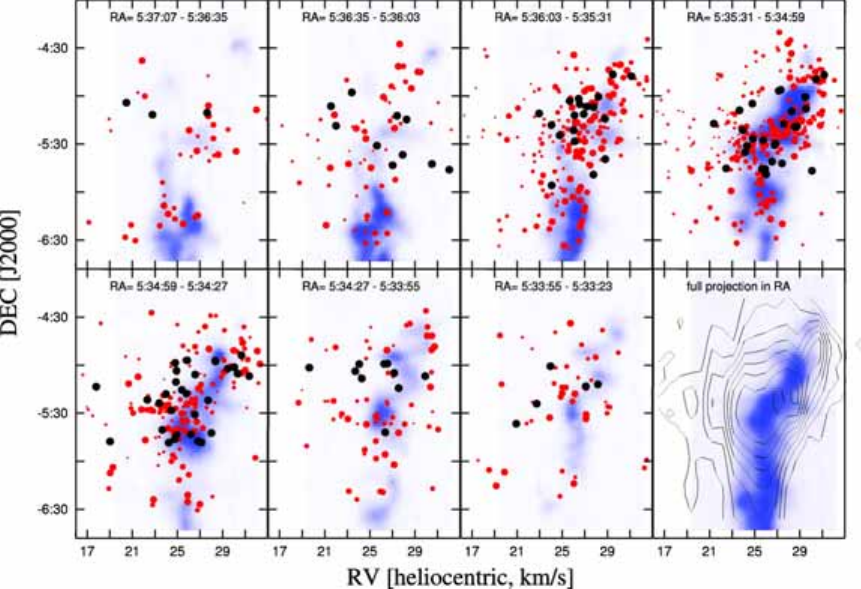}
	\caption{Distribution of (heliocentric) radial velocity as a
	function of declination for stars (dots) and gas
	\citep[${}^{13}\mbox{CO}$ as the blue background,
 from][]{1987ApJ...312L..45B} in and around the {Orion Nebula.}
	Reproduced from \citet{2008ApJ...676.1109F}. Each panel
	represents a different range of right ascension; the color and
	sizes of the dots correspond to the net precision of their
	radial velocity measurements (larger or black dots being the
	best quality); the last panel compares the stellar and gas
	distribution integrated across the cloud. \label{fig:onc:rv}}
\end{figure}

\subsubsection{Radial Velocity Studies} The difficulty of acquiring
radial velocities for a large number of young stars has been recently
overcome using  new multi-fiber echelle spectrographs. While earlier
works such as \citet{1965ApJ...142..964J} or
\citet{1983ApJ...271..642W} were able to observe tens of stars, the
recent surveys published by \citet{2005AJ....129..363S} and
\citet{2008ApJ...676.1109F} include a total of about 1300 stars around
 the {Orion Nebula} (and the Northern Orion A Molecular Cloud). They
find a bulk velocity dispersion of $\sim2.3\,-\,3$~km~s$^{-1}$ for all
the members. This dispersion should be interpreted carefully
considering the existence of a large $\sim5$~km~s$^{-1}$ gradient
across the Nebula, which is most evident in the top right panel of
 Figure \ref{fig:onc:rv}. Indeed across the {Northern Orion A} cloud,
\citeauthor{2008ApJ...676.1109F} find that the radial velocities of
the stars appear to correlate  strongly with the radial velocity of
the molecular gas cloud.

\subsubsection{Structure} The Orion Nebula Cluster is elongated
parallel to the Orion A Molecular Cloud with a centroid from
 elliptical star count fitting falling just North of {$\theta^{1}$~Ori~C}
\citep{1998ApJ...492..540H}. No substructure is evident in the
optically revealed stars \citep{1998ApJ...492..540H,
  2002MNRAS.334..156S}, which has been interpreted to suggest that the
cluster may have formed from a fairly large number of small
clusterings that quickly dispersed. Some evidence for subclustering at
the $<0.1\mbox{ pc}$ scale appears when one studies the youngest, most
embedded members \citep{2000AJ....120.3162L, 2004AJ....128.1254L,
  2005ApJS..160..530G}.  Hillenbrand \& Hartmann also found that the
cluster displays a radial profile that is well fit by a King profile,
having a core radius of 0.1-0.2~pc and a central stellar density
exceeding $10^{4}\mbox{ stars}\,\cdot\,\mbox{pc}^{-3}$
\citep{1994AJ....108.1382M, 1998ApJ...492..540H}. It has been argued
that this mass distribution is primordial as it seems unlikely that
the cluster achieved a King like radial profile as the result of
dynamics given its youth. A similar, primordial explanation is given
regarding evidence for mass segregation of the highest mass stars to
the cluster's core and a skew of lower mass stars to the outer parts
of the cluster \citep{2000ApJ...540..236H}. Investigations into the
spatial distributions of stars and brown dwarfs have been made
\citep{2007A&A...471L..33K}, but that field of study was very small
and perhaps more interesting results await much wider field studies
that cover more than the central 0.5~pc and include radial velocity
studies of the faintest members.

\section{The Most Massive Stars in the Orion Nebula} \label{sec:ob}

 {The Orion Nebula} is an H\textsc{II} region created (primarily) by the
 ionizing photons of the central O star, {$\theta^{1}$~Ori~C.} There are
a total of 20 optically revealed O or B type primaries, which are
listed in Table~\ref{tab:ob1}; this tabulation is a merger of Orion 1
subgroup ``d'' association members from \citet{1996PhDT........78B}
 and the \citet{1997AJ....113.1733H} {ONC} census.

\begin{table}
	\caption[]{Massive (O or B spectral type) stars of the Orion
     Nebula: Names, Cross-References and Positions. Positions of the
     primary star are given in the mean epoch of their observation
     except where marked (``$*$''). Reference catalogs: ``2MASS''
     \citet{2006AJ....131.1163S}; ``COUP''
     \citet{2005ApJS..160..319G}; ``HIP'' (Hipparcos)
     \citet{1997A&A...323L..49P}; ``MLLA''
     \citet{2002ApJ...573..366M}; ``PPM'' \citet{1988A&AS...74..449R};
     ``TYC'' (Tycho-2) \citet{2000A&A...355L..27H} \label{tab:ob1}}
	\begin{center}
		{\scriptsize
		\begin{tabular}
			{c@{\hskip3pt}l@{\hskip3pt}c@{\hskip3pt}c@{\hskip3pt}l@{\hskip3pt}l@{\hskip3pt}l@{\hskip3pt}l@{\hskip3pt}l}
			\tableline \noalign{\smallskip} \multicolumn{4}{c}{Designations} & \multicolumn{5}{c}{Positional Information} \\
			\noalign{\smallskip} Parenago & Common & Draper & Brun & \multicolumn{2}{c}{Reference Catalog} & \multicolumn{2}{c}{Equatorial (ICRS)} & Epoch \\
			\noalign{\smallskip} (1954) & & (HD) & (1935) & Name & \# & RA & DEC & \ \ \ (yr) \\
			\noalign{\smallskip}
			\tableline \noalign{\smallskip} 1539 & & & 328 & 2MASS & 05343999-0510070 & 83.666636 & -5.168633 & 2000.9\\
			\noalign{\smallskip} 1605 & V~327~Ori & 36917 & 388 & TYC & 4774-809-1 & 83.69575750 & -5.57072639 & 1991.7 \\
			\noalign{\smallskip} 1660 & & 36939 & 442 & TYC & 4744-823-1 & 83.73038972 & -5.50614139 & 1991.7 \\
			\noalign{\smallskip} 1744 & & 36981 & 502 & TYC & 4744-915-1 & 83.77582389 & -5.20442611 & 1991.7 \\
			\noalign{\smallskip} 1772 & LP~Ori & 36982 & 530 & TYC & 4774-849-1 & 83.79098528 & -5.46478444 & 1991.7 \\
			\noalign{\smallskip} 1865 & $\theta^{1}$~Ori~A & 37020 & 587 & HIP & 26220 & 83.81592896 & -5.38731536 & 1991.25 \\
			\noalign{\smallskip} 1863 & $\theta^{1}$~Ori~B & 37021 & 595 & COUP & 778 & 83.817270 & -5.385220 & 2003.04 \\
			\noalign{\smallskip} 1891 & $\theta^{1}$~Ori~C & 37022 & 598 & TYC & 4774-931-1 & 83.81860444 & -5.38969611 & 1991.8 \\
			\noalign{\smallskip} 1889 & $\theta^{1}$~Ori~D & 37023 & 612 & HIP & 26224 & 83.82166016 & -5.38768076 & 1991.25 \\
			\noalign{\smallskip} 1892 & $\theta^{1}$~Ori~F & & 603 & MLLA & 388 & 83.819625 & -5.390222 & 2000.3\\
			\noalign{\smallskip} 1956 & V~1230~Ori & & 655 & PPM & 702316 & 83.83632500 & -5.36236111 & 2000$(*)$ \\
			\noalign{\smallskip} 1993 & $\theta^{2}$~Ori~A & 37041 & 682 & TYC & 4774-933-1 & 83.84540917 & -5.41606333 & 1991.7 \\
			\noalign{\smallskip} 2031 & $\theta^{2}$~Ori~B & 37042 & 714 & TYC & 4774-934-1 & 83.86000444 & -5.41687306 & 1991.7 \\
			\noalign{\smallskip} 2085 & $\theta^{2}$~Ori~C & 37062 & 760 & PPM & 188231 & 83.88098334 & -5.42122222 & 2000$(*)$ \\
			\noalign{\smallskip} 2074 & NU~Ori & 37061 & 747 & TYC & 4774-906-1 & 83.88068417 & -5.26738778 & 1991.7 \\
			\noalign{\smallskip} 2284 & & 37114 & 920 & TYC & 4774-867-1 & 83.99391639 & -5.37537417 & 1991.7 \\
			\noalign{\smallskip} 2271 & & 37115 & 907 & TYC & 4778-1369-1& 83.97533639 & -5.62839861 & 1991.6 \\
			\noalign{\smallskip} 2366 & & 37150 & 980 & TYC & 4778-1378-1& 84.06261139 & -5.64792000 & 1991.6 \\
			\noalign{\smallskip} 2387 & & 37174 & 992 & TYC & 4774-855-1 & 84.11326917 & -5.40870417 & 1991.7 \\
			\noalign{\smallskip} 2425 & & & 1018 & TYC & 4774-873-1 & 84.15876833 & -5.47638250 & 1991.7\\
			\noalign{\smallskip}
		\tableline
\end{tabular}
		}
	\end{center}
\end{table}

\begin{figure}[t*]
	\centering
   \vspace{0.5mm}
   \newlength{\deffboxrule}
   \setlength{\deffboxrule}{\fboxrule}
   \newlength{\deffboxsep}
   \setlength{\deffboxsep}{\fboxsep}

   \setlength{\fboxrule}{1.5pt}
   \setlength{\fboxsep}{3pt}

  \framebox[0.51\textwidth]{
	\centering
	\includegraphics[angle=0, width=0.5\textwidth]{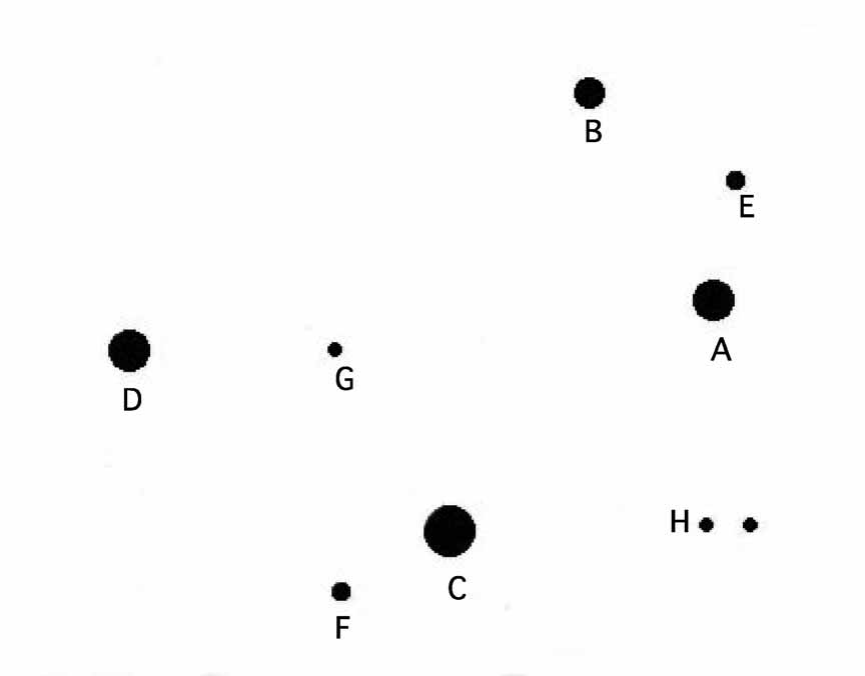}
   }
   \setlength{\fboxrule}{\deffboxrule}
   \setlength{\fboxsep}{\deffboxsep}

 \caption{Finder chart for the {Trapezium}
 and the inner {Orion
 Nebula.} North is up and East is to the
	left. \label{fig:trap_finder}}
 \vspace{-2mm}
\end{figure}

 The collective Bayer designation for the {Orion Nebula} is
 $\theta$~Orionis. The nomenclature of members of the inner {Orion
 Nebula} break down further into $\theta^{1}$ and $\theta^{2}$
designations, where $\theta^{1}$~Ori is the Bayer designation for the
 famous {``Trapezium''} asterism. A finder chart for the {Trapezium} is
given in Figure \ref{fig:trap_finder}. The principal components of the
 {Trapezium} are further labeled A, B, C, D, E, F; e.g.,
 {$\theta^{1}$~Ori~C;} subcomponents of these ``A,B,C...'' designations
are marked with subscripts, e.g., $\theta^{1}$~Ori~C$_{1}$. Five of
these six primaries in $\theta^{1}$~Ori are O or B type stars. The
 {$\theta^{2}$~Ori} designation corresponds to a string of three OB
stars, e.g. the A,B,C components, lying near the bright bar; they are
easily identified in Fig.\ref{fig:Figures_ONCtreasury_reduced2} just
 southeast of the {Trapezium.} Additional cross-references and observed
positions for the unresolved primaries of these systems are given in
Table~\ref{tab:ob1}. The identifiers from the
\citet{1935POLyo...1...12B} catalog especially useful for interpreting
older texts. For completeness the star names 41 and 43~Ori are the
 Flamsteed designations for $\theta^{1}$~Ori and {$\theta^{2}$~Ori~A}
 sources, respectively. Finally, the {Trapezium} is also frequently found
listed by its catalog entry in the \citet{1932QB821.A43......},
 {ADS~4186.}

\subsection{Basic Properties} \label{sec:ob:basic}

\citet{2006A&A...448..351S} derive MK spectral types for 5 of the OB
members in the central nebula and provide stellar properties (R, T, M,
L, g, $v\sin{i}$) and extensive analysis of oxygen abundances through
comparison to synthetic stellar models and template OB stars. An
additional source of basic stellar properties can come from the
multi-spectral flux ratios of the various binary components. For
example, \citet{2007A&A...466..649K} used flux ratios from $V$ to $K$
bands for the $\theta^{1}$~Ori~C$_{1,2}$ binary to constrain the
components to have spectral types of $\sim$O5.5 and O9.5.  Papers
surveying rotational and radial velocities (to the extent permitted by
the high multiplicity) of these objects include
\citet{1970AJ.....75.1095A}, \citet{1991ApJ...367..155A} and
\citet{2004ApJ...601..979W}.

\begin{figure}[ht*]
	\centering
	\includegraphics[width=\textwidth]{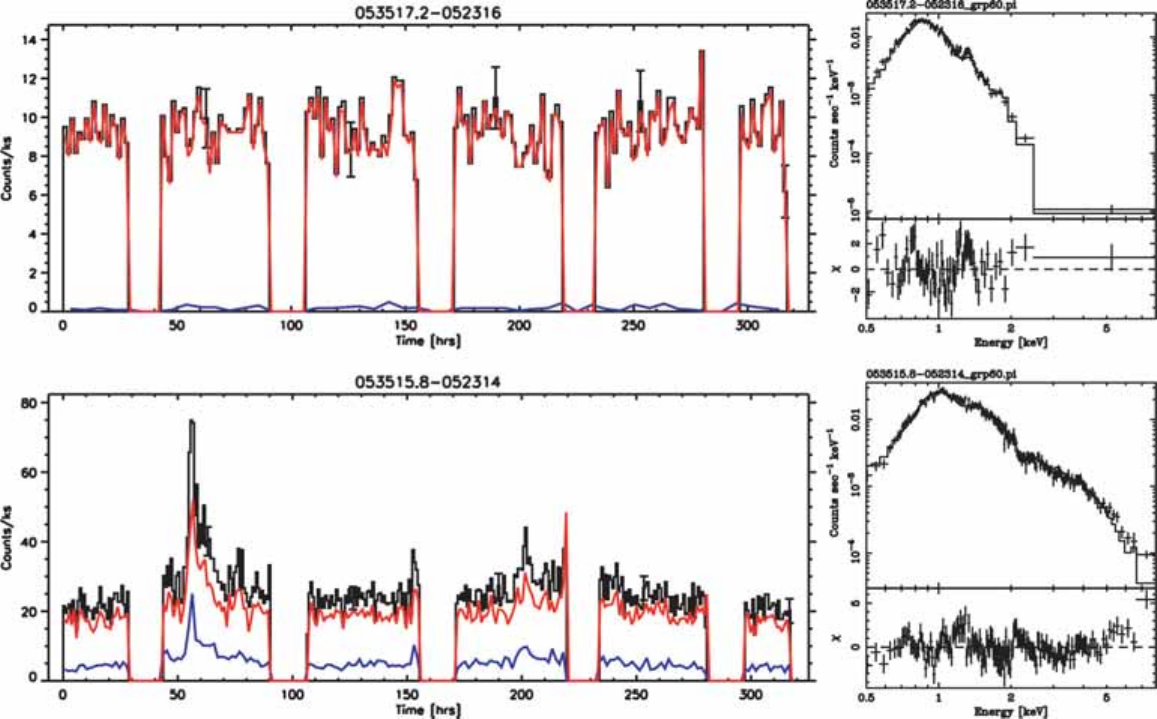}%
	\caption{Top: lightcurve and spectrum for $\theta^1$~Ori~D
	with the soft constant emission, as a signature of many small
	wind shocks. Bottom: in contrast to $\theta^1$~Ori~D the
	lightcurve and spectrum of $\theta^1$~Ori~A show hard flaring
	emission possibly due to the confinement of the wind by a
	strong stellar magnetic field. In the lightcurves, the black
	line indicates the total $(0.5-8.0)$ keV energy band, red line
	- soft $(0.5-2.0)$ keV band, and blue - hard $(2.0-8.0)$ keV
	band. \textit{Chandra} X-ray spectra and lightcurves from the
	COUP. Reproduced from
	\citet{2005ApJS..160..557S}. \label{xray:fig6}}
\end{figure}

X-rays from the OB stars were thought to be generated in a myriad of
tiny shocks in their powerful winds. But this model predicts that the
X-ray spectrum will be soft and the emission will be constant. The
\textit{Chandra} Orion Ultradeep Project (COUP) shows that the spectra
of the Trapezium OB stars often have a hard component and can exhibit
rapid flares \citep{2005ApJS..160..557S}. Of the 10 unobscured COUP
 sources earlier than B4, only three ($\theta^1$~Ori~D, {NU Ori,} and
 possibly {$\theta^1$~Ori~B)} show the expected signature of many small
wind shocks, while most show flares and/or hard spectral
components. This suggests that the winds of these OB stars are, at
least in part, trapped by magnetic fields, resulting in large scale
shocks and production of hard X-ray emission
\citep{1997A&A...323..121B}. Figure \ref{xray:fig6} compares X-ray
 emission from the steady-wind source {$\theta^{1}$~Ori~D} with that of
 the magnetically active {$\theta^{1}$~Ori~A.} COUP also confirms that
the X-ray emission from intermediate-mass stars with spectral types
B5-A9 is attributable to lower mass companions rather than the
intermediate-mass star itself.  Additional X-ray studies of the OB
 stars in the {Orion Nebula} include \citet{1994ApJ...432..386C},
\citet{1995A&A...299...39G}, \citet{1996A&A...312..539S},
\citet{2000ApJ...545L.135S}, \citet{2001ApJ...549..441S},
\citet{2003ApJ...586.1441S} and \citet{2003ApJ...595..365S}.

\subsection{Kinematics} \label{sec:ob:velocity}

 Compact stellar groups like the {Trapezium} are expected to be
intrinsically unstable \citep{1955Obs....75...72A}. Therefore, the
kinematics of such a stellar grouping could display systematic
expansion or contraction, display evidence for disintegrating multiple
systems or provide an origin for runaway stars
\citep{1967BOTT....4...86P}. Regarding signatures of disintegrating
systems, \citet{1988AJ.....95.1744V} reported that the two O stars in
 the central {Orion Nebula,} {$\theta^1$~Ori~C} \& {$\theta^2$~Ori~A,} were
observed to have proper motions relative to other bright $(V<12)$
stars that were large enough to carry them out of the center of the
Nebula in less than 1 Myr.  However, the relative proper motions for
 the {Trapezium} stars derived by \citet{1974RMxAA...1..101A} and updated
in \citet{2004RMxAC..21..195A} are in general very small, which calls
 into question whether {$\theta^1$~Ori~C} is actually be ejected relative
to the other Trapezium OB stars.

 The dynamics of two more systems related to the {Orion Nebula} have
garnered much more detailed attention. The first is the existence of
two high velocity OB stars whose origins appear to coincide near the
 {Orion Nebula.} \citet{1953BAN....12...76B} described the high velocity
 and apparent space motion of {AE~Aurigae} away from Orion, while
\citet{1954ApJ...119..625B} made this interesting connection for the
 star {$\mu$~Columbae} as well. While each is over $25\deg$ from the
Nebula, they have proper motion vectors corresponding to space motions
$>100$~km~s$^{-1}$ and Blaauw \& Morgan (1954) estimated that they
both originated in a dissolution event occurring near to the current
Nebula about 2.7 Myr ago. Subsequent analysis included
\citet{1961BAN....15..265B}, while \citet{1986ApJS...61..419G}
 proposed adding the O star {$\iota$~Ori,} which is just South of the
Nebula, into the dissolution.  \citet{2000ApJ...544L.133H} and
\citet{2001A&A...365...49H} performed the integration of the orbits
including improved Hipparcos data, radial velocities and a Galactic
potential field; their results further constrained the dissolution
event to have occurred approximately 2.5~Myr ago and could have been
located nearer the center of the cluster even than the location of
 {$\iota$~Ori} today. Numerical modeling of this event has been performed
most recently by \citet{2004MNRAS.350..615G}.
\begin{figure}[t*]
	\centering
	\begin{minipage}[c]{0.48\textwidth} \centering
		\includegraphics[angle=0,width=\textwidth]{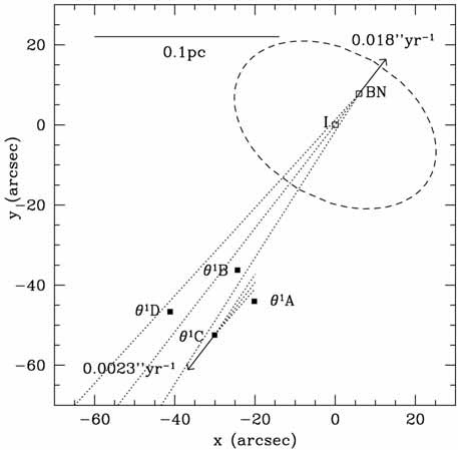}%
	\end{minipage}%
	\hspace{0.04\textwidth}%
	\begin{minipage}[c]{0.48\textwidth} \centering
		\includegraphics[angle = 0,width=\textwidth]{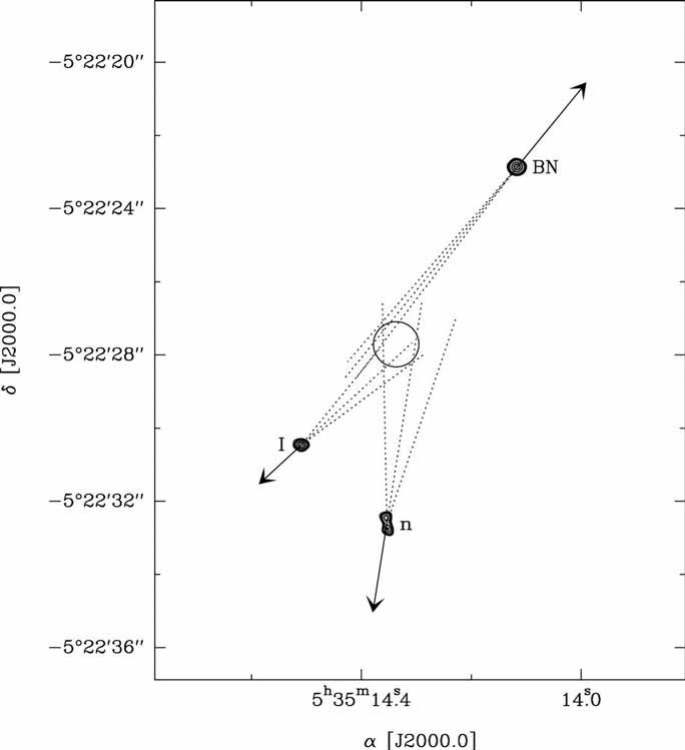}%
	\end{minipage}
 \caption{Proper motion histories for observations relating to the origin of the {B-N} object as a runaway B star. Left: the disintegration of the {B-N} object and $\theta^1$~Ori~C 4,000 years ago as described by \citet{2004ApJ...607L..47T}; Right: the disintegration of the protostellar triple {B-N,} source ``i'' and source ``n'' 500 years ago as described by \citet{2005ApJ...635.1166G}. \label{fig:bn_running}}
\end{figure}

The second apparently disintegrating high mass system concerns the
 relationship between the {B-N} object and other high mass protostars or
O stars in the central cluster. The relatively high proper motion of
 the {B-N} object relative to the cluster was derived first by
\citet{1995ApJ...455L.189P}; monitoring of the proper motion of this
object using published \citep{2005ApJ...627L..65R,2005ApJ...635.1166G}
and unpublished data have led to two hypotheses on the origin of its
motion. Figure \ref{fig:bn_running} compares proper motion diagrams
 tracing the {B-N} object back to an origin in two different formation
scenarios. \citet{2004ApJ...607L..47T} argued this is a recoil motion
 from the {B-N} object being ejected from {$\theta^{1}$~Ori~C} $\sim4000$
years ago, while \citet{2005ApJ...635.1166G} argue for a dissolution
 of the {B-N} object from the protostellar sources ``i'' and ``n'' a few
hundred years ago. Either scenario claims to be consistent with the
 production of the explosive outflow in the {K-L} region 500-1000 years
ago.(see O'Dell et~al., Part II).  While this outflow is a consequence
 of the dissolution of the {B-N,} ``i'' and ``n'' system in the Gomez
 et~al. scenario, the hypothesis of Tan is that the {B-N} triggered the
outflow by passing extremely close to source ``i.''

Recent high $(<0.1\arcsec)$ optical/near-IR observations have allowed
the detailed orbits of the many multiple high mass systems (see next
subsection) to be constrained.  These include the recent works by
\citet{2003A&A...402..267S}, \citet{2003ApJ...599..537C},
\citet{2007A&A...466..649K}, and \citet{2008arXiv0801.2584P}. There
is, however, an almost complete lack of such high resolution imaging
 or monitoring of high mass objects outside of the immediate {Trapezium}
core; such observations could be very valuable for better
understanding the kinematic history of the high mass stars, their
interactions, and the dissolution of higher order systems.

\subsection{Multiplicity} \label{sec:ob:multi}

\begin{figure}[t*]
	\centering
	\includegraphics[angle=0,width=\textwidth]{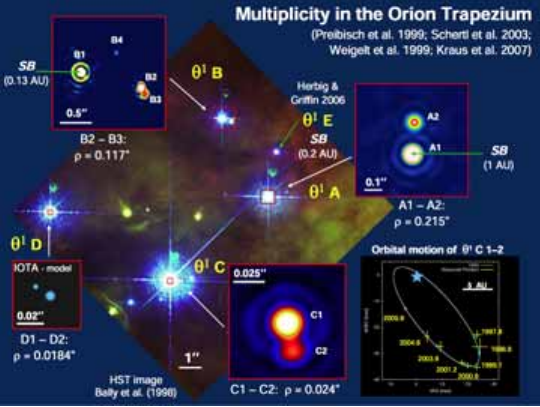}
 \caption{The multiplicity of the stars in the {Trapezium}
	systems as revealed by infrared
	interferometry. \label{trapezium.fig}}
\end{figure}

 The massive members of the {ONC} have been the target of several studies
concerning the multiplicity of these stars. Light curves revealed a
number of eclipsing systems
\citep[e.g.][]{1947QB835.P23......,1969PASP...81..771H,1975IBVS..988....1L,1994ExA.....5...61W},
which have been further analyzed to identify additional companions
\citep[e.g][]{2000AstL...26..529V}. Searches for spectroscopic
companions were presented by \citet{1991ApJ...367..155A} and
\citet{1991ApJS...75..965M}, who found spectroscopic binary
frequencies of 20\% -- 30\%. Studies that detected visual companions
 to massive {ONC} members include \citet{1997ApJ...477..705P},
\citet{1998ApJ...500..825P}, and
\citet{1999AJ....117.1375S}. \citet{1999NewA....4..531P} performed a
 systematic survey for multiple systems among 13 bright {ONC} members of
spectral type O or B with the technique of near-infrared bispectrum
speckle interferometry, and complemented the results with information
about known spectroscopic companions. In the speckle images, which
have a resolution of $0.075\arcsec$, eight visual companions were
found (see Figure~\ref{trapezium.fig}). Stellar masses of the
companions were estimated from the observed near-infrared flux
ratios. The properties of these multiple systems are summarized in
Table~\ref{multiplicity.tab}.

The results of this multiplicity survey allowed general conclusions
 about the multiplicity of the massive stars in the {ONC.} The mean
number of \textit{known companions} in the sample of the 13 observed
OB stars is $15/13 = 1.2$ per primary star. Since it is highly likely
that there are even more, still undetected companions, this is clearly
a strict \textit{lower limit} for the \textit{true multiplicity}. The
multiplicity of these stars is thus considerably (at least about 2
times) higher than that among low-mass primary stars in the general
 field as well as in the {ONC.} Multiplicity seems to be a function of
spectral type: based on the literature summarized above stars of
spectral type B3 or later have, on average, 0.6 known companions per
primary, while the stars of spectral type O to B2 have at least 1.5
companions per primary. This comparison clearly suggests a
mass-dependence of the multiplicity, which is in qualitative agreement
with the generally observed trend of increasing multiplicity with
increasing stellar mass seen among low- and intermediate-mass
stars. While most low-mass stars may be single
\citep{2006ApJ...640L..63L}, the data suggest that essentially all
high-mass stars are members of higher-order multiple systems (triples,
quadruples).

These results point towards a fundamental difference between high- and
low-mass stars and support the assumption that the formation of
massive stars is \textit{not} simply a scaled-up version of low-mass
star formation \citet{2007ARA&A..45..481Z}. Theoretical scenarios for
the formation of these massive stars, including stellar interactions,
the capture of companions and perhaps even stellar mergers scenarios
are discussed in \citet{2004MNRAS.349..735B, 2005MNRAS.362..915B},
\citet{2006ApJ...653..437M,2007ApJ...661L.183M,2007ApJ...656..275M},
\citet{2006MNRAS.370.2038D}, and \citet{2005AJ....129.2281B}.

In the near future, long-baseline interferometric surveys of the ONC
OB stars will allow resolution of companions as close as
$0.001\arcsec$ (0.45~AU). This will largely fill the still existing
gap between currently detectable visual and spectroscopic companions
and provide much more complete information about the multiplicity of
these stars.
\begin{table}[tb]
	\caption[]{Massive stars of the Orion Nebula:
	multiplicity. Adapted from the survey of
	\citet{1999NewA....4..531P}. The Table lists the name,
	spectral type, and mass of the primary component, and then the
	nature (visual or spectroscopic companion), separation, and
	mass ratio of the known companions. If there is a constraint
	on the existence of a binary companion, then that limit and
	its reference are given. References (Column~Ref) correspond by
	number to the following sources: 1:
	\citet{1999NewA....4..531P}; 2: \citet{1999A&A...347L..15W};
	3: \citet{1998ApJ...500..825P}; 4:
	\citet{1989A&A...222..117B}; 5: \citet{1999AJ....117.1375S};
	6: \citet{1991ApJ...367..155A}; 7:
	\citet{1976PASP...88..712L}; 8: \citet{2007A&A...466..649K}; 9
	\citet{2001AstL...27..581V}; 10: \citet{2006A&A...458..461K}
	\label{multiplicity.tab}} \smallskip
	\begin{center}
		{\footnotesize
		\begin{tabular}
			{llrr|lrlr}

			\tableline \noalign{\smallskip} Parenago & Other & $M_p$ & $M_t$ & Companion & $\rho\;\;\;$ & $\;\;\;\;M/M_p$ & Ref \\
			& & $[M_\odot]$ & $[M_\odot]$ & & [AU] & & \\
			\noalign{\smallskip}
			\tableline \noalign{\smallskip}


			1605 & V~327~Ori & 3.5 & 7 &-2 (spec)& & $\sim 0.9\!-\!1.0$& 7\\
			\noalign{\smallskip}

			1660 & & & & --- & $<60$ & --- & 10 \\
			\noalign{\smallskip}

			1744 & & & & --- & $<60$ & --- & 10 \\
			\noalign{\smallskip}

			1772 & LP~Ori & & & --- & $<100$ & --- & 1 \\
			\noalign{\smallskip}

			1865 & $\theta^{1}$~Ori~A & 14 & 21 &-2 (vis)    & 100 & $\sim 0.25$ & 2,3 \\
			     &                    &    &    &-3 (spec)   &   1 & $0.19-0.33$ & 4,9 \\
			\noalign{\smallskip}

			1863 & $\theta^{1}$~Ori~B &  7 & 12 &-2 (vis)    & 430      & $\sim 0.22\,(<\!0.71)$ & 1,2\\
			     &                    &    &    &-3 (vis)    & 460      & $\sim 0.10\,(<\!0.50)$ & 1,2\\
			     &                    &    &    &-4 (vis)    & 260      & $\sim 0.03\,(<\!0.29)$ & 1,5\\
			     &                    &    &    &-5 (spec)   &   0.13   & $\sim 0.39$ & 6\\
			\noalign{\smallskip}

			1891 & $\theta^{1}$~Ori~C & 34 & 50 &-2 (vis)    & 18       & $\sim 0.45$& 2,8\\
			\noalign{\smallskip}

			1889 & $\theta^{1}$~Ori~D & 16 & $>16$ &-2? (vis) & 8 & & 8\\
			\noalign{\smallskip}



			%
			1993 & $\theta^{2}$~Ori~A & 25 & $40$ &-2 (vis)& 173 & $\sim 0.28\,(<\!0.32)$ & 1\\
			& & & &-3 (spec)& 0.47 & $\sim 0.35$ &6\\
			\noalign{\smallskip}

			2031 & $\theta^{2}$~Ori~B & & & --- & $<100$ & --- & 1 \\
			\noalign{\smallskip}


			%
			2074 & NU~Ori & 14 & $17$ &-2 (vis)& 214 & $\sim 0.07\,(<\!0.28)$ & 1\\
			& & & &-3 (spec)& 0.35 & $\sim 0.2$ &6\\
			\noalign{\smallskip}

			2284 & & & & --- & $<60$ & --- & 10 \\
			\noalign{\smallskip}

			2271 & HD~37115 & 5 & $7$ &-2 (vis)& 400 & $\sim 0.29\,(<\!0.96)$ & 1 \\
			\noalign{\smallskip}

			2366 & & & & --- & $<100$ & --- & 1 \\
			\noalign{\smallskip}


			%
			2425 & & 4 & $5$ &-2 (vis)& 388 & $\sim 0.04\,(<\!0.35)$ & 1\\
			\noalign{\smallskip}

		\tableline
\end{tabular}
		}
	\end{center}
\end{table}

\subsection{Notes on Individual Trapezium $(\theta^{1}\mbox{Ori})$ Stars} \label{sec:ob:notes}

\subsubsection{$\theta^{1}$~Ori~A; Parenago 1865}
 {(V1016~Ori;} {HR~1893;} {HD~37020)}
 The bright $(V = 6.7)$ westernmost member of the {Trapezium,}
of spectral type B0.5V \citep{2006A&A...448..351S}, was discovered as
recently as 1974 to be an eclipsing binary \citep{1975IBVS..988....1L}
with a period of 65.43 days and deep primary minima ($\Delta V
\sim$1); no secondary minima have been seen with certainty. Early
attempts to model the system include \citet{1989A&A...222..117B},
\citet{1998AstL...24..296V} and \citet{1999AstL...25..179V}. Optical
and ultraviolet spectra are analyzed in \citet{2000AstL...26..104V},
\citet{2001AstL...27..809V} and
\citet{2001AstL...27..581V}. \citet{1999Obs...119...16S} review and
analyze existing data up to that time, deriving masses of 12 and 3
$M_\odot$ for the eclipsing pair. \citet{2000AstL...26..452B} analyzed
new light curves, and \citet{2001AstL...27..581V} suggested that the
components have masses of 21~$M_\odot$ and 3.9~$M_\odot$. A primary
mass of $14\,M_\odot$ and a mass ratio $q\sim0.2$ seems a reasonable
summary of recent findings. A third star (A$_2$) at $0.2\arcsec$
separation has been found next to $\theta^{1}$~Ori~A$_{1,3}$ by
\citet{1998ApJ...500..825P}; in the near-IR its flux ratio to the
eclipsing primary suggests a reddened A or F type pre-main sequence
 star. Radio continuum observations of {V~1016~Ori} at 2 and 6~cm
\citep{1987ApJ...321..516C,1989A&A...217..179F,1991A&A...248..453F,1993A&AS..101..127F}
reveal a variable non-thermal source now confidently associated with
this visual secondary \citep{1998ApJ...500..825P,2007MmSAI..78..362P};
the system is unresolved in \textit{Chandra} data. Various
explanations for this radio emission have been proposed, including
wind-wind collisions. Orbital motion between A$_2$ and the eclipsing
primary A$_{1,3}$ has been observed through multi-epoch observations
over a 7 year period \citep{2003ApJ...599..537C,2003A&A...402..267S}.

\subsubsection{$\theta^{1}$~Ori~B; Parenago~1863}
 {(BM~Ori;} {HR~1894;} {HD~37021)}
This source consists of at least 5 stars but possibly as
many as 7. The long-known visual companion, at a separation of about
$1\arcsec$ from the primary $\theta^{1}$~Ori~B$_{1}$, is clearly
resolved into a close ($0.117\arcsec$) binary ($\theta^1{\rm
  Ori}$\,B$_{2,3}$). At least one member of the B$_{2,3}$ system is a
proplyd, source identifier 160-307 (note that
\citet{1996AJ....111..846O} gave this source the identifier 161-307
and considered it stellar); both appear extended in the bispectrum
band images of \citet{2003A&A...402..267S}; one of them drives a
 microjet {(HH~508)} \citep{2000AJ....119.2919B}; in the thermal
infrared, the $\theta^{1}$~Ori~B$_{2,3}$ system dominates the B$_{1}$
primary \citep{2004AJ....128.1254L,2005AJ....130.1763S}. This system
is also the dominant X-ray source and exhibits all the features of
typical solar mass T~Tauri stars including multiple flares
\citep{2005ApJS..160..557S}. Another faint visual companion,
$\theta^{1}$~Ori~B$_{4}$, is located $0.578\arcsec$ from the primary
star \citep{1999AJ....117.1375S}. The primary,
$\theta^{1}$~Ori~B$_{1}$, is itself an eclipsing binary with a period
of 6.46 days; early observations are discussed by
\citet{1921AN....212..229H, 1921AN....212..383H},
\citet{1947QB835.P23......} and \citet{1948AN....276..144S}. A near-IR
light curve is shown in Figure \ref{fig:var1}
(Sect.~\ref{sec:variables:oir}).  This system is also a double-lined
spectroscopic binary, so masses and radii of the two components can be
derived. The primary is a B3 star. \citet{1969PASP...81..771H}
obtained a detailed light curve, demonstrating that the primary
eclipse is total with a duration of 9 hours, and finding a shallow
second minimum. Spectra obtained by \citet{1976ApJ...205..462P}
revealed that the secondary component is a late A-type star. A primary
mass of $\sim7M_\odot$ and a mass ratio $q\sim0.4$ is adopted for this
system. New light curves obtained by \citet{1994ExA.....5...61W} were
analyzed by \citet{2000AstL...26..529V}, who suggested that there is a
third (B type) star in the primary $\theta^{1}$~Ori~B$_{1,5}$ system
that is not involved in the eclipse, but affects the spectra
observed. \citet{2006Ap.....49...96V} found a fourth, late-type
component in the primary system based on a radial velocity anomaly
\citep{1976ApJ...205..462P, 2004Ap.....47..169V}. Ultraviolet spectra
by \citet{2006ARep...50..392V} detected high-velocity outflowing gas
 in the system. The position of the COUP X-ray source {(COUP~778)} is
adopted for $\theta^1{\rm Ori}$\,B$_{1}$ because the \textit{Chandra}
images provide the best-resolution observations with a wide field
astrometric reference frame.

\subsubsection{$\theta^{1}$~Ori~C; Parenago~1891}~(HR~1895; HD~37022)
 The most massive star in the {ONC,} is an extremely close visual binary
system with an initial discovery of the companion at a separation of
$0.033\arcsec$ by
\citet{1999A&A...347L..15W}. \citet{2003A&A...402..267S} subsequently
reported the detection of orbital motion. \citet{2007A&A...466..649K}
and \citet{2008arXiv0801.2584P} report recent multi-epoch observations
with visual and near-infrared bispectrum speckle and near-infrared
long-baseline interferometry.  The current data trace the orbital
motion of the companion over a more than 10-year period and cover a
significant part of its orbit. \citet{2007A&A...466..649K} derived a
highly eccentric $(e \sim 0.91)$ orbit with a period of 10.9~yrs and a
total mass of $\sim50M_{\odot}$ with the primary being a
$34\,M_{\odot}$ O5.5 star and the companion a $15.5\,M_{\odot}$ O9.5
star. The addition of 6 epochs over a baseline of a year by
\citet{2008arXiv0801.2584P} suggest a longer period
$(\sim26\mbox{yr})$, a less eccentric orbit $(e \sim 0.16)$ and a
somewhat lower total mass $(\sim40M_{\odot})$.

 Another source projected very close to {$\theta^{1}$~Ori~C} is a
mid-infrared source found $\sim2\arcsec$ West of the O star. A
ring-like structure around this object is evident in thermal and
mid-IR images \citep{2004AJ....128.1254L,2005AJ....130.1763S} and it
is a VLA radio source. Source identifiers for this object include
VLA-16 \citep{1993A&AS..101..127F}, SC3 \citep{1994ApJ...433..157H},
the $L$ band source \#268 \citep{2004AJ....128.1254L}, and the mid-IR
identifiers MAX~106 \citep{2005AJ....129.1534R} and 163-323
\citep{2005AJ....130.1763S}. The spectral energy distribution of
VLA-16 was modeled as an irradiated proplyd by
\citet{2002ApJ...578..897R}, though it is not entirely clear where the
source lies along the line of sight.

Additional radial velocity components that do not turn out to
correspond to this binary pair were reported by
\citet{2002AstL...28..324V}. Understanding such radial velocity
 variations in {$\theta^{1}$~Ori~C} is complicated by the source's strong
magnetic activity, its stellar winds and a 15 day periodicity apparent
in radio and radial velocity monitoring. These winds and strong
magnetic field have been investigated by \citet{1994ApJ...425L..29W},
\citet{1996A&A...312..539S}, \citet{2002MNRAS.333...55D},
\citet{2005ApJ...628..986G}, \citep{2005ApJS..160..557S},
\citep{2006A&A...448..351S}, \citet{2006A&A...451..195W}, and most
recently \citet{2008arXiv0805.0701S}.

\subsubsection{$\theta^{1}$~Ori~D; Parenago~1889}~(HR~1896; HD~37023)
 The second most massive primary in the {Trapezium} has a B0.5V type and
a mass of $16\,M_{\odot}$ \citep{2006A&A...448..351S}. The source
displays a constant soft X-ray spectrum (lacking flares or hybrid
spectra) \citep{2005ApJS..160..557S}, suggesting it is neither
magnetic nor contains a low mass $(q<0.2)$ companion. While
\citet{2007A&A...466..649K} found evidence in the UV plane for a
companion at 8 AU separation, having $10\%$ of the luminosity of the
primary, they also point out this could be the result of a disk's
 inner edge. {$\theta^{1}$~Ori~D} is surrounded on 2 sides by the Ney \&
Allen nebula \citep{1969ApJ...155L.193N, 1990ASPC...14..301M}, a 10-20
micron diffuse structure whose origin is
unclear. \citet{2005AJ....129.1534R} present evidence that this dust
arc cannot originate simply from the ionizing photons of
 {$\theta^{1}$~Ori~C} but requires a wind from {$\theta^{1}$~Ori~D} and a
substantial reservoir of dust, which they infer to arise in a disk. On
the other hand, \citet{2005AJ....130.1763S} propose that the dust
 shell is the result of the winds of {$\theta^{1}$~Ori~D} interacting
with the PDR as it plows down into the remnant molecular cloud \citep{
  2001ARA&A..39...99O}.

\subsubsection{$\theta^{1}$~Ori~E; Parenago~1864}  The fifth brightest
 member of the {Trapezium} is {$\theta^1$~Ori~E,} a very strong X-ray
source, the second-strongest in the central cluster after
 {$\theta^1$~Ori~C}
\citep{1982Sci...215...61K}. \citet{1993A&AS...98..137F,1993A&AS..101..127F}
found the star to be a bright and variable non-thermal radio continuum
source at 2 and 6 cm. In a new detailed study,
\citet{2006AJ....132.1763H} \citep[see also][]{2006IAUC.8669....2C}
 discovered that {$\theta^1$~Ori~E} is a double-lined spectroscopic
binary consisting of two essentially identical mid-G-type components
orbiting with a period of 9.89 days. This is a revision of the lower
quality G+B5-B8 spectral type tabulated for component E in the
\citet{1954TrSht..25....1P}, \citet{1997AJ....113.1733H}, and
\citet{2000ApJ...540.1016L} catalogs but originating from observations
by \citet{1950ApJ...111...15H}. Herbig \& Griffin also summarize clues
that suggest significant optical variability of this system over the
past 200 years.

\subsubsection{$\theta^{1}$~Ori~F; Parenago~1892} Component F is a
bright point source $4.5\arcsec$ SE (PA=$120\deg$) of
 {$\theta^{1}$~Ori~C,} appearing as a blue star in Figure
\ref{trapezium.fig} (see also \ref{fig:trap_finder}); it was assigned
a spectral type B8 by \citet{1950ApJ...111...15H}. There are, however,
no subsequent published observations to confirm the spectral type of
this star. Existing high angular resolution studies imply this star is
solitary down to a separation of 60 AU
\citep{1998ApJ...500..825P,1999AJ....117.1375S}, which could perhaps
explain why it is X-ray quiet \citep{2005ApJS..160..557S}.

\subsubsection{Regarding Components G, H} During nineteenth century
 observations of the {Trapezium,} two additional sources within the
 {Trapezium} received designations as components of $\theta^{1}$~Ori
. Components G and H are now known to be proplyds (See O'Dell et~al.,
Part II) corresponding to sources 2 and 3 of
\citet{1979A&A....73...97L}.

\section{Variable Stars} \label{sec:variables}

Photometric variability is a traditional technique for identifying
young members of a known star forming region. Other methods such as
kinematics (proper motions and radial velocities), presence of lithium
in the photosphere, and stellar/circumstellar activity exhibited as
X-ray emission, UV and optical emission lines, and/or infrared excess
have also been used. This section concentrates on variability as a
young star selection technique applied to Orion, in order to both
provide historical perspective and connect to modern questions.
\begin{table}
	[!htb]

	\caption{Modern variability surveys of the Orion Nebula. In
	addition to publication details, the Table lists details of
	the observations such as whether photometry was published for
	the sources, the number of periodic stars recorded, and the
	observatory used. Column entries with the value \textit{n.p.}
	indicate that quantity was not published.} \smallskip
	\begin{center}
		{\small
		\begin{tabular}
			{llcc@{\hskip3pt}c@{\hskip3pt}rrcl}

			\tableline \noalign{\smallskip} Year & Author & Region & Filter & Phot? & $N_{\star}$ & $N_{P}$ & P(days) & Obs. \\
			\noalign{\smallskip}
			\tableline \noalign{\smallskip}

			1990 & Walker & ONC & V & Y & 5 & 4 & 0.4-3 & Lick \\
			1991 & Mandel & Trap. & $I_C$ & N & $150$ & 7 & 6-14 & VVO \\
			1992 & Attridge & $R<0.25\deg$ & $I_C$ & N & 525 & 35 & 2-17 & VVO \\
			1995 & Eaton & Trap. & $I_C$ & N & 126 & 11 & 2-35 & VVO \\
			1996 & Choi & $R<0.25\deg$ & $I_C$ & N & 525 & 50 & 2-20 & VVO \\
			1999 & Stassun & $R<1\deg$ & $I_C$ & Y & 2279 & 254 & 0.5-10 & Multiple \\
			2000 & Herbst & $R<0.25\deg$ & $I_C$ & N & 500 & 134 & 2-35 & VVO \\
			2001 & Carpenter & Orion~A & $JHK_S$ & Y & 17808 & 233 & 2-12 & 2MASS(S) \\
			2001 & Rebull & $R>0.25\deg$ & $I_C$ & Y & 3585 & 281 & 0.5-20 & MacDonald \\
			2002 & Herbst & $R<0.25\deg$ & $m816$ & Y & 1562 & 369 & 1-22 & La Silla \\
			2006 & Stassun & COUP & $BVRI$ & N & 814 & \textit{n.p.} & \textit{n.p.} & Multiple \\
			2007 & Marilli & Orion & $BV$ & Y  & 40 & 39 & 0.5-13 & Multiple \\
			2007 & Irwin & $R<0.25\deg$ & $Vi$ & Y & 2500 & \textit{n.p.} & \textit{n.p.} & INT \\
         \tableline
		\end{tabular}

		}%
	\end{center}
	\label{tab:var1}
\end{table}

\subsection{Optical \& Infrared Variable Stars} \label{sec:variables:oir}
\begin{figure}[tp*]
	\centering
	\begin{minipage}[c]{0.48\textwidth} \centering
		\includegraphics[angle=0,width=\textwidth]{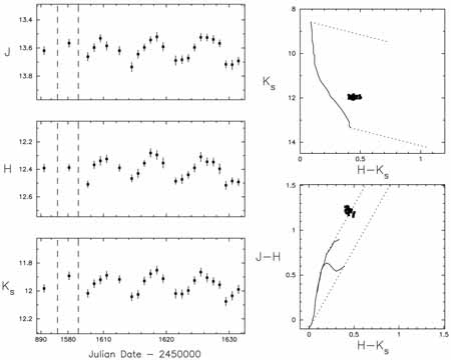}%
	\end{minipage}%
	\hspace{0.04\textwidth}%
	\begin{minipage}[c]{0.48\textwidth} \centering
		\includegraphics[angle=0,width=\textwidth]{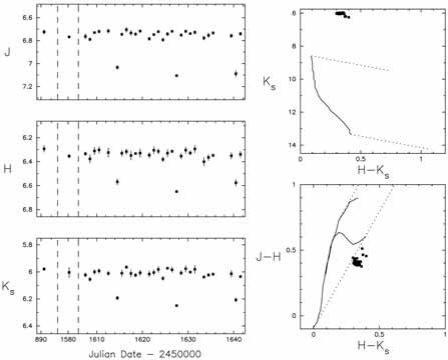}%
	\end{minipage}
 \caption{Infrared light curves of variables in the {Orion
 Nebula} from \citet{2001AJ....121.3160C}. Left: star
 {2MASS~J05342437-0452524} with a period of 8 days; Right: the
 eclipsing B star $\theta^{1}$~Ori~B$_{1,5}$ {(BM~Ori);} see also
	Sect.~\ref{sec:ob:notes}. \label{fig:var1}}

\vspace{5mm}
	\begin{minipage}[c]{0.48\textwidth} \centering
		\includegraphics[angle=0,width=\textwidth]{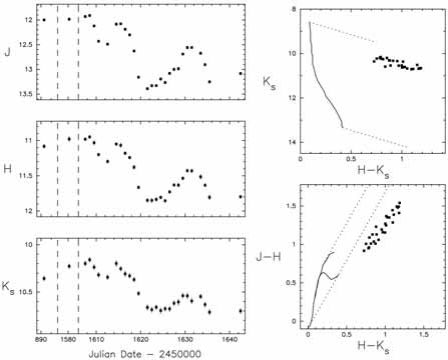}%
	\end{minipage}%
	\hspace{0.04\textwidth}%
	\begin{minipage}[c]{0.48\textwidth} \centering
		\includegraphics[angle=0,width=\textwidth]{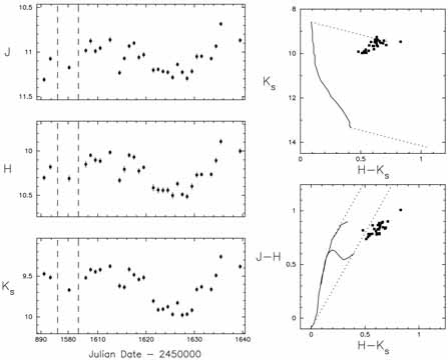}%
	\end{minipage}
	\caption{Infrared light curves of irregular, long term
 variables in the {Orion Nebula} from
 \citet{2001AJ....121.3160C}. {Left: Parenago~2171} {(AO~Ori);}
 Right: Parenago~1617 {(YY~Ori).} \label{fig:var2}}
\end{figure}%

\subsubsection{Historic Studies} The Orion Nebula Cluster is a rich
collection of variable stars. According to \citet{1982NYASA.395...64H}
 the first variable star identified within the {Orion Nebula} was found
 by W.C. Bond in or around 1848, and is now known as {AF Ori}
\citep{1982NYASA.395...64H}. Herbig notes in 1982 that in addition to
 the known eclipsing binary members of the {Trapezium} itself
 {($\theta^{1}$~Ori~A} and {$\theta^{1}$~Ori~B)} there were 17 named
 variables within the inner {Orion Nebula Cluster} {(Trapezium} region) and
several hundred within the larger nebula. This area is known today to
contain several thousand young stars, the vast majority of which are
known variables based on modern optical CCD or infrared array
monitoring surveys with the cadence and sensitivity to detect them.

The first extensive catalog of variable stars in (and near) the Nebula
was derived (pretty much solely) by H.~S. Leavitt using a large number
($\sim20$) of plates in the Harvard collection. The catalog was
published subsequently in \citet{1904ApJ....19..289P}; it included
approximately 70 variables and 30 suspected variables with minimum and
maximum brightness quoted. Additional notable variable identification
work was performed by \citet{1935POLyo...1...12B},
\citet{1946POBol...5a...3R, 1956MmSAI..27..335R} and by
\citet{1947QB835.P23......}\footnote{Another significant study
performed by P.~P.~Parenago was the development of the General Catalog
of Suspected Variables with B.~V.~Kukarkin}. These and other stars
were followed up spectroscopically, by
e.g. \citet{1946PASP...58..366G}, \citet{1950ApJ...111...15H}, and
\citet{1953ApJ...117...73H}. The analogy to ``the objects of Joy"
(1945), variable emission-line stars associated with bright or dark
nebulosities and now known as T~Tauri stars, was debated in these
early works on the Orion variables, and eventually they were accepted
as such, especially following photoelectric photometry studies such as
that of \citet{1969ApJ...155..447W}.

With improved technology, the variability could not only be identified
and crudely characterized, but monitored in
detail. \citet[][]{1990PASP..102..726W} undertook the first
comprehensive study of rotational flux modulation in the Orion
population, following up his earlier work on rotational broadening of
spectral lines. He sought differences between the ``weak" and
``classical" emission stars as well as to understand the periods and
nature of the surface disturbances causing the rotational
modulation. He reported results on 5 stars.

\subsubsection{Modern Monitoring Surveys} With the advent of CCD detectors variability could be monitored and quantified for large numbers of members rather than just individual stars. In particular, there was interest in searching for periodic variables. Soon following the \citet{1990PASP..102..726W} paper was a series of papers by Herbst and collaborators \citep{1991ApJ...383L..75M,1992ApJ...398L..61A,1995AJ....110.1735E, 1996AJ....111..283C}. Table \ref{tab:var1} documents and compares the results from these modern survey beginning with Walker (1990).  These works built up a database of periods in the Orion Nebula Cluster and presented the angular velocity distribution for the young stellar population. Of great interest was an apparent bimodal period distribution, which was interpreted in the context of ``disk locking" whereby the slow rotators (with periods around 8 days) were assumed to be kept rotating slowly due to interaction between the stellar magnetosphere and the Keplerian accretion disk, while the more rapid rotators (with periods around 3 days) were assumed free of such disk interactions. The gap or valley in the period distribution was interpreted as rapid evolution between the disk-regulated to disk-free scenarios.

Additional work on optical
\citep{1999AJ....117.2941S,2000AJ....119..261H,2001AJ....121.1676R,
2001ApJ...554L.197H,2002A&A...396..513H, 2006ApJ...649..914S,
2007ApJ...660..704S, 2007A&A...463.1081M, 2007MNRAS.380..541I} and
infrared \citep{2001AJ....121.3160C} variability broadened the
discussion considerably. Included were not only the periodic stars
with either cool spots (due to photospheric inhomogeneities) or hot
spots (due to accretion columns), but also eclipsing systems, and
irregular variables dominated by accretion or extinction
effects. \citet{2002A&A...396..513H} claim that essentially every star
(of 1500 monitored) is optically variable at the $>$1\% level with
half of the stars having peak-to-peak brightness variables at I-band
larger than 0.2 mag. \citet{2001AJ....121.3160C} found the same mean
peak-to-peak amplitudes at near-infrared $(JHK)$ wavelengths.

The mass dependence and the activity/disk dependence of the
periodicity was discussed explicitly in several of the above works as
well as in
\citet{2001AJ....121.1003S,2004AJ....127.3537S,2006ApJ...649..914S}
and \citet{2006ApJ...646..297R}. As a result of the young age and
 large amount of available rotation data, the {ONC} has become the de
facto cluster setting the initial conditions for models of stellar
angular momentum evolution.  A compendium of the literature for the
 {Orion Nebula} and flanking field periodic stars is provided by
\citet{2006ApJ...646..297R}.  Their results on variation in near-IR
color with periodicity and long term irregularity provide clues to the
origin of the variability, whether it be from time variable hot spots,
from changes in line of sight extinction or due to changes in the
geometry and rate of mass accretion. Near-IR light curves that
illustrate different kinds of longer term variability are given in
Figures \ref{fig:var1} and \ref{fig:var2}.

\begin{figure}
	[ht*]

	\centering
	\includegraphics[angle=0.,totalheight=3in]{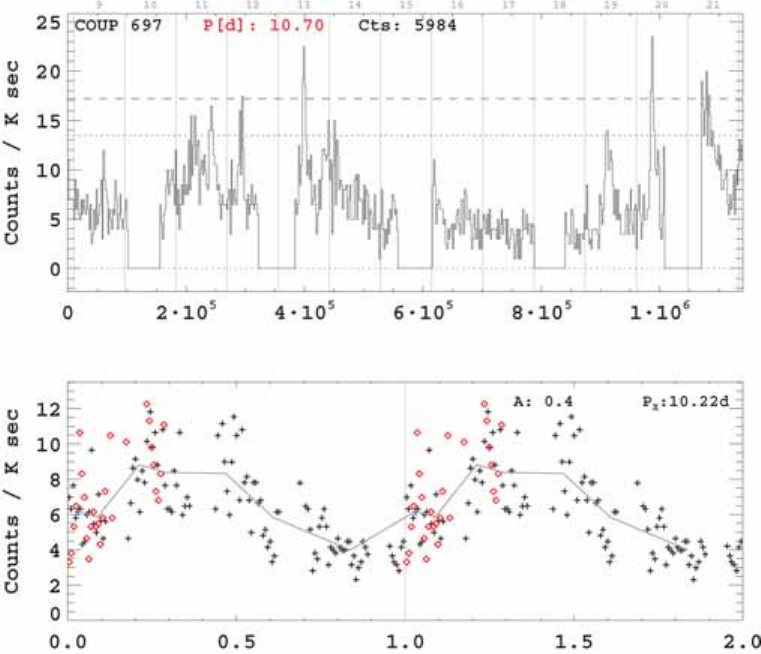}

 \caption{Top: The lightcurve of the {COUP 697} source, for which
	a modulation analysis yields an X-ray period similar to 10.2
	days, as seen for the visible light starspots. Bottom: Here is
	the same lightcurve, after the flares are removed, shown
	folded with the 10.2 day period. Reproduced from
	\citet{2005ApJS..160..450F}. \label{xray:fig8}}

\end{figure}
\subsection{X-ray Flares and Rotational Modulation of Variable Stars in Orion} \label{sec:variables:xray}

Concerning the geometry of the magnetic structures producing X-ray
emission of T~Tauri stars, solar-type coronal loops associated with
multi-polar fields rooted in the stellar surface are probably the
dominant source of the observed X-ray emission. The analyses of
simultaneous optical and X-ray data for about 800 COUP stars show that
optical and X-ray variability are very rarely time correlated
\citep{2006ApJ...649..914S}, but the strong correlation between
optical variability and X-ray luminosity is present
\citep{2007ApJ...660..704S}. This fits into the picture in which
sights of optical variability may represent footprints of X-ray
emitting coronal magnetic structures of complex magnetic
topologies. Studying 233 Orion-COUP PMS stars with known rotational
periods, \citet{2005ApJS..160..450F} detected X-ray rotational
modulation in 16 stars, indicating that the stellar surface has
similar inhomogeneities in their photospheres and their coronae; it
also suggests that the coronae in these cases are relatively compact
$(\ll R_{\star}$). An example of X-ray rotational modulation with the
optical period is shown in Figure \ref{xray:fig8} for
{COUP source \#697.}
In 7 other cases, \citet{2005ApJS..160..450F} find X-ray periods
equal to half of the optical periods, suggesting two bright
hemispheres in the X-ray corona. But for a number of COUP stars with
very powerful and hot (peak temperatures $> 100$ MK) flares,
\citet{2005ApJS..160..469F} derive the length of the magnetic
structures to be much greater than the stellar radius. These
structures are probably too large to be stable, particularly to
centrifugal forces as the star rotates, in a solar-type geometry. They
most likely extend from the star to the inner edge of a protoplanetary
disk.

\vspace{0.5cm}

{\bf
Acknowledgements. } Bo Reipurth provided numerous suggestions and
comments that added to the breadth and detail of this review. We thank
David Mihalyfy for Russian translations of the Parenago text and Laura
Nasrallah for translations of early 20th century German texts. We
acknowledge Robert Gendler for permitting the reproduction of his
large scale image of the northern Orion A cloud shown in Figure 1.
This research has made use of NASA's Astrophysics Data System, the
SIMBAD database, operated at CDS, Strasbourg, France and SAOImage DS9,
developed by the Smithsonian Astrophysical Observatory. \\


\end{document}